\documentclass[twocolumn,tighten,times]{aastex61}

\usepackage{latexsym}
\usepackage{amssymb}
\usepackage{amsmath}
\usepackage{graphics}
\usepackage{graphicx}
\usepackage{natbib}
\usepackage{subfigure}
\usepackage{soul}

\usepackage{bm} 
\usepackage{url}

\received{8 Aug 2017}
\accepted{28 Apr 2018}

\submitjournal{ApJ}

\shorttitle{GalMod}
\shortauthors{Pasetto et al.}

\begin{document}

\title{GalMod: a Galactic synthesis population model \\
\href{www.galmod.org}{www.GalMod.org}}

\correspondingauthor{Stefano Pasetto}
\email{spasetto@carnegiescience.edu}

\author{Stefano Pasetto}
\affil{The Observatories of the Carnegie Institution for Science, 813 Santa Barbara St., Pasadena, CA 91101, United States of America}

\author{Eva K. Grebel}
\affiliation{Astronomisches Rechen-Institut, Zentrum f\"{u}r Astronomie der Universit\"{a}t Heidelberg, M\"{o}nchhofstrasse 12-14, D-69120 Heidelberg, Germany}

\author{Cesare Chiosi}
\affiliation{Department of Physics \& Astronomy, "Galileo Galilei", University of Padua, Vicolo dell'Osservatorio 2, Padua, Italy}

\author{Denija Crnojevi\'c}
\affiliation{Department of Physics \& Astronomy, Texas Tech University, Box 41051, Lubbock, TX 79409-1051, United States of America}

\author{Peter Zeidler}
\affiliation{Space Telescope Science Institute (STScI), 3700 San Martin Dr, Baltimore, MD 21218, United States of America}

\author{Giorgia Busso}
\affil{Institute of Astronomy, University of Cambridge, Madingley Road, Cambridge CB3 0HA, United Kingdom}

\author{Letizia P. Cassar\`{a}}
\affiliation{National Observatory of Athens, Institute for Astronomy, Astrophysics, Space Applications and Remote Sensing, Ioannou Metaxa and Vasileos Pavlou GR-15236, Athens, Greece}

\author{Lorenzo Piovan}
\affiliation{Department of Physics \& Astronomy, "`Galileo Galilei"', University of Padua, Vicolo dell'Osservatorio 2, Padua, Italy}

\author{Rosaria Tantalo}
\affiliation{Department of Physics \& Astronomy, "`Galileo Galilei"', University of Padua, Vicolo dell'Osservatorio 2, Padua, Italy}

\author{Claudio Brogliato}
\affiliation{Clover-lab, Valene rd, Salo, I-25087 Brescia, Italy}

\begin{abstract}
We present a new  Galaxy population synthesis Model (GalMod). GalMod is a star-count model featuring an asymmetric bar/bulge as well as spiral arms and related extinction. The model, initially introduced in \citet{2016MNRAS.461.2383P}, has been here completed with a central bar, a new bulge description, new disk vertical profiles and several new bolometric corrections.

The model can generate synthetic mock catalogs of visible portions of the Milky Way (MW), external galaxies like M31, or N-body simulation initial conditions. At any given time, e.g., a chosen age of the Galaxy, the model contains a sum of discrete stellar populations, namely bulge/bar, disk, halo.
These populations are in turn the sum of different components: the disk is the sum of spiral arms, thin disks, a thick disk, and various gas components, while the halo is the sum of a stellar component, a hot coronal gas, and a dark matter component.
The Galactic potential is computed from these population density profiles and used to generate detailed kinematics by considering up to the first four moments of the collisionless Boltzmann equation. The same density profiles are then used to define the observed color-magnitude diagrams in a user-defined field of view from an arbitrary solar location. Several photometric systems have been included and made available on-line and no limits on the size of the field of view are imposed thus allowing full sky simulations, too. Finally, we model the extinction adopting a dust model with advanced ray-tracing solutions.

The model's web page (and tutorial) can be accessed at \href{www.galmod.org}{www.GalMod.org} and support is provided at Galaxy.Model@yahoo.com
\end{abstract}


\section{Introduction}\label{Intro}
Stars are one of the key visible constituents of our Universe. To study what governs the structure and evolution of our Galaxy, the Milky Way (MW), we need to observe and understand the processes that govern the formation, evolution, and motion of its stars over their evolutionary time-scales. This process of research implies the comprehension of stellar energy feedback (by UV emission, stellar winds, or supernova explosions), the yields of chemically-enriched material into the interstellar medium, the stellar end-products (i.e., white dwarfs, neutron stars, and black holes) and what determines stellar motions in space.

The process of collecting detailed data on our Galaxy represents the first step in this work of archaeological research, and we are now living in a "golden era" for Milky Way archeology. The many surveys that scan the sky to unveil the MW's secrets nowadays provide data covering the largest possible spectrum of frequencies: from the radio continuum \citep[e.g., ][]{1982A&AS...47....1H,1995MNRAS.277...36D}, to the HI/21-cm emission line \citep[e.g.,][]{1986A&AS...66..373K} to molecular ${{H}_{2}}$ \citep[e.g.,][through CO observations]{1995MNRAS.277...36D} up to X-rays \citep[e.g.,][]{1997ApJ...485..125S} or $\gamma $-rays  \citep[][]{1999ApJS..123...79H}.

However, it is in the optical and infrared bands where we are going to focus our attention in this work because of the growing interest in these bands (e.g., for the forthcoming Large Synoptic Survey Telescope, or the James Webb Space Telescope) and the tight connection with phase-space studies (e.g., thanks to the Gaia satellite).

We propose a model aimed to extract the most relevant information from optical and/or infrared surveys. We look at the data products of past and present projects or surveys such as SDSS/APOGEE \citep{2015ApJS..219...12A,2014ApJS..211...17A}, the $2\mu m$ All-Sky Survey \citep[2MASS,][]{2006AJ....131.1163S}, the UKIRT Infrared Deep Sky Survey \citep[UKIDSS,][]{2007MNRAS.379.1599L}, the Visible and Infrared Survey Telescope for Astronomy \citep[VISTA,][]{2010Msngr.139....2E} or VISTA-Variables in the Via Lactea \citep{[VVV][]{2010NewA...15..433M}}, the Galactic Archaeology with HERMES survey \citep[GALASH,][]{2015MNRAS.449.2604D}, the AAVSO Photometric All-Sky Survey \citep[APASS,][]{2014CoSka..43..518H}, the Optical Gravitational Lensing Experiment \citep[OGLE,][]{2015AcA....65....1U}, the RAdial Velocity Experiment \citep[RAVE,][]{2006AJ....132.1645S} but also to future projects as Gaia and Gaia-ESO Survey \citep{2016A&A...595A...2G, 2012Msngr.147...25G}, the Large Sky Area Multi-Object Fiber Spectroscopic Telescope \citep[LAMOST,][]{2012RAA....12.1197C}, the 4-meter multi-object spectroscopic telescope \citep[4MOST,][]{2012SPIE.8446E..0TD}, and so forth.

Because of the possibility to fine-tune the Galaxy model for other spiral galaxies, we will briefly mention the opportunity to model M31 surveys like the Pan-Andromeda Archaeological Survey \citep[PAndAS,][]{2009Natur.461...66M} as well. At the same time, we will base our phase-space description mostly on our current understanding of the MW from star count surveys, radial velocity maps, standard candle distance indicators and the most recent phase-space data provided by Gaia \citep[e.g.,][]{2016A&A...595A...2G}. This is because the best available data are indeed those for the MW.

It is in this context of increasing complexity of the MW's picture that we want to develop up-to-date analytical instruments that help us to extract accurate theoretical information from these optical/infrared, chemical, and phase-space surveys.

We introduce here an advanced mock catalog generator and relative model (hereafter, GalMod) that we make freely available to the scientific community through a dedicated web interface at \href{www.galmod.org}{www.GalMod.org}.

GalMod is a galaxy modeler software, highly tunable, that generates synthetic catalogs of the MW. It consists of a synthetic color-magnitude diagram (CMD) generator, a kinematical model, and a MW global potential generator to provide photometry, stellar parameters, proper motions, radial velocities, and indirect MW global potential indicators.

The model makes extensive use of the concept of multiple stellar populations (see Sec.\ref{Theory} for a brief review) to define the MW as a non-linear superposition of discrete components: bulge, bar, thin disks, spiral arms, thick disks, stellar halo, dark matter, and coronal halo. We define each population with a set of parameters that characterize its mass distribution, its metallicity distribution, and its phase-space distribution at a given instant in time. The model outputs a mock catalog directly in the space of observations for any field of view (FoV) desired (even full-sky), thus simulating real observational data.

The origin of this type of modeling rests on an approach using the fundamental equation of stellar statistics, the star-count equation \citep[e.g.,][]{1898PA......5..544S,1953stas.book.....T}, pioneered by \citet[][]{1980ApJS...44...73B} \citep[see also ][]{1984ApJS...55...67B,1984ApJ...276..169B,1985ApJS...59...63R}. In these early and fundamental works, the concept of the stellar population relates to the photometry alone without phase-space treatment. The first works attempting a global model generalization can be traced to \citet[][]{1986A&A...157...71R,1990ApJ...357..435C} and \citet[][]{1996AJ....112..655M}. Finally, the first attempt to account for the MW potential constraints on a star-count type modeling technique is due to \citet{1987A&A...180...94B}. In the latter study we can find both the first use of the Poisson equation for exponential density profiles \citep[e.g., in the form presented by][]{1986ApJ...309..472Q} related to the moments of the collisionless Boltzmann equation, and the core of the consistency cycle of the Besan\c{c}on model \citep[see][Sec. 8]{2016MNRAS.461.2383P}.

\subsection{Why a new star-count model?}
Currently, the only Galactic model available in the literature including both a photometric and phase-space description are the  Besan\c{c}on model \citep{2014A&A...564A.102C} and the Galaxia \citep{2011ApJ...730....3S} model (a Besan\c{c}on 
based model with extended capabilities). An extensive analysis of the Besan\c{c}on model in comparison with GalMod has been presented already in \citet{2016MNRAS.461.2383P}. Here we want to describe how GalMod was developed to try to surpass some of the Besan\c{c}on model limits. GalMod has no limit on the size of the field generated (even full sky is allowed), a feature that can be especially appreciated in the era of the current and upcoming wide-field surveys and already explored in Besan\c{c}on-based codes as Galaxia \citep[][]{2011ApJ...730....3S}. GalMod includes non-axisymmetric features such as spiral arms as well as a tilted bar (see next sections). Finally, following the approach used by codes as Trilegal model \citep[e.g.,][]{2005A&A...436..895G} and Galaxia, GalMod includes a wider range of photometric bands than Besan\c{c}on, thus allowing the generation of CMDs with assigned passbands instead of forcing the user to adopt transformation equations between photometric systems \citep[e.g.,][]{2006A&A...460..339J}. In this context, we need to remark that in the Besan\c{c}on model the vertical scale height is constrained by the vertical velocity dispersion using an iterative procedure. In GalMod the scale height does not depend on the adopted vertical velocity dispersion, as GalMod does not attempt to establish dynamical equilibrium in the vertical direction. However, GalMod through its web interface offers the user the freedom to specify their own scale heights, in other words, the user can impose their own consistency conditions.

The Trilegal model has an exceptionally large number of photometric systems but MW potential and kinematics are not implemented. This exposes the user to the risk of generating MW models that produce a good CMD fit in a given direction, but whose density profiles correspond to a MW potential that generates unrealistic rotation curves, unrealistic mass distributions or Oort constants, or other unrealistic diagnostic parameters. GalMod computes the gravitational potential for the density profiles adopted and provides the user with a complete set of diagnostic parameters on the Galactic model realized (e.g., rotation curve, Oort function, etc.). This is an important extra feature that GalMod (and Besan\c{c}on) offers with respect to Trilegal.

Furthermore, it is worth to distinguish the tools available through web-interface at \href{www.galmod.org}{www.GalMod.org} from the galactic model introduced in \citet{2016MNRAS.461.2383P} and in the present paper. In the GalMod model, convergence to observational data is based on machine learning techniques that are not available through a web interface. In particular, in GalMod we employed genetic algorithms \citep[see Sec. 5 in ][]{2002A&A...392.1129N,2016MNRAS.461.2383P} which seek convergence to a set of data by use of a few dynamical estimators directly connected to the global galactic potential (e.g, the rotation curve, the vertical force on the plane, the Oort function, etc.) and synthetic distributions of, e.g., radial velocities, proper motions, color-magnitude diagrams, $T_{\text{eff}}$, etc.(\footnote{In this respect we need to rephrase a sentence in Sec. 8, first col. pg. 2405 of \citet{2016MNRAS.461.2383P}: "The dynamical consistency is clearly poorer than that in our kinematical model." and upgrade it with "The approaches used by GalMod and Besan\c{c}on models are quite different, and it is not easily quantifiable whether the genetic codes used by GalMod and based on several dynamical estimators can lead to better or worse consistency than the iterative cycle used by Besan\c{c}on."}). In this paper, we will introduce only the GalMod features available through web-interface, i.e. no data fitting procedure is presented.

Finally, not forcing the modeled galaxy at the center of any coordinate system, we can use GalMod to simulate external objects like M31, or a dwarf galaxy. At the moment, GalMod can be used to initialize phase-space information and star formation histories for N-body simulations. We will detail these new features of GalMod in Sec.\ref{Test}.

The organization of this paper is as follows. In Sect.\ref{Theory} we will briefly review our concept of stellar populations; in Sect.\ref{StellarProfiles} we review the stellar population model adopted in \citet[][]{2016MNRAS.461.2383P} and extensively discuss the bulge model introduced in GalMod here; in Sect.\ref{Test} a few test cases are presented, and in Sec.\ref{Conclusions} we conclude.

\section{Theory of multiple composite stellar populations}\label{Theory}
The stars in a galaxy can be approximated by discrete units, which evolve alone, in couples, or in groups, interacting with the interstellar medium (ISM) and under the influence of a common gravitational potential. Here we  proceed to describe them within a framework introduced in \citet[][]{2012A&A...545A..14P} and \citet[][]{2016MNRAS.461.2383P}, and completed in the companion paper \citet[][in prep.]{Pasetto2018T}.

\begin{figure*}
\plotone{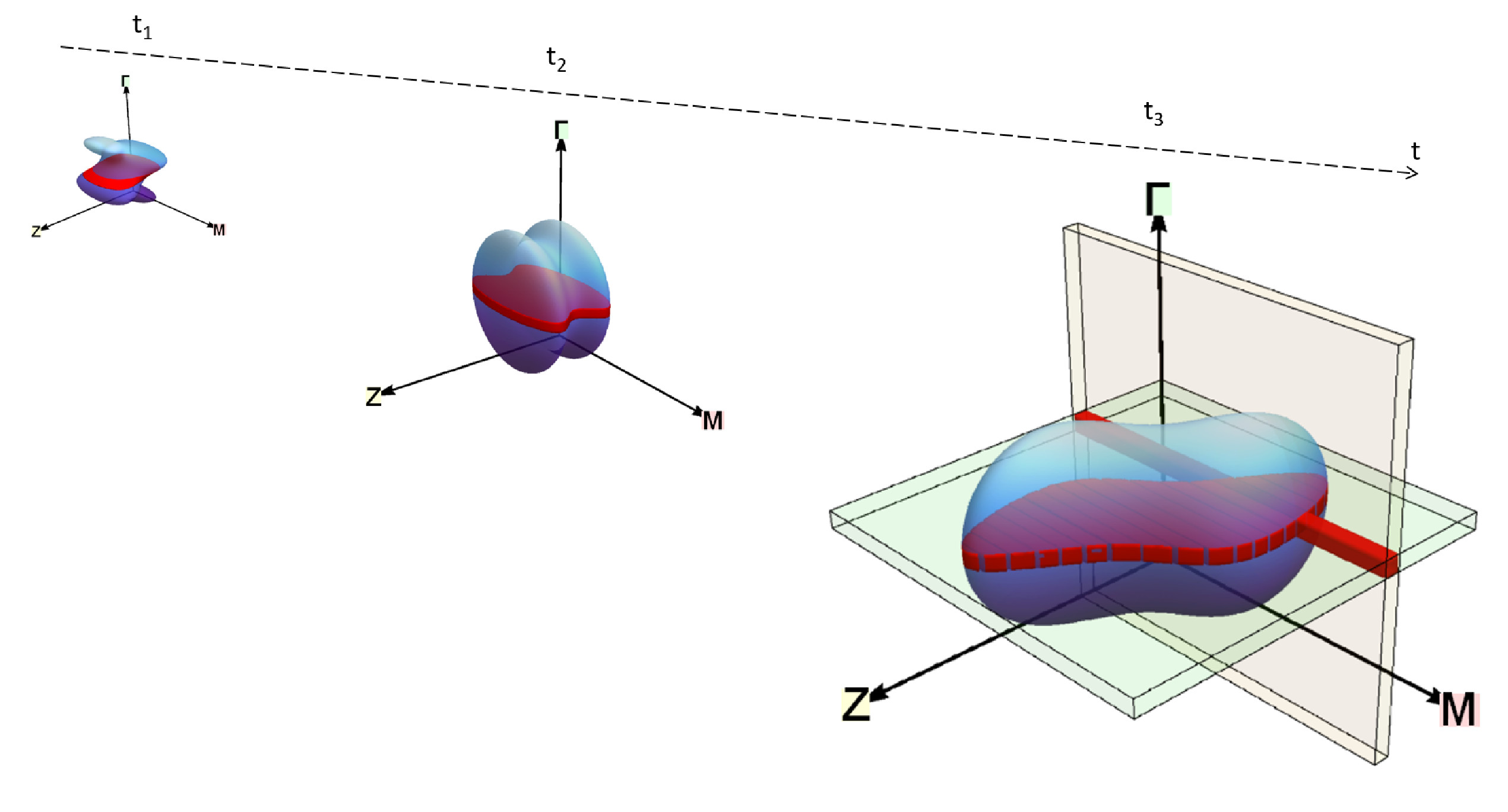}
\caption{SSPs foliate the CSPs at every time ${{t}_{1}}$ , ${{t}_{2}}$ ,..., ${{t}_{n}}$. In blue, the manifold of a CSP, e.g., a galaxy is depicted. In red at ${{t}_{1}}$ , ${{t}_{2}}$ , ..., ${{t}_{n}}$ the CSP of a stellar population, e.g., the thin disk is plotted. As the time passes the CSP of the thin disk evolves as does the CSP of the galaxy. The CSP of the thin disk represents a single point in $\bm{\gamma}$ with a distribution in $Z$ and $M$, i.e., a slab orthogonal to the $\bm{\gamma}$ axis and intersecting the blue manifold at constant $\bm{\gamma}$  in this cartoon. This slab can be foliated in orthogonal disjointed SSPs shown as red cubes in the $t={{t}_{3}}$. On the instant $t={{t}_{3}}$, an elemental SSP unit, i.e., a leaf of the foliation, is highlighted for constant $\bm{\gamma}$ and $Z$.}
\label{EvolTemp}
\end{figure*}

We are interested only in the modeling of (quasi)-stationary states, i.e., dynamical-equilibrium states of a galaxy at a fixed time $t$. A composite stellar population, or simply CSP, is a set of stars born at different times $t$, positions $\bm{x}$, with different velocities $\bm{v}$, masses $M$, and chemical compositions $Z$. We describe the CSPs with a continuous and differentiable distribution function defined in its domain (say e.g., its existence space manifold) $\mathbb{E}\equiv M\times Z\times \bm{\gamma }$, given by the Cartesian product of the space of stellar mass values, $M\subset \mathbb{R}_{0}^{+}$, metallicity $Z\subset \mathbb{R}_{0}^{+}$, and phase-spaces $\bm{\gamma }\subset {{\mathbb{R}}^{6}}$ for a collisionless galaxy system ($\mathbb{R}_{0}^{+}$ refers to positive real numbers including the zero). Following \citet[][]{2012A&A...545A..14P} \citep[see also][Sec.2]{2016MNRAS.461.2383P},  we foliate the existence space $\mathbb{E}$ of a CSP at each time $t$ in elemental units called single (or simple) stellar populations (SSP). A SSP, i.e., a ``leaf'' of the foliation, is a subset of $\mathbb{E}$ at constant $\bm{\gamma}$ and $Z$. A CSP can be foliated in SSPs, i.e., the existence manifold $\mathbb{E}$ of the CSP can be described at every time $t$ as a union of disjoint parallel sub-manifolds, the SSPs. Fig.\ref{EvolTemp} shows a cartoon representing the concept of this formalism.

The relative contributions of two or more stellar populations to the total galactic mass depend at every instant on their relative density distribution and their star formation history. Specifically, the distribution of the total mass of each stellar population in the configuration space depends on its density profiles. If we consider an arbitrary portion of the configuration space, then the relative amount of mass due to a specific stellar population (e.g., halo, disk, and so forth) depends on the relative importance of the different density profiles at that position. Moreover, within each of the density profiles for a single CSP (e.g., the thin disk), the amount of mass is distributed among the stars depending on their underlying star formation history and initial mass function. As time passes, the stellar population evolves in its existence space $\mathbb{E}$ (see Fig.\ref{EvolTemp}). This motion is due to the stars moving in $\mathbb{E}$, i.e., contemporaneously in the phase-space and in the mass-metallicity space.

The way in which these stars are distributed in $\mathbb{E}$ at every time $t$ determines their number in each observed FoV, a result that is obtained in a completely analytical way as a corollary of the multiple stellar population consistency theorem \citep[hereafter MSP-CT, see][]{Pasetto2018T}. This theorem grants the existence of a solution for the classical condition of consistency between the gravitational mass/potential (ruling dark matter, ISM and stellar dynamics) and the total stellar mass (as distributed in each stellar CSP by initial mass function and star formation laws), when at least two CSPs are considered in a galaxy.

In the assumption that we can split the present-day mass function $\hat{\xi }=\hat{\xi }\left( M;t \right)$ into an initial mass function (IMF), $\xi =\xi \left( M \right)$, univocally dependent on the mass $M$, and a star formation rate (SFR), $\psi =\psi \left( t \right)$, carrying the temporal dependence, we can write $\hat{\xi }\left( M;t \right)=\xi \left( M \right)\psi \left( t \right).$ Hence, we can prove \citep[MSP-CT ,][]{Pasetto2018T} that the requirement that the total stellar dynamical mass of the galaxy
\begin{equation}\label{(2.2)}
		{{M}_{\text{tot}}}=\sum\limits_{c}^{{}}{\int_{{}}^{{}}{{{\rho }_{c}}\left( \bm{x} \right)d\bm{x}}},
\end{equation}
(where ${{\rho }_{c}}$ is the density of the $c^{\text{th}}$-CSP) equals at every instant $T$ the sum of the stellar masses given by a present-day mass function, i.e.,
\begin{equation}\label{(2.3)}
	{M_{\text{tot}}}=\sum\limits_{c}^{{}}{\int_{{{M}_{l}}}^{{{M}_{u}}}{dM}M{{\xi }_{c}}\left( M \right)\int_{{{t}_{0}}}^{T}{{{\psi }_{c}}\left( t \right)dt}},
\end{equation}
can \textit{always be fulfilled }once the normalization coefficients ${{\xi }_{0,c}}$  and ${{\psi }_{0}}$ of the IMF and the SFR (say $\xi ={{\xi }_{0}}{{I}_{\Xi ,c}}\left( M \right)$ and $\psi ={{\psi }_{0,c}}{{I}_{\Psi ,c}}\left( t \right)$, where ${{I}_{\Xi ,c}}\left( M \right)$ and ${{I}_{\Psi ,c}}\left( t \right)$ are functions only of mass and time, respectively) are provided by(\footnote{Note that we do not normalize the IMF to 1, hence a normalization factor $\xi_0$ is necessary.}):
\begin{equation}\label{(2.4)}
	{{\xi }_{0}}=\frac{{{M}_{\text{tot}}}}{\sum\limits_{c}^{{}}{{{\psi }_{0,c}}{{I}_{\Psi ,c}}{{I}_{\Xi ,c}}}},
\end{equation}
and(\footnote{Note that $c$, $i$, $j$ are mute indexes running always from 1 to $N_p$, thus we can safely omit their extremes.})
\begin{equation}\label{(2.5)}
		{{\psi }_{0,c}}=\frac{{{M}_{c}}\prod\limits_{j\ne c}^{{}}{{{I}_{\Xi ,j}}{{I}_{\Psi ,j}}}}{\sum\limits_{i}^{{}}{{{M}_{i}}\prod\limits_{j\ne i}^{{}}{{{I}_{\Xi ,j}}{{I}_{\Psi ,j}}}}}.	
\end{equation}
For the IMFs the normalization coefficient is one, ${{\xi }_{0}}$, because the total mass of all the CSPs is a single value; the normalization coefficients of the SFR, ${{\psi }_{c}}$, are different for each CSP to allow for varying SFRs depending on the different star formation histories of the different CSPs.

We provide GalMod with four profiles for the SFR and four for the IMF; furthermore we  assume that outside the time interval pertinent to each CSP the star formation profile is identically null(\footnote{In \citet{Pasetto2018T}, we present the exact IMF and SFR profiles implemented in GalMod. They differ from what is shown here for the explicit presence of a function that nullifies the integrals outside a range of interest (called Tori-function); such a feature is necessary but omitted here for the sake of simplicity. The limits of the integrals are changed accordingly assuming that for each CSP the $t_1$ and $t_2$ can be different.}). These profiles define the integral functions ${{I}_{\Xi ,c}}\left( t \right)$  and ${{I}_{\Psi ,c}}\left( t \right)$ implicitly:
\begin{itemize}
	\item \textbf{Constant SFR}. This profile represents a constant star formation between two arbitrary instants, i.e., $t\in \left[ {{t}_{1}},{{t}_{2}} \right]$ with ${{t}_{1}}>0$ and ${{t}_{2}}<{{t}_{G}}=13.8$ Gyr  (the age of the Universe):
	      \begin{equation}\label{(2.6)}
	      	\psi\left( t \right) = {{\psi }_{0}} \Psi \left( t \right)={{\psi }_{0}}\times 1={{\psi }_{0}}=\text{const}\text{.}
	      \end{equation}
	      The integral of Eq.\eqref{(2.3)} between two arbitrary times hence reads
	      \begin{equation}\label{(2.7)}
	      	{\psi _0}{I_\Psi } ={\psi _0} \int_{{t_1}}^{{t_2}} {dt\Psi \left( t \right)}  = {\psi _0}\left( {{t_2} - {t_1}} \right).
	      \end{equation}
	      	
	\item \textbf{Exponential SFR}. We allow star formation profiles that permit to model increasing or decreasing phases of star formation activity in the galaxy or exponential bursts. The utility of these patterns goes beyond the modeling of the MW: they can be used to test peculiar synthetic CMDs, to model N-body simulations, to model M31, or any dwarf galaxy. For these reasons, we introduce the profile:
	      \begin{equation}\label{(2.8)}
	      	\psi \left( t \right)={{\psi }_{0}}e^{  -\frac{t}{{{h}_{\tau }}} }, 	
	      \end{equation}
	      where ${{h}_{\tau }}\in \mathbb{R}\backslash \left\{ 0 \right\}$ is the non-null time scale length of an exponentially in/decreasing profile. We find for the integral in Eq.\eqref{(2.3)} (with ${{t}_{2}}>{{t}_{1}}>0$):
	      \begin{equation}\label{(2.9)}
	      	{\psi _0}{I_\Psi } = {\psi _0}\int_{{t_1}}^{{t_2}} {dt\Psi \left( t \right)}  = {\psi _0}{h_\tau }\left( {{e^{ - \frac{{{t_1}}}{{{h_\tau }}}}} - {e^{ - \frac{{{t_2}}}{{{h_\tau }}}}}} \right).
	      \end{equation}
	      	
	\item \textbf{Linear SFR}. To allow the exploration of a wide range of parameters, we propose a linear pattern for the SFR between two assigned times, i.e., a SFR profile given by
	      \begin{equation}\label{(2.10)}
	      	\psi \left( t \right)={{\psi }_{0}}\left( \frac{{{\psi }_{{{t}_{2}}}}-{{\psi }_{{{t}_{1}}}}}{{{t}_{2}}-{{t}_{1}}} \right)\left( t-{{t}_{1}} \right)+{{\psi }_{{{t}_{1}}}}. 	
	      \end{equation}
	      Eq.\eqref{(2.3)} (with ${{t}_{2}}\ne {{t}_{1}}$, ${{\psi }_{{{t}_{2}}}},{{\psi }_{{{t}_{1}}}}$ all positive numbers and ${{\psi }_{{{t}_{i}}}}$ corresponding to the SFR at ${{t}_{i}}$) is then integrated entirely analytically as
	      \begin{equation}\label{(2.11)}
	      	{\psi _0}{I_\Psi } = \int_{{{t}_{1}}}^{{{t}_{2}}}{dt{{\psi }_{0}}\Psi \left( t \right)}=\frac{{{\psi }_{0}}}{2}\left( {{\psi }_{{{t}_{1}}}}+{{\psi }_{{{t}_{2}}}} \right)\left( {{t}_{2}}-{{t}_{1}} \right).	
	      \end{equation}
	      Considered that GalMod allows us to input several stellar disk components, we can combine these linear profiles to virtually achieve any composite disk SFR profile and related age-metallicity relation.
	      	
	\item \textbf{Rosin-Rammler SFR}. Considering chemical models of the MW \citep[e.g.,][]{1980A&A....83..206C,2012ceg..book.....M,2012A&A...548A..60G}, it is of interest to study a SFR family of profiles that allows a rapid increase of the SFR with time, followed by its shallow decline to the present day. This can be easily achieved by a functional form of the type (\footnote{The name is taken from the popular statistical Weibull distribution originally used by Rosin \& Rammler to describe a particle size distribution  \citep{rosin1933laws}. The integral of the Rosin \& Rammler function relates easily to the gamma function.}):
	      \begin{equation}\label{(2.12)}
	      	\psi \left( t \right)={{\psi }_{0}}{{t}^{\beta }}{{e}^{-\tfrac{t}{{{h}_{\tau }}}}}, 	
	      \end{equation}
	      under the condition that ${{t}_{2}}>{{t}_{1}}>0$ ($1\ne \beta >0$ and ${{h}_{\tau }}>1$), and where $h_\tau$ is a timescale and $\beta$ the power-law exponent. The integrals needed in Eq.\eqref{(2.3)} read simply:
	      \begin{equation}\label{(2.13)}
	      	\begin{aligned}
	      		\int_{{t_1}}^{{t_2}} {dt{\psi _0}\Psi \left( t \right)} & = {\psi _0}h_\tau ^{\beta  + 1}\left( {\gamma\left( {\beta  + 1,\frac{{{t_1}}}{{{h_\tau }}}} \right)} \right. \hfill \\
	      		                                                        & \left. { - \gamma\left( {\beta  + 1,\frac{{{t_2}}}{{{h_\tau }}}} \right)} \right), \hfill                            \\
	      	\end{aligned}
	      \end{equation}
	      where with $\gamma\left( a,z \right)=\int_{z}^{\infty }{dt{{e}^{t}}{{t}^{a-1}}}$ we indicate the incomplete gamma function.
\end{itemize}

Moreover, we will consider four IMF profiles with  mass limits of $M\in \left[ {{M}_{l}},{{M}_{u}} \right]=\left[ 0.08,100.0 \right]{{M}_{\odot }}$:
\begin{itemize}
	\item \textbf{Single power law}. The prototype of this IMF is the work of \citet[][]{1955ApJ...121..161S}. We assume the functional form
	      \begin{equation}\label{(2.14)}
	      	\xi \left( M \right)={{\xi }_{0}}{{\Xi }}\left( M \right)={{\xi }_{0}}{{M}^{-\alpha }},
	      \end{equation}
	      with $\alpha =\text{const}\text{.}$ to yield
	      \begin{equation}\label{(2.15)}
	      	{\xi _0}{I_\Xi }=\int_{{{M}_{l}}}^{{{M}_{u}}}{dMM{{\xi }_{0}}{{\Xi }}\left( M \right)}={{\xi }_{0}}\frac{M_{u}^{2-\alpha }-M_{l}^{2-\alpha }}{\alpha -2},
	      \end{equation}
	      with $M_l$ and $M_u$ the smallest and largest mass considered.
	\item \textbf{Piecewise linear.} Piecewise linear function IMFs are considered from the works of \citet[][]{2001MNRAS.322..231K} and \citet[][]{1986FCPh...11....1S}. We fix a single normalization factor ${{\xi }_{0}}$ for the global IMF and compute the different coefficients ${\xi }_{0,{{\alpha }_{i}}}$ to grant continuity between the slopes $\alpha_j$ of the function in each ${{j}^{\text{th}}}$-section  of the mass intervals $M\in \left[ {{M}_{\text{sep},j}},{{M}_{\text{sep},j+1}} \right[$ (i.e., including the mass of  separation between two contiguous mass intervals ${M}_{\text{sep},j}$ and excluding ${M}_{\text{sep},j+1}$). The IMF has the following form:
		\begin{equation}\label{(2.16)}
			\xi \left( M \right)=\sum\limits_{i=1}^{3}{{{\xi }_{0,{{\alpha }_{i}}}}{{M}^{-{{\alpha }_{i}}}}},		
		\end{equation}
		and it is null outside the mass interval of interest. In this case, for the integrals involved in Eq.\eqref{(2.3)} we get
		\begin{equation}\label{(2.17)}
			\begin{aligned}
				{\xi _0}{I_\Xi } & =\int_{{M_l}}^{{M_u}} {dMM{\xi _0}{\Xi }\left( M \right)}                                                                                                                                 \\
				                 & = \sum\limits_{i = 1}^{3} {\prod\limits_{j = 1}^{i - 1} {M_{{\text{sep}},j}^{{\alpha _{j + 1}} - {\alpha _j}}  } }  \frac{{M_u^{2 - {\alpha _i}} - M_l^{2 - {\alpha _i}}}}{{\alpha  - 2}}, \\
			\end{aligned}
		\end{equation}
where $M_l$ and $M_u$ are the smallest and largest mass considered within the mass range of interest, which can be different for each CSP (a proof of this relation is in \citealt{Pasetto2018T}).
		
\item \textbf{Log-Normal IMFs}. Following \citet[][]{2003PASP..115..763C} or \citet[][]{1979ApJS...41..513M}, we implement in GalMod a commonly used parametric family of IMF profiles for stellar systems (often used in combination with the power laws mentioned above) of the form:
		\begin{equation}\label{(2.18)}
			\xi \left( M \right) = \frac{{{\xi _0}{C_a}}}{M}\exp {\left( { - \frac{1}{{\sqrt 2 {\sigma _M}}}\log \frac{M}{{{M_l}}}} \right)^2},
		\end{equation}
		with $C_a$, $\sigma_M$ as normalization constants and $M_l$ the smallest mass considered.
\end{itemize}
These four profiles of SFR and IMF \citep[three IMF profiles initialized with four sets of constants from ][respectively]{1955ApJ...121..161S,2001MNRAS.322..231K,1986FCPh...11....1S,1979ApJS...41..513M} represent all the tools necessary to run GalMod or merely to predict the number of stars in any given FoV. They have all been implemented on the GalMod web page, thus offering 16 combinations for each of the CSPs adopted, and providing a flexible and fast tool for the scientific investigation of stellar populations.

\section{Stellar population profiles}\label{StellarProfiles}
Besides the functional profiles that are necessary to describe the SFR and IMF of each CSP in the framework presented in \citet[][]{2012A&A...545A..14P}, we need to equip GalMod with the CSP density-potential pairs that are a solution to the Poisson equation. GalMod implements these profiles using values tuned to the MW. \citet[][]{2016MNRAS.461.2383P} introduced the GalMod spiral arms model, and in this work, we describe the updated treatment of the central areas of the galaxy (bulge and bar) together with an  investigation of the adopted CSP density profiles (Sec.\ref{Theoretical}).

\subsection{Disks}\label{Disks}
The density profiles used to define the CSPs in the phase-space section of   were recently presented in \citet[][]{2016MNRAS.461.2383P} and are briefly reviewed in Appendix \ref{ImplNotes}. In this work, we introduce a second density profile vertically to the Galactic plane, based on the ${{\mathop{\rm sech}\nolimits} ^2}$ function. This profile was studied a long time ago in the context of the MW potential modeling \citep[see, e.g.,][]{1989MNRAS.239..605K,1989MNRAS.239..651K,1989MNRAS.239..571K}. This profile has recently been proven to be very successful in reproducing the projected MW phase-space from Gaia data release 1 \citep[][hereafter Gaia DR1]{2016A&A...595A...2G} and the Radial Velocity Experiment \citep[][hereafter RAVE]{2006AJ....132.1645S}  by, e.g., \citet[][]{2017arXiv170406274R}. We adopt this formula in a star count approach very similar to \citet{2017arXiv170406274R}. Instead of applying the solution proposed in \citet[see Eq.(A15) of][based on the binomial series]{1989MNRAS.239..571K}, we use the experience gained in \citet[][]{2016MNRAS.461.2383P} to work out a solution based on a hypergeometric function (see Appendix A).

The proposed density profile for a symmetric Galaxy disk in cylindrical coordinates reads as follows:
\begin{equation}\label{(3.1)}
		\rho \left( R,z \right)={{\rho }_{0}}{{e}^{-\frac{R}{{{h}_{R}}}}}{{\operatorname{sech}}^{2}}\left( \frac{z}{2{{h}_{z}}} \right),
\end{equation}
for each CSP, where ${{\rho }_{0}}$ is the central density, ${{h}_{R}}$ the scale length, and $h_z$  the scale height. We can obtain the relative potential by solving the Poisson equation through a Hankel transformation \citep[e.g.,][]{1989MNRAS.239..571K,2016MNRAS.461.2383P} to obtain the potential $\Phi$ analytically (for details see Appendix A):
\begin{equation}\label{(3.2)}
		\Phi \left( R,z \right)=-4\pi G{{\Phi }_{R}}\left( R \right){{\Phi }_{z}}\left( z \right) 	
\end{equation}
with
\begin{equation}\label{(3.3)}
		\begin{aligned}
&{\Phi _R}\left( R \right) = \int_0^\infty  {dk\frac{{{J_0}\left( {kR} \right)}}{{{{\left( {h_R^{ - 2} + {k^2}} \right)}^{3/2}}}}} \\
&{\Phi _z}\left( z \right) = \frac{{{e^{ - 2k\left| z \right|}}}}{{{e^{^{\frac{{\left| z \right|}}{{{h_z}}}}}} + 1}}{\left( { - {e^{ - \frac{{\left| z \right|}}{{{h_z}}}}}} \right)^{{h_z}k}}\left[ {\frac{{{e^{2k\left| z \right|}} + 1}}{{{e^{\left| z \right|/{h_z}}}}}} \right.\\
& + {e^{k\left| z \right|}}\left( {\pi {h_z}k\left( {{e^{\frac{{\left| z \right|}}{{{h_z}}}}} + 1} \right)\csc (\pi {h_z}k) + 2\cosh (kz)} \right)\\
&\left. { - {h_z}k\left( {{e^{\frac{{\left| z \right|}}{{{h_z}}}}} + 1} \right)\left( {{{\hat B}_{{h_z}k}} - 2{e^{k\left| z \right|}}\cosh (kz){{\hat B}_{{h_z}k + 1}}} \right)} \right]\\
& - {{\mkern 1mu} _{\left( {1, - {h_z}k} \right)}}{{\hat F}_{\left( {1 - {h_z}k} \right)}},
	\end{aligned}
\end{equation}
where $J_0$ indicates the Bessel $J_0$ function, $_{\left( {\bullet,\bullet} \right)}{\hat F_{\left( \bullet \right)}} \equiv {\,_2}{F_1}\left( {\bullet,\bullet;\bullet; - {e^{ - {\textstyle{\frac{\left| z \right|}{{h_z}}}}}}} \right)$ describes the common $_{2}{{F}_{1}}\left( \bullet \right)$ hypergeometric function, and ${\hat B_\bullet} \equiv B\left( {\bullet,0, - {e^{ - {\textstyle{\frac{\left| z \right|}{{h_z}}}}}}} \right)$ is the incomplete Beta function.

Eq.\eqref{(3.2)} reduces to a 1D integral, which has already been discussed in the literature \citep[e.g., ][]{1986ApJ...309..472Q, 1987A&A...180...94B,2016MNRAS.461.2383P} but for it, no explicit analytical formulation has been found yet (see Appendix C for our numerical implementation details). The only differences with respect to the solution presented in \citet[][]{2016MNRAS.461.2383P} are the determination of the vertical force ${{F}_{z}}\left( {{R}_{\odot }},z \right)$ at the Solar/observer position and the proportionality factor for the tilt of the velocity ellipsoid in the vertical and radial direction, $\lambda =\lambda \left( {{R}_{\odot }},z \right)$. The latter can be obtained by the derivative of Eq.\eqref{(3.3)}.

This new treatment of the central galactic zones substitutes entirely the model presented in \citet[][]{2016MNRAS.461.2383P}.

\subsection{Bulge/bar model: observational constraints}\label{Observational}
The MW bulge is the most complex structure that we model. Its position with respect to the Sun makes it difficult to observe. The non-axisymmetric nature of the bulge challenges the models. Its kinematic characteristics, including a rotating bar embedded in a spherical bulge, are difficult to understand and the stars with ages of $>5$~Gyr span a wide range in metallicity (from sub to super solar). For a recent review see, e.g., \citet{2016ARA&A..54..529B}. We give a brief introduction to a few observational works related to the development of our theoretical model. We do not intend to give a complete review but rather justify our bulge/bar modeling approach and the assumptions made in GalMod.

Some of the oldest photometric studies of the bulge date back to Arp's works \citep[Arp 1959, quoted in][]{1988ARA&A..26...51F}. For a recent review on this topic, we refer to \citet{2016ARA&A..54..529B}. The existence of a bar was for the first time observed by \citet[][]{1991ApJ...379..631B}, together with the peanut shapes \citep[e.g.,][]{1998ApJ...492..495F} while modeling COBE/DIRBE images. Since then, several projects have been carried out over the last decades to improve our understanding of the bulge, such as the Bulge Radial Velocity Assay \citep[BRAVA, ][]{2008ApJ...688.1060H, 2012AJ....143...57K}, the VISTA Variables in the Via Lactea (VVV) survey \citep[e.g.,][]{2010NewA...15..433M,2013Msngr.152...23G}, the UKIRT Infrared Deep Sky Survey \citep[UKIDSS,][]{2007MNRAS.379.1599L}, the Galactic bulge survey ARGOS \citep[e.g.,][]{2013MNRAS.430..836N,2013MNRAS.428.3660F}, and the GIRAFFE Inner Bulge Survey (GIBS)  \citep[][]{2014A&A...562A..66Z,2015A&A...584A..46G}.

In these photometric studies, star-counting techniques, such as the one used in GalMod, played an important role. Star counts in the 2MASS and OGLE-III surveys \citep[][]{2006AJ....131.1163S,2003AcA....53..291U} confirmed the X-shaped bulge with a bifurcation in the red clump \citep[][]{2006AJ....131.1163S,2010ApJ...724.1491M,2010ApJ...721L..28N,2011AJ....142...76S}. Recently, the VVV ESO public survey \citep[][]{2010NewA...15..433M} was used by \citet[][]{2013MNRAS.435.1874W} and \citet[][]{2016A&A...587L...6V} to constrain the first 3D density model of the bulge. Similar work was done by \citet[][]{2017arXiv170403325R} with the Gaia-ESO Survey and \citet[][]{2016ApJ...832..132Z} with APOGEE II (both are based on star-count type studies).

When we consider the time evolution, the major contribution to our understanding of the evolution of galaxy bulges comes from $\sim40$ years of N-body simulations. From the pioneering N-body works of \citet[][]{1981A&A....96..164C}, \citet[][]{1980A&A....92...33C}, \citet[][]{1983A&A...127..349A}, \citet[][]{1990A&A...233...82C}, to orbit investigations \citep[][]{2015MNRAS.448..713P,2013MNRAS.428.3478L,2013MNRAS.432.3062H,2013MNRAS.435.3437W} of X-shape banana orbits to orbital-based investigation with made-to-measure (M2M) based techniques by \citet[e.g.,][]{2014MNRAS.438.3275G,2015MNRAS.447.1535N,2014MNRAS.445.3525P,2014MNRAS.445.3546P,2015ApJ...808...75Q,2005MNRAS.358.1477A}. The aim was to deduce the bulge mass \citep[][]{2017MNRAS.465.1621P,2012MNRAS.421..333S,2012MNRAS.427.1429W}, its origin  \citep[][]{2014MNRAS.437.1284Q,2014MNRAS.445.1339L,2014ApJ...787L..19N} or connection with the disks \citep[][]{2016arXiv161109023D,2016ApJ...830..108G,2017MNRAS.465.1621P}. The technique as the one used in GalMod is able to combine first-order kinematic data to predict the radial velocity distribution of the bulge/bar stars, thus offering a simple model for surveys such as BRAVA \citep[][]{2012AJ....143...57K,2007ApJ...658L..29R,1994MNRAS.269..753H}.

\subsection{Bulge/bar model: theoretical model}\label{Theoretical}
From the analysis of observed data and N-body simulations, we understand that GalMod needs to be equipped with a flexible galaxy bulge model that can sustain the various investigative aims and the future challenges that have been posed by past and future surveys. To achieve this flexibility, GalMod will give the opportunity to explore a much larger parameter space than the ones in existing MW models. For example, the bar outside the bulge \citep[e.g.,][]{1994MNRAS.269..753H} can be represented as either a double bar system \citep[e.g.,][]{2005ApJ...630L.149B,2008A&A...491..781C}, or as a smooth system connected with the bulge \citep[e.g.,][]{2015MNRAS.450.4050W}. The features we include in GalMod allow the user to investigate both scenarios.

Last but not least, GalMod is able to sustain a larger parameter space investigation than the one needed for the MW alone. GalMod can be used as an N-body initial condition (i.c.) generator, as well as a model to investigate M31 by using an M31 parameter model instead that of the MW's (see Sec.\ref{Test}).

\subsubsection{Free functional forms in the DWT}\label{sec:free-functional-forms-in-the-dwt}
To treat spiral arms or bar instabilities coherently, i.e., accounting for the density profiles and the gravitational potential simultaneously, a robust framework is represented by the density wave theory (DWT) \citep[e.g.,][]{2014dyga.book.....B}. The DWT as developed by \citet[][]{1964ApJ...140..646L}, and \citet[][]{1969ApJ...155..721L} works on the quite restrictive assumption of tightly wound spiral arms, which does not hold firmly for the MW. Nevertheless, in the following, we will adopt the DWT framework and explore the parameter space allowed by our model while keeping in mind that the tightly-wound approximation does not hold well in the full parameter space. Moreover, the DWT includes a few arbitrary functions chosen for practical analytics. With this in mind, we made GalMod accessible to a rather broad parameter space.

To a first order approximation, the orbits of stars in the Galactic disk are mostly circular. The spiral arms define the locus of points at any given time (i.e., an isochrone in the configuration submanifold of $\mathbb{E}$) among a family of these circular orbits as a result of an evolving pattern \citep[e.g.,][]{1964ApJ...140..646L} or a dynamical modal structure \citep[e.g.,][]{1989ApJ...338...78B,1989ApJ...338..104B}. The simplest of such loci is traditionally the logarithmic spiral structure described by an isochrone (formally a shape function, Fig.\ref{ShapeFunk}) that reads:
\begin{figure}
\includegraphics[width=9cm]{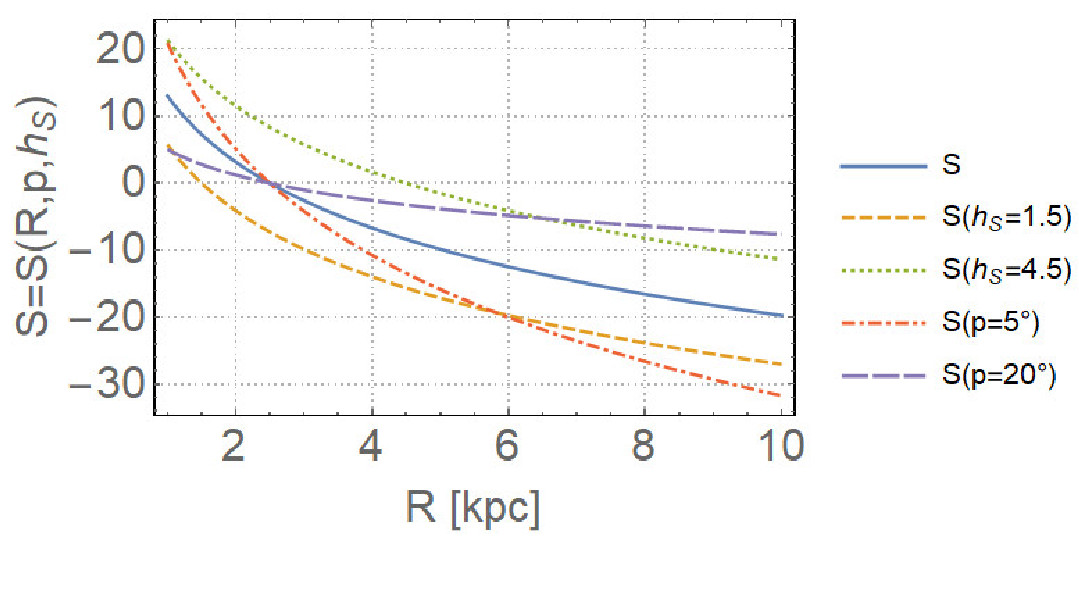}
\caption{Shape functions (Eq.\eqref{(3.4)}) for the set of allowed parameters in GalMod. The reference values of $p=8{}^\circ $ and ${{h}_{s}}=2.5\ \text{kpc}$  reported in Table 1 are represented as blue solid line.}
\label{ShapeFunk}
\end{figure}
\begin{equation}\label{(3.4)}
		S\left( R,p,{{h}_{S}} \right)=-2\cot p\lg \left( \frac{R}{{{h}_{S}}} \right),
\end{equation}
with $p$ being the pitch angle and $h_S$ the scale length of the shape function.

\begin{figure}
\includegraphics[width=9cm]{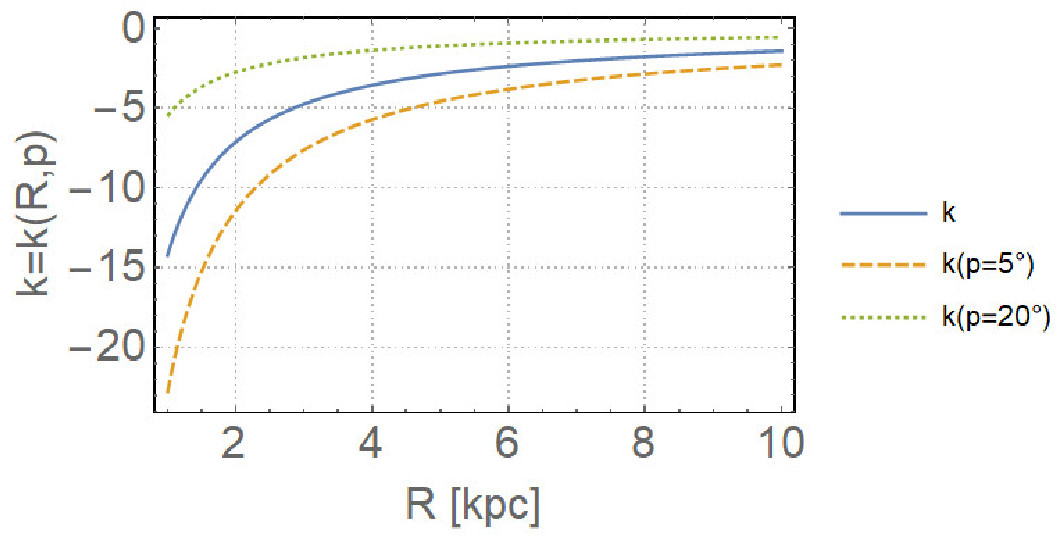} 
\caption{Wave number for the same pitch angles as in Fig.\ref{ShapeFunk}. The blue solid line refers to $p=8{}^\circ $ and the parameters presented in Table 1.}
\label{waveNum}
\end{figure}

We allow in GalMod both positive and negative values of the wave number, $k$, but following \citet[][]{1969ApJ...155..721L} the MW is a trailing spiral galaxy, and accordingly $k=k\left( R,p \right)<0$ (see Sec.1 in \citet[][]{2016MNRAS.461.2383P} for an extended review on this subject, Fig.\ref{waveNum}).

\begin{figure}
\includegraphics[width=9cm]{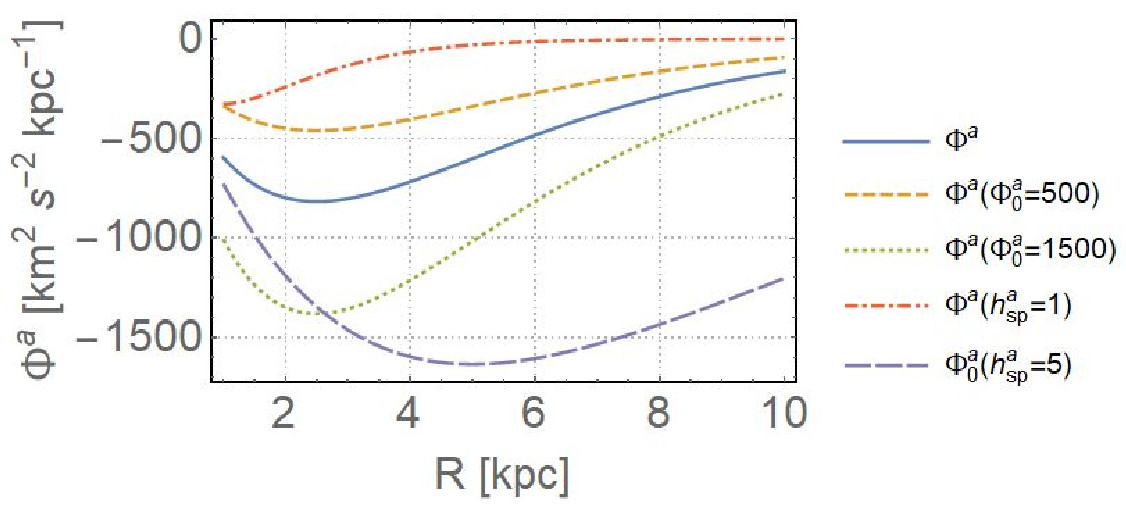}
\caption{Spiral arm amplitude ${{\Phi }^{a}}$ as a result of Eq.\eqref{(3.5)} for the parameter space boundary values allowed in GalMod. The blue solid line refers to the parameters of Table 1.}
\label{Amplit}
\end{figure}

The amplitude of the potential is also an arbitrary function. Popular choices from the literature based  on the flexibility of the Rosin-Rammler distribution function (already encountered in Eq.\eqref{(2.12)}) are entirely arbitrary. For example, following \citet[][]{1986A&A...155...11C} we can set:
\begin{equation}\label{(3.5)}
		{{\Phi }^{a}}\left( \Phi _{0}^{a},h_{\text{sp}}^{a} \right)=\Phi _{0}^{a}R{{e}^{-\frac{R}{h_{\text{sp}}^{a}}}},
\end{equation}
or alternatively, from \citet[][]{1977LNP....69.....R}:
\begin{equation}\label{(3.6)}
		{{\Phi }^{a}}=-\frac{2\pi G{{\Sigma }_{0}}}{\left| k\left( R \right) \right|}.
\end{equation}
Both functions work to enhance the spiral arm strength to a maximum and release it throughout the disk outskirts but with different profiles of the normalization amplitude $\Phi _{0}^{a}$, of the scale profile with scale length $h_{\text{sp}}^{a}$, and central surface density ${{\Sigma }_{0}}$, used in Eq.\eqref{(3.5)} and Eq.\eqref{(3.6)}. We implemented this Eq.\eqref{(3.5)} in GalMod because it is more popular but not because any observational data to date is able to support it better than Eq.\eqref{(3.6)}. The boundaries of the parameter space allowed in GalMod are plotted in Fig.\ref{Amplit}, where the reference model for the MW assumes $\Phi _{0}^{a}\cong 887\ \text{k}{{\text{m}}^{2}}{{\text{s}}^{-2}}\text{kp}{{\text{c}}^{-1}}$.

Finally, the last assumed shape function is the shape of the potential, whose real part reads:
\begin{equation}\label{(3.7)}
		\begin{aligned}
&{\Phi _{{\rm{sp}}}}\left( {R,\phi ,\Phi _0^a,h_{{\rm{sp}}}^a,m,{\Omega _p},t,p,{h_S}} \right)\\
 &= {\mathop{\rm Re}\nolimits} \left( {{\Phi ^a}\left( {\Phi _0^a,h_{{\rm{sp}}}^a} \right){e^{\iota \left( {m{\Omega _p}t - m\phi  + S\left( {R,p,{h_S}} \right)} \right)}}} \right)\\
 &= \Phi _0^a{e^{ - \frac{R}{{h_{sp}^a}}}}R\cos \left( {2\cot p\log \frac{R}{{{h_S}}} - m(\phi  - {\Omega _p}t)} \right),
	\end{aligned} 	
\end{equation}
with $m$ being the number of spiral arms and $\Omega_p$ the pattern speed.
For each location $R=\hat{R}$ and $\phi =\hat{\phi }$ on the galaxy plane, ${{\Phi }_{\text{sp}}}$ depends on seven parameters. An extensive investigation that we performed revealed the importance of the pitch angle as well as the amplitude function ${{\Phi }^{a}}\left( \Phi _{0}^{a},h_{\text{sp}}^{a} \right)$, while the dependence on the integration times is weak or almost null (tested between  $t=\left[ 0.1,10 \right[\times 0.250\ \text{Gyr}$) where the function almost entirely degenerates with the degrees of freedom associated with the Sun/observer location's azimuthal position. The possibility to move the Sun/observer in the azimuthal direction ${{\phi }_{\odot }}$ is degenerate with the rigid rotation of the pattern, i.e., with the origin of the reference frame for the coordinate $\phi$.

\subsubsection{Bar structure}\label{BarStruct}
Probably the most widely known instability in stellar disk systems is the bar instability. Hence, the simplest way to realize the bar comes from the possibility to exploit the analysis of the functions in Sec.\ref{sec:free-functional-forms-in-the-dwt} in a formula that easily connects the density profiles for these instability modes found in Eq. (56) of \citet[][]{2016MNRAS.461.2383P}:
\begin{equation}\label{(3.8)}
		\rho =\frac{{{\Sigma }_{0}}}{2{{h}_{z}}}\left( 1-\frac{{{\Phi }_{\text{sp}}}}{\sigma _{RR}^{2}}\frac{{{X}^{2}}}{1-\nu }\Re  \right){{e}^{-\frac{z}{{{h}_{z}}}}},\
\end{equation}
with $h_z$ being the vertical scale length of the unperturbed surface density $\Sigma_0$, $\Re $ as the reduction factor (Appendix A of \citet[][]{2016MNRAS.461.2383P}), and $X=\frac{k}{\kappa }{{\sigma }_{RR}}$ being the Toomre number, given by the ratio between wavenumber $k$ and the radial epicyclic frequency $\kappa $ times the fully radial element of the velocity dispersion tensor ${{\sigma }_{RR}}$. Finally $\nu =\frac{\Omega -m{{\Omega }_{p}}}{\kappa }$ with $\Omega $ as  the angular speed, ${{\Omega }_{p}}$ as the pattern speed, and $m$ as the number of spiral arms.

The dependence of $\phi $ on the density profiles is only incorporated in Eq.\eqref{(3.7)}. It is evident that the max and min of ${{\Phi }_{\text{sp}}}$ at any given radius $R$ can be easily located at $\phi =\pi \mathbb{Z}$ for all $\cos \left( m(\phi -{{\Omega }_{p}}t)+2\cot p\lg \frac{R}{{{h}_{S}}} \right)=\cos \left( 2\varphi  \right)$, i.e., for $\varphi =\frac{1}{2}\left( m\phi -mt{{\Omega }_{p}}+2\cot p\log \frac{R}{{{h}_{S}}} \right)$.
This condition is satisfied for $\phi =t{{\Omega }_{p}}-\frac{2\cot p}{m}\log \frac{R}{{{h}_{S}}}+2\pi \mathbb{Z}$. We want these maxima to match a given direction, say $\phi =0\wedge \phi =\pi $, which will represent the major axis of the bar, and we minimize them in the orthogonal direction $\phi =\frac{\pi }{2}\wedge \phi =\frac{3\pi }{2}$. In this way, we obtain a natural bar from the DWT and a
consistent gravitational potential with coherent kinematics.
Note that, while we can model the MW as a multi-armed spiral galaxy by varying $m=m_{\text{sp}}$, we have to force $m=m_{\text{bar}}=2$ for the realization of the bar alone.

\begin{figure*}
\plotone{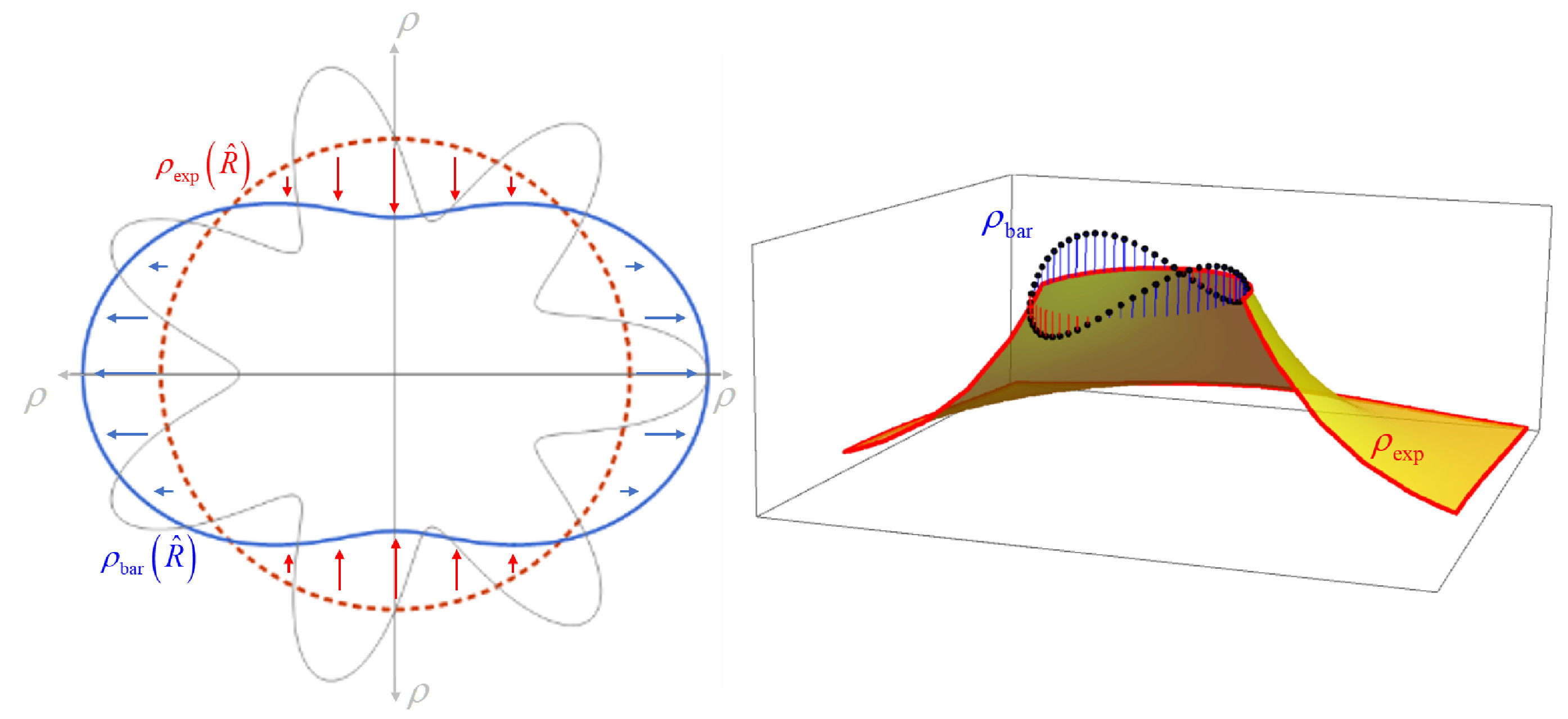}
\caption{Cartoon of the bar as modeled in GalMod. On the left, a polar plot of the density is shown. The blue line represents the minima and maxima of the sinusoidal curve at varying $\phi$  and arbitrary but fixed $R=\hat{R}$ in the central volume of the galaxy. The perturbation enhances the density (blue arrows) or decreases it (red arrows) over or below the reference level of the unperturbed exponential profile (dashed red circle) at the given radius. On the horizontal gray axis, the density is marked as the red dashed reference line for the unperturbed density and on the vertical gray axis the density is reduced under the red dashed reference line. The angular dependence is tuned to have precisely two orthogonal maxima and minima over any exponential reference line (red dashed line). The other gray curve is another example of enhancing or reducing density over a reference radius shown as an example; clearly, the method cannot work if $m_{\text{bar}} \neq 2$.  On the right, the concept of the left panel is shown in 3D: the enhancement or decrease for our ${{\rho }_{\text{bar}}}$ is represented by vertical bars connected to the unperturbed arbitrary level of reference ${{\rho }_{\exp }}$. The same colors are used as in the right panel, and the thick red line marks the border of ${{\rho }_{\exp }}$ plotted for just a section to show the idea of the DWT and used here to create density profiles.}
\label{BarModel}
\end{figure*}
We gain a better insight into these basic concepts of the DWT from Fig.\ref{BarModel}. This approach was already presented in Fig.\ref{ShapeFunk} of \citet[][]{2016MNRAS.461.2383P} with an example of the cone of view, and we do not repeat that figure here.

\begin{figure*}
\plotone{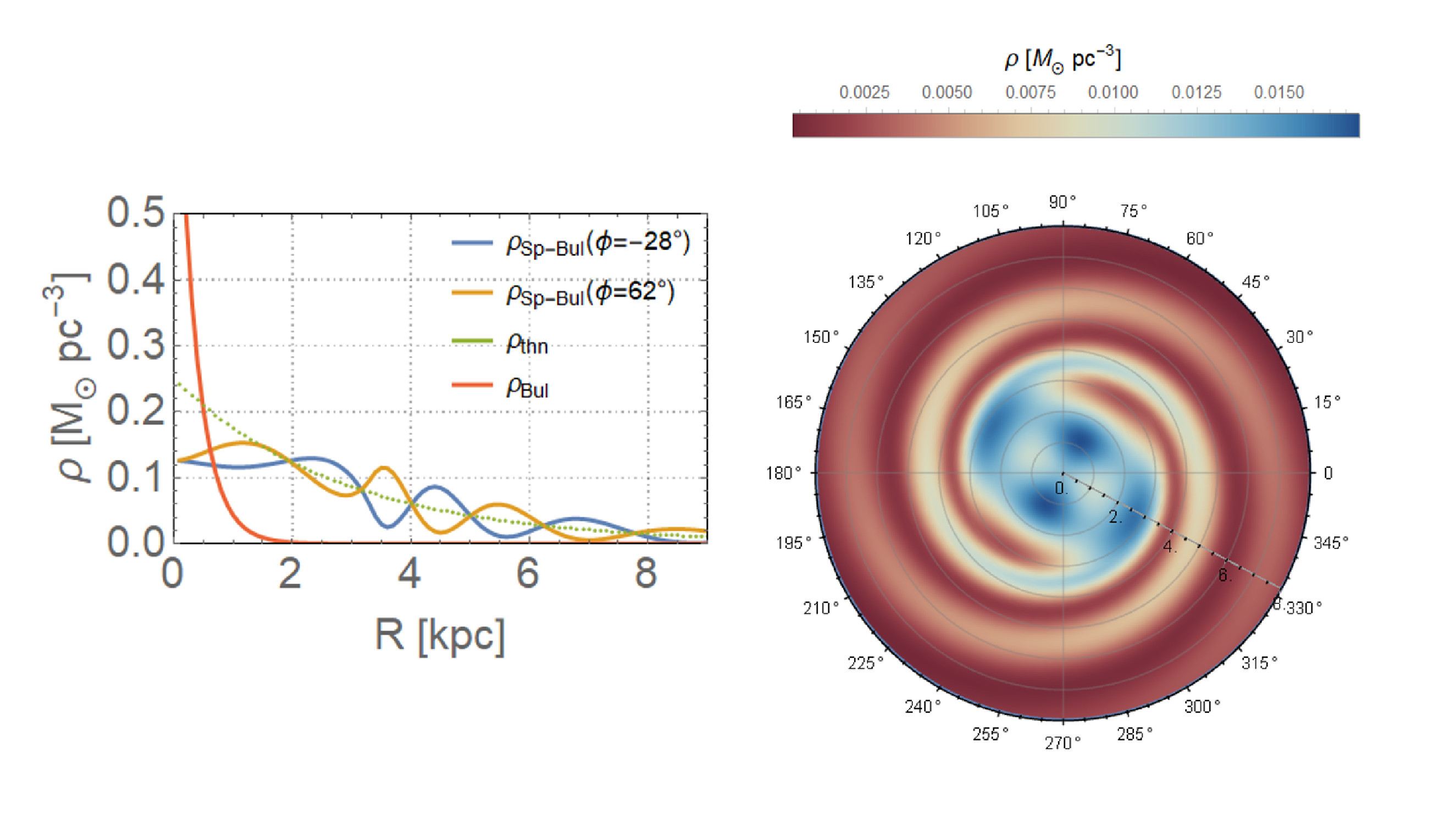}
\caption{(Left panel) Density profiles for the spiral/bar population along the major axis (blue line) and orthogonal to it (brown line). For comparison, the density profile for the thin disk component of Eq.\eqref{(3.5)} is shown as a green dotted line together with the spherical bulge component (red line). (Right panel) Isocontour at $z=0$ of the density profiles for Eq.\eqref{(3.5)}. Only the central 8~kpc are shown, the Sun is located at ${{\left\{ R,\phi  \right\}}_{\odot }}=\left\{ 8.0,0.0 \right\}\ \text{kpc}$, and the radial bar direction is detected automatically and aligned with the major axis of the bar (shown in the plot as a line).}
\label{BarMW}
\end{figure*}

The DWT developed by \citet[][]{1964ApJ...140..646L} and \citet[][]{1969ApJ...155..721L} has been artificially modified in \citet[][]{2016MNRAS.461.2383P} to cover the phase-space discontinuities (i.e., the Lindblad resonances) with a tailored scheme that grants continuity to the moments up to the order two and cumulants up to the order four (see Appendix B of \citet[][]{2012A&A...547A..70P} and \citet[][]{2012A&A...547A..71P}) of the perturbed distribution function (see Sec.6.1.2 in \citet[][]{2016MNRAS.461.2383P}). This is a convenient 4th-order polynomial scheme standing on a single free parameter that we can fix by minimizing the total mass difference between the DWT-perturbed distribution and the non-perturbed density profile. Positivity of the underlying distribution function (DF) is required to grant physical meaning to the emerging perturbed DF. For example, the density of the spiral arms ${{\rho }_{\text{sp}}}\left( \bm{x} \right)={{\rho }_{\text{sp}}}\left( R,\phi ,z \right)$ at the resonance radius $R={{\hat{R}}_{\text{res}}}$ is divergent, and hence, the distribution function cannot be ''filled'' by a finite number of stars in the GalMod star-count modeling approach, i.e., $\underset{R\to {{{\hat{R}}}_{\text{res}}}}{\mathop{\lim }}\,\rho \left( R,\phi ,z \right)=\infty \forall \left\{ \phi ,z \right\}$. We covered this discontinuity with a polynomial that cover this discontinuity around the inner/outer Lindblad resonance (ILR/OLR) neighborhoods, e.g., at $R={{R}_{\text{ILR}}}\pm \varepsilon$, where $\varepsilon$ is fixed so that the mass of the continuity extension of $\rho$, say ${{\hat{\rho }}_{\text{sp}}}$, is as close as possible to the total mass of ${{\rho }_{\text{sp}}}$. We keep on adopting the same polynomial scheme with an explicit correction for the central part of the MW. When the density profile parameters adopted for ${{\rho }_{\text{sp}}}$ lead to an ILR very close to the center, such as ${{R}_{\text{ILR}}}<\varepsilon$, because of $R\in \mathbb{R}_{0}^{+}$ we arbitrarily fixed the central value to $\hat{\rho }$, the unperturbed density value at the resonance to avoid unphysical negative radii or negative DF values. We provide better insight on the density profiles obtained in this way in Fig.\ref{BarMW}. Note how in the left panel of the figure the profiles are along the major axis. The tilt of the bar is about $-28{}^\circ$ with respect to the direction of the Sun.

\begin{figure*}
\plotone{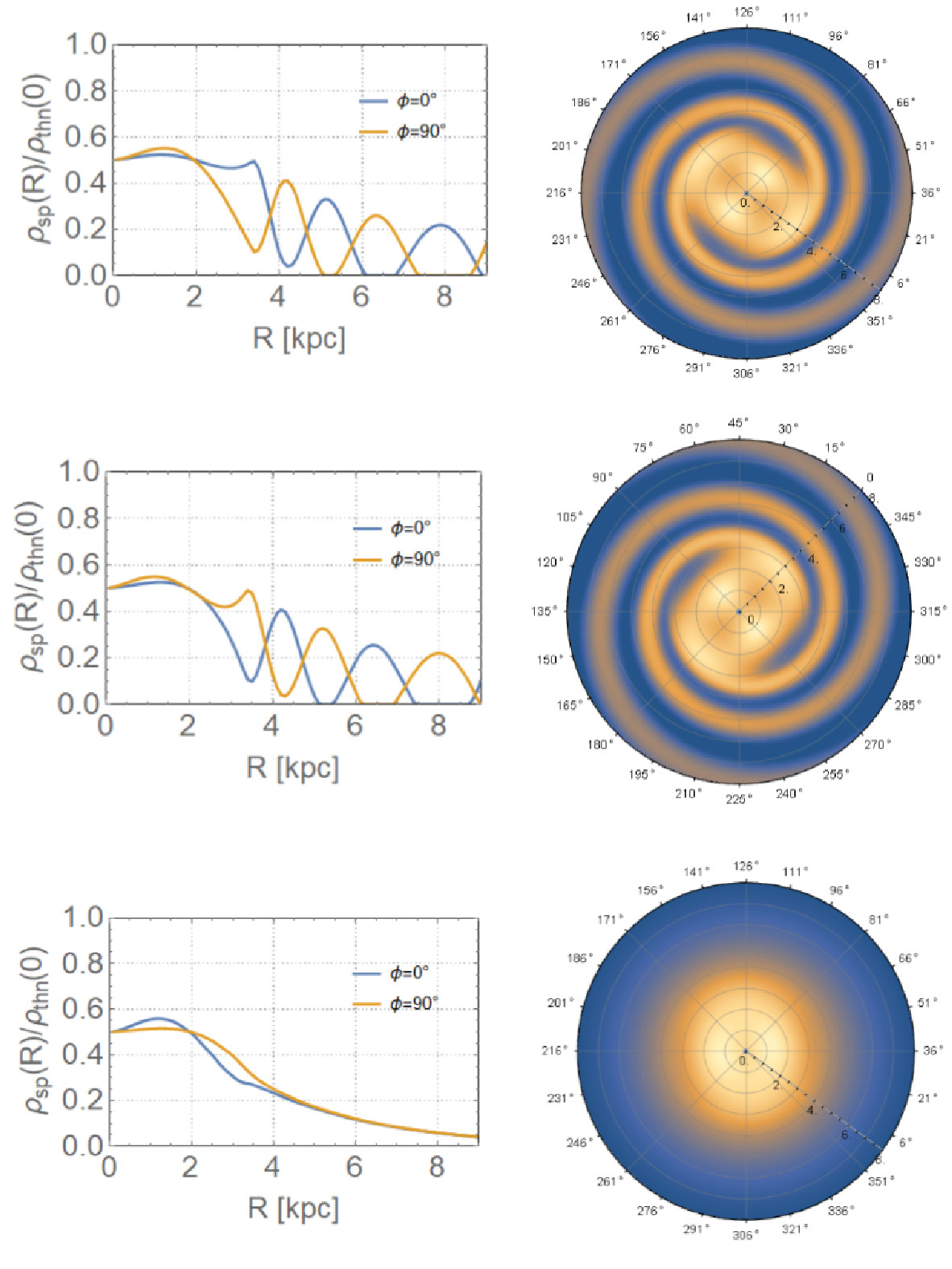}
\caption{Six panels are showing the implemented model capabilities. The density profiles are normalized to the central value of the unperturbed exponential density profile. In the left panels, the density profiles on the major and minor axis are shown. In the right panels, the corresponding contour plots are presented. See text for details.}
\label{Panel}
\end{figure*}

We point out that the GalMod user is entirely free to explore a galaxy with no bulge, a completely bar-dominated one, with a weak bar, or an entirely bulge-dominated model; the scheme works for the whole parameter space proposed as shown in Fig.\ref{Panel}. In Fig.\ref{Panel}, we selected a few examples from an extended parameter space investigation that we performed. The top row model is a case of a strong bar instability with an axis ratio on the plane of $3:2$ obtained for $\Phi _{0}^{a}\cong 1500\ \text{k}{{\text{m}}^{2}}{{\text{s}}^{-2}}\text{kp}{{\text{c}}^{-1}}$,  $h_{\text{sp}}^{a}\cong 5\ \text{kpc}$, ${{\Omega }_{p}}\cong 40\ \text{km}\ {{\text{s}}^{-\text{1}}}\ \text{kp}{{\text{c}}^{-1}}$, $t=1.76\ \text{Gyr}$, $p=25{}^\circ$ ${{h}_{S}}=5$ in Eq. \eqref{(3.7)}. As evident from the left panel, in this case the structure of the bar is very flat (only small density fluctuations are visible) with a pronounced sharp cut at $3.8\ \text{kpc}$ in the direction of the long axis and a slighter decline along the orthogonal direction. The profiles are normalized to the unperturbed thin disk component at its central value ${{\rho }_{\text{thn}}}\left( R=0,z=0 \right)$. In the second row of Fig.\ref{Panel}, we chose to represent a model with the major axis of the bar tilted by 90 deg with respect to the observer located at ${{\left\{ R,\phi  \right\}}_{\text{obs}}}=\left\{ 8.0,0.0 \right\}\ \text{kpc}$ obtained by setting $\Phi _{0}^{a}\cong 1369\ \text{k}{{\text{m}}^{2}}{{\text{s}}^{-2}}\text{kp}{{\text{c}}^{-1}}$,  $h_{\text{sp}}^{a}\cong 4.5\ \text{kpc}$, ${{\Omega }_{p}}\cong 40\ \text{km}\ {{\text{s}}^{-\text{1}}}\ \text{kp}{{\text{c}}^{-1}}$, $t=0.75\ \text{Gyr}$, $p=25{}^\circ$ ${{h}_{S}}=1\ \text{kpc}$ in Eq.\eqref{(3.7)}. Considering the literature reviewed in Sec.\ref{Observational}, this model can hardly represent the MW, if at all.

Finally, in the two bottom plots of Fig.\ref{Panel} we provide another extreme example obtained from the parameters $\Phi _{0}^{a}\cong 1500\ \text{k}{{\text{m}}^{2}}{{\text{s}}^{-2}}\text{kp}{{\text{c}}^{-1}}$,  $h_{\text{sp}}^{a}\cong 5\ \text{kpc}$, ${{\Omega }_{p}}\cong 21\ \text{km}\ {{\text{s}}^{-\text{1}}}\ \text{kp}{{\text{c}}^{-1}}$, $t=0.35\ \text{Gyr}$, $p=5{}^\circ$, ${{h}_{S}}=5\ \text{kpc}$ (i.e., in this case, we have $\left| {kR} \right| \sim 22$). We do not see a bar anymore, and the central zone of the galaxy shows almost an unperturbed symmetry.

\begin{figure}
\plotone{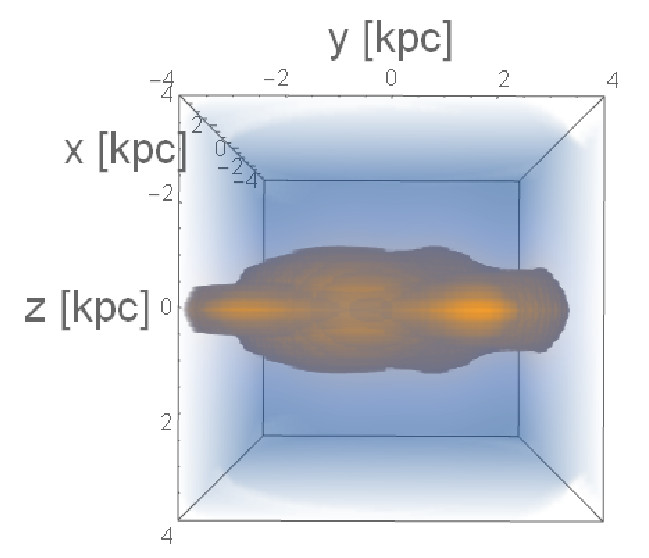}
\caption{Edge-on view of the bar model: the spherical bulge, exponential disk, and stellar halo are omitted. The slicing of the density distribution at 4kpc evidences the density profile of a spiral arm passing in front of the FoV (i.e., the overdensities at about  ${z,y}={0,-3}$ kpc and ${z,y}={0,+1.7}$ kpc in this view). The point of view of the stereographic projection is at $x=+8$ kpc, but only the 4 kpc inside the cube are shown. The Galactic center is located at the origin of the coordinate system shown in the cube.}
\label{Vert3D}
\end{figure}

In Fig.\ref{BarMW} we treated the face-on view of the MW, but it is in the vertical structure description where we obtain the most exciting features from the implemented bar/bulge model. In Fig.\ref{Vert3D} we highlight some of the most exciting features of the secular bar-instability framework, as applied in our model.

The present-day understanding of the vertical structure of the bar is made difficult by deprojection effects (e.g., \citet[][]{2014ApJ...791...11Z} for a review on the danger of the projection effects), nevertheless, a box-peanut-shape is visible in Fig.\ref{Vert3D} where the 3D-isodensity contours of Fig.\ref{BarMW} are plotted as seen from the Solar location. We evidenced a double-peaked profile naturally, for suitably chosen observer positions. We introduced a detailed study of the star counts along the line of sight (l.o.s.) for this model in the paper \citep{Pasetto2018T}, where the different relative numbers of stars in a direction passing through one spiral arm and passing through the peanut structure for longitudinal direction $l>0$  and $l<0$ are discussed. Here we assumed our best fit parameter model of Table \ref{Table1}. The Solar position in the figure is at ${{\left\{ R,\phi ,z \right\}}_{\odot }}=\left( 8.0,0.0,0.02 \right)\ \text{kpc}$ and a bar tilted by about $\cong -29\ \deg$ with respect to the Sun's direction is assumed. The characteristic shape of the bulge can be directly compared with red clump star isodensity plots in, e.g., \citet[][]{2016A&A...587L...6V} even if our model is not finely-tuned to reproduce red clump stars in the Galactocentric (GC) direction.

Another significant advantage of the scheme used in GalMod is that we do not need a fully non-symmetric solver for the Poisson equation. We can resolve all the significant non-axisymmetric structures of the MW with the DWT-linear response framework. The adopted approach confers an elegant first-order coherence to the model. First-order perturbation theory is used for the spiral arm density, for the central bar density, but also for their velocity distribution so that both configuration and velocity space, are treated as a perturbation of the same order. The unperturbed densities have associated velocity distributions treated with the Jeans equations, i.e., with the first few orders of the DF velocity moments (see Appendix C and \citet[][]{2016MNRAS.461.2383P} Sec.6). Still, it is worth to remark that the resulting kinematics cannot be dominated by rotational motion at all if the bar component is left to be dominated in mass by the spherical bulge component.

Such a novel approach of describing the non-axisymmetric central galaxy features in star count models also comes with two significant drawbacks. The first is that we assign virtually no free parameters to the bar. In this unified bar/disk model we fix the spiral arm component parameters in the extended Solar neighborhood and the bar component results automatically. This was the reason why in Table 1 of \citet[][]{2016MNRAS.461.2383P} two separate spiral components were introduced: to give the freedom to choose different pattern speeds for the bar and disks. Nevertheless, in our approach, the rotation speed profile $\Omega =\Omega \left( R \right)=\frac{{{v}_{c}}\left( R \right)}{R}$ is common to the bar and spiral arms so that to have different pattern speeds for the bar and spiral arms, ${{\Omega }_{p\text{,sp}}}\ne {{\Omega }_{p\text{,bar}}}$, two distinct components are necessary. A single component is not necessarily a problem for the age and chemical composition of the MW's central CSP since this CSP is dominated in mass by a second spherical component (the bulge). The drift of the stars of a CSP in an inside-out model of disk formation is, to date, a popular scenario in the vast majority of cosmologically motivated N-body simulations \citep[e.g.,][]{2017MNRAS.467.2430M}.

The second drawback is related to the previous arguments on the velocity space. The kinematics of the bar is given by the linear superposition of rotational density waves on the spherical symmetry kinematics of the bulge component. The amount and the relative orbit type are weighted by the mass assigned to the bulge or the bar. It is sufficient to review the literature of bulge orbits (analytical, numerical, and perturbation techniques) in classic textbooks on stellar dynamics such as \citet[][]{2002ocda.book.....C} to grasp the complexity of the orbital superposition supposed to coexist in the MW central area. We do not claim that our underlying orbit representation in GalMod, based just on the first moments of the DF implemented of the DWT, correctly captures this complexity. Nevertheless, we think the projected star counts are a valuable alternative to orbital integration and a benchmark to test the different formation scenarios of spiral arms \citep[e.g.,][]{2014PASA...31...35D}.

\subsubsection{Spherical bulge}
To present a flexible model,  we proceed to implement in GalMod an entirely spherical bulge given by the density-potential couple solution of the Poisson equation in spherical coordinates, $r$:
\begin{equation}\label{(3.9)}
\begin{aligned}
\rho  &= {\rho _{{\rm{0,blg}}}}{e^{ - \frac{r}{{{h_{{\rm{blg}}}}}}}} \Leftrightarrow \\
{\Phi _{{\rm{blg}}}}\left( r \right) &= 4\pi Gh_{{\rm{blg}}}^2{\rho _{{\rm{0,blg}}}}\left( {{e^{ - \frac{r}{{{h_{{\rm{blg}}}}}}}} - {\rm{Ei}}\left( { - \frac{r}{{{h_{{\rm{blg}}}}}}} \right)} \right),
\end{aligned}
\end{equation}
with ${\rho _{{\rm{0,blg}}}}$ being the central bulge density, ${h_{{\rm{blg}}}}$ being the radial scale length, and $\text{Ei}(\bullet)$ representing the exponential integral function (see Appendix B).

In particular, in relation to the MW modeling, we want to point out that self-standing models of the MW central regions without bulges (as a pure disk) have recently been presented in the literature \citep[e.g.,][]{2012ApJ...757L...7L,2008ApJ...688.1060H,2010ApJ...720L..72S,2010ApJ...715L.176K,2012ApJ...756...22N,2013ApJ...766L...3M,2013MNRAS.430.2039S}. These scenarios imply a secular-instability formation where the pseudo-bulge deploys from the inner-disk material. This pseudo-bulge suggestion was already supported  in the first chemical abundance analysis (e.g., of K/M-giants) in the inner Galactic disk \citep[e.g.,][]{2010A&A...516L..13B,2012ApJ...746...59R} and found to be in partial agreement with photometric analyses. Conversely, spectroscopic alpha-enhanced gradients were favored for the classical component of the bulge (e.g., \citet[][]{2010ApJ...724.1491M}, \citet[][]{2011ApJ...732..108J}, \citet[][]{2011A&A...530A..54G}, \citet[][]{2012ApJ...749..175J}, \citet[][]{2012A&A...546A..57U}, \citet[][]{2013NewAR..57...80F}, \citet[][]{2014AJ....148...67J}).
A possible solution for this apparent contradiction between the kinematic evidence of a bar and the existence of a metallicity gradient may exist in the proposition of a diverse mix of two populations. One possible configuration is a metal-rich population that presents bar-like kinematics, and a metal-poor population that shows kinematics corresponding to an old spheroid or a thick disc as one moves away from the Galactic plane \citep[e.g.,][]{1997A&AS..121..301C, 2002ApJ...569..245N,  2002A&A...391..195G, 2014A&A...564A.102C, 2004A&A...422..205G, 2012A&A...538A.106R, 2013MNRAS.430..836N}. GalMod offers the possibility to model both the components with an independent chemical enrichment.

\section{A few scenarios for GalMod applicability}\label{Test}

In \citet[][]{2016MNRAS.461.2383P} we presented an extensive comparison between GalMod and the Besan\c{c}on model \citep[e.g.,][]{1987A&A...180...94B,2014A&A...564A.102C,2012A&A...538A.106R}. In this work, we perform a different comparison of our model with works of more observational nature.
For this reason, we decided to extend the number of photometric bands available to GalMod to a few photometric systems of general interest. An extensive description of  the implemented synthetic pseudo-bolometric corrections is in \citet[][]{1997A&AS..121..301C}, \citet[][]{2002A&A...391..195G}, and \citet[][]{2004A&A...422..205G} to which we refer the interested readers.

We describe then six examples where we highlight the most useful features of our modeling. In the next sections we will present:
\begin{enumerate}
\item a comparison between a GalMod mock catalog and a large-scale photometric catalog based on the SDSS photometry;
\item a study of the radial velocity distribution in the MW central regions;
\item two studies of contamination by MW foreground stars for a FoV containing an in-plane MW cluster and for a FoV containing a galaxy outside of the Local Group;
\item an example of an N-body i.c. generation;
\item an example on how to generate M31 models.
\end{enumerate}

\begin{table*}
	\caption{Kinematic and dynamical properties of the MW components. Because a map of the metallicity gradients ${\nabla _{\bm{x}}}\left[ {\frac{{Fe}}{H}} \right]$ is still uncertain, no standard default values are assumed and the gradients are used as free parameters. A uniform distribution is assumed between the indicated values.
		For the equations defining the parameters in this table see \citet{2016MNRAS.461.2383P}. Here we recall that ${M_B}$ and ${h_{r,B}}$ are the total bulge mass and radial scale length, $ {\rho _D,h_R,h_z,\Phi _0^a,h_{{\text{spr}}}^a,m,{\Omega _p},p,h_S}$ are the disk central density, scale length, scale height, perturbation amplitude, spirals/bar scale length, total number of spiral arms, angular pattern speed, pitch angle, and shape function scale length respectively. All the parameters are assumed for disks exponential profiles and interstellar medium profiles as indicated. ${{\rho _{0,H*}},h_{r{H^*}},\alpha }$ are the stellar halo central density, scale length, and density slope, respectively, and $ {{v_0},h_{r,DM},q}$ are the scale velocity, scale length, and flattening factor of the dark matter profile. Finally, ${{\bm{\sigma }}_{ii}}_ \odot $ are the velocity dispersion tensor normalization values for the CSP considered along the principal axis of the system of reference of the population at the observer position, IMFs are reported with reference to the equations used as well as the SFRs equation with reference to the temporal interval indicated in col 3.}
	\label{Table1}
	\begin{tabular}{lllllll}
		\hline
		Components         & Scale parameters                                                                                                                                                                                & $\Delta t$     & $\left[ {Fe/H} \right]$ & ${{\bm{\sigma }}_{ii}}_ \odot $  & IMF            & SFR            \\
		                   &                                                                                                                                                                                                 & $[\text{Gyr}]$ & [dex]                   & [${\rm{km}}\;{{\rm{s}}^{ - 1}}$] & Eq.            & Eq.            \\ \hline
		                   & $\{{M_B},{h_{r,B}}\}$                                                                                                                                                                           &                &                         &                                  &                &                \\
		                   & $\left[ {{{\rm{M}}_ \odot }\;{\rm{,kpc}}} \right]$                                                                                                                                              &                &                         &                                  &                &                \\
		Bulge pop          & $9.3 \times {10^{9}},0.32$                                                                                                                                                                      & [6.0,12.0[     & [-0.40,+0.30[           &                                  & \eqref{(2.17)} & \eqref{(2.12)} \\
		                   &                                                                                                                                                                                                 &                &                         &                                  &                &                \\
		                   & $  {\rho _D,h_R,h_z,\Phi _0^a,h_{{\text{sp}}}^a,m,{\Omega _p},t,p,h_S} $                                                                                                                        &                &                         &                                  &                &                \\
		                   & $\left[ {{\rm{k}}{{\rm{m}}^{\rm{2}}}{{\rm{s}}^{{\rm{ - 2}}}}{\rm{kp}}{{\rm{c}}^{{\rm{ - 1}}}}{\rm{,kpc,km}}\;{{\rm{s}}^{{\rm{ - 1}}}}{\rm{kp}}{{\rm{c}}^{{\rm{ - 1}}}}{\rm{,deg,kpc}}} \right]$ &                &                         &                                  &                &                \\
		Bar pop            & $20.5\times 10^6,2.71,0.33,887.82, 2.5, 2, 35.77, 0.13, 2.6$                                                                                                                                    & [5.0,12.0[     & [-0.70, 0.05[           & 57.0,41.0,27.0                   & \eqref{(2.17)} & \eqref{(2.12)} \\
		                   &                                                                                                                                                                                                 &                &                         &                                  &                &                \\
		Thin disk pop 1    & $12.5\times 10^6,2.71,0.11,887.82, 2.5, 2, 35.77, 0.13, 2.6$                                                                                                                                    & [0.1, 0.5[     & [-0.70, 0.05[           & 27.0,15.0,10.0                   & \eqref{(2.18)} & \eqref{(2.6)}  \\
		(spr)              &                                                                                                                                                                                                 &                &                         &                                  &                &                \\
		                   & $ \{{{\rho _D},{h_R},{h_z}}\}_\odot $                                                                                                                                                           &                &                         &                                  &                &                \\
		                   & $ \left[ {{{\rm{M}}_ \odot }\;{\rm{kp}}{{\rm{c}}^{{\rm{ - 3}}}}{\rm{,kpc}}{\rm{,kpc}}} \right] $                                                                                                &                &                         &                                  &                &                \\
		Thin disk pop 2    & $0.75\times 10^6,2.00,0.14$                                                                                                                                                                     & [0.5, 0.9[     & [-0.70, 0.05[           & 30.0,19.0,13.0                   & \eqref{(2.17)} & \eqref{(2.6)}  \\
		Thin disk pop 3    & $1.57\times 10^6,2.00,0.15$                                                                                                                                                                     & [0.9, 3.0[     & [-0.70, 0.05[           & 41.0,24.0,22.0                   & \eqref{(2.17)} & \eqref{(2.6)}  \\
		Thin disk pop 4    & $1.04\times 10^6,2.00,0.18$                                                                                                                                                                     & [3.0, 7.5[     & [-0.70, 0.05[           & 48.0,25.0,22.0                   & \eqref{(2.17)} & \eqref{(2.6)}  \\
		Thin disk pop 5    & $14.0\times 10^6,4.00,0.28$                                                                                                                                                                     & [7.5, 10.0[    & [-0.70, 0.05[           & 52.0,32.0,23.0                   & \eqref{(2.17)} & \eqref{(2.6)}  \\
		                   &                                                                                                                                                                                                 &                &                         &                                  &                &                \\
		Thick disk         & $2.95\times 10^6,2.09,1.10$                                                                                                                                                                     & [10.0,12.0[    & [-1.90,-0.60[           & 51.0,36.0,30.0                   & \eqref{(2.17)} & \eqref{(2.6)}  \\
		                   &                                                                                                                                                                                                 &                &                         &                                  &                &                \\
		ISM                & $22.63\times 10^6,4.51,0.20$                                                                                                                                                                    &                &                         &                                  &                &                \\
		                   &                                                                                                                                                                                                 &                &                         &                                  &                &                \\
		                   & $\{{{\rho _{0,H*}},h_{r{H^*}},\alpha }\} $                                                                                                                                                      &                &                         &                                  &                &                \\
		                   & $ \left[ {{{\rm{M}}_ \odot }\;{\rm{kp}}{{\rm{c}}^{{\rm{ - 3}}}}{\rm{,kpc}}{\rm{,kpc}}} \right] $                                                                                                &                &                         &                                  &                &                \\
		Stellar halo pop 1 & $3.1 \times {10^4},1.23,-2.44$                                                                                                                                                                  & [12.0,13.0[    & $<-1.90$                & 151.0,116.0,95.0                 & \eqref{(2.17)} & \eqref{(2.6)}  \\
		                   &                                                                                                                                                                                                 &                &                         &                                  &                &                \\
		                   & $\{ {{v_0},h_{r,DM},q} \}$                                                                                                                                                                      &                &                         &                                  &                &                \\
		                   & $\left[ {{\rm{km}}\;{{\rm{s}}^{-1}}{\rm{,kpc}}} \right]$                                                                                                                                        &                &                         &                                  &                &                \\
		Dark matter        & $195.6,1.23,0.78 $                                                                                                                                                                              &                &                         &                                  &                &                \\ \hline
	\end{tabular}
\end{table*}

To set up a model of the MW to use for the first four examples, we computed the MW potential (Appendix C) with a set of parameters representative of the major MW constraints as in Table \ref{Table1}.

\begin{figure}
	\plotone{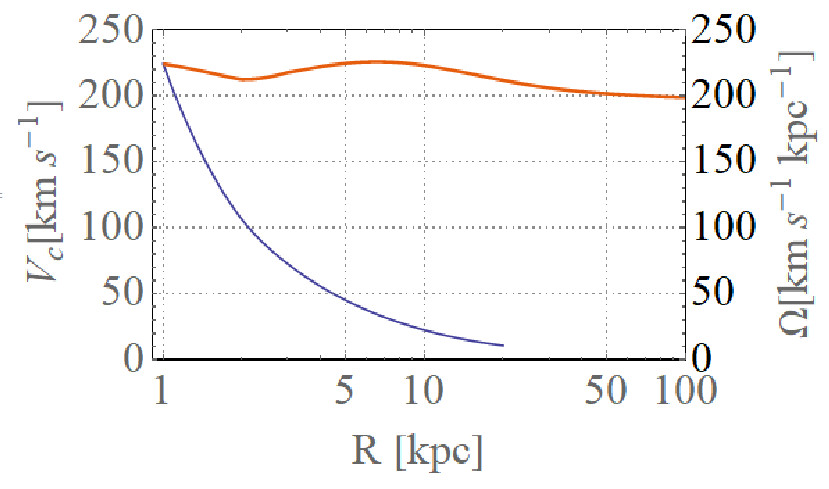}
	\caption{Rotation curve (orange) and angular speed (blue) for the model of  Table \ref{Table1}.}
	\label{RotC}
\end{figure}

The rotation curve for the reference model in Table \ref{Table1} is shown in Fig.\ref{RotC} (where $\Omega \left( R \right)=\frac{{{v}_{c}}\left( R \right)}{R}$ is set to $\Omega \left( {{R}_{\odot }} \right)=35.5\ \text{km}\ {{\text{s}}^{-\text{1}}}\ \text{kp}{{\text{c}}^{-1}}$) to prove the capability of the Poisson solver integrator introduced in \citet[][]{2016MNRAS.461.2383P}. From the solution of the resonance equation $\nu \left( R,m,{{\Omega }_{p}} \right)=\frac{m\left( {{\Omega }_{p}}-\Omega \left( R \right) \right)}{\kappa \left( R \right)}=\pm 1$
we get the inner and outer Lindblad’s resonances (e.g., for two or four spiral arms) at ${{R}_{\text{OLR}}}\left( m=2 \right)=10.6\ \text{kpc}$, ${{R}_{\text{OLR}}}\left( m=4 \right)=8.6\ \text{kpc}$, ${{R}_{\text{ILR}}}\left( m=2 \right)=1.9\ \text{kpc}$, and ${{R}_{\text{ILR}}}\left( m=4 \right)=4.0\ \text{kpc}$, respectively.

This set of parameters is not meant to optimize any FoV, rather to provide a simple global MW potential model. With these values and the equations for the MW potential in \citet{2016MNRAS.461.2383P} or Appendix C, we obtain for the total mass of the MW within 100 kpc, ${{M}_{\text{100}}}\cong 1.11\times {{10}^{12}}\ {{\text{M}}_{\odot }}$. The rotation curve at the solar location is then ${{v}_{c}}\left( {{R}_{\odot }} \right)=225.2\ \text{km}\ {{\text{s}}^{-1}}$, the fraction of spiral component over the disk mass is $\frac{{{M}_{\text{sp}}}}{{{M}_{D}}}\cong 0.14$, the fraction of thick disk density over the thin disk component is ${{\left. \frac{{{\rho }_{\text{thkD}}}}{{{\rho }_{\text{thnD}}}} \right|}_{\odot }}\cong 0.09$, the vertical force on the plane is $\frac{{{F}_{z}}}{2\pi G}\left( {{R}_{\odot }},z=1.1\ \text{kpc} \right)\cong 68.8$, $\frac{{{F}_{z}}}{2\pi G}\left( {{R}_{\odot }},z=2.0\ \text{kpc} \right)\cong 92.3$, and the Oort constants  are
${{O}^{+}}\left( {{R}_{\odot }} \right)=14.9\ \text{km}\ {{\text{s}}^{-1}}\ \text{kp}{{\text{c}}^{-1}}$ and ${{O}^{-}}\left( {{R}_{\odot }} \right)=-13.6\ \text{km}\ {{\text{s}}^{-1}}\ \text{kp}{{\text{c}}^{-1}}$ \citep[e.g.,][]{2016ARA&A..54..529B}. For completeness we compute also ${{O}^{\times }}\left( {{R}_{\odot }} \right)=-2.1\ \text{km}\ {{\text{s}}^{-1}}\ \text{kp}{{\text{c}}^{-1}}$  and ${{O}^{\div }}\left( {{R}_{\odot }} \right)=-2.2\ \text{km}\ {{\text{s}}^{-1}}\ \text{kp}{{\text{c}}^{-1}}$ as defined in \citet{1942psd..book.....C} whose observational constraints are compatible with \citet{2017MNRAS.468L..63B}.

\subsection{Large scale FoV: an SDSS photometry based example}\label{SDSS}

\begin{figure*}
	\includegraphics[width=19cm]{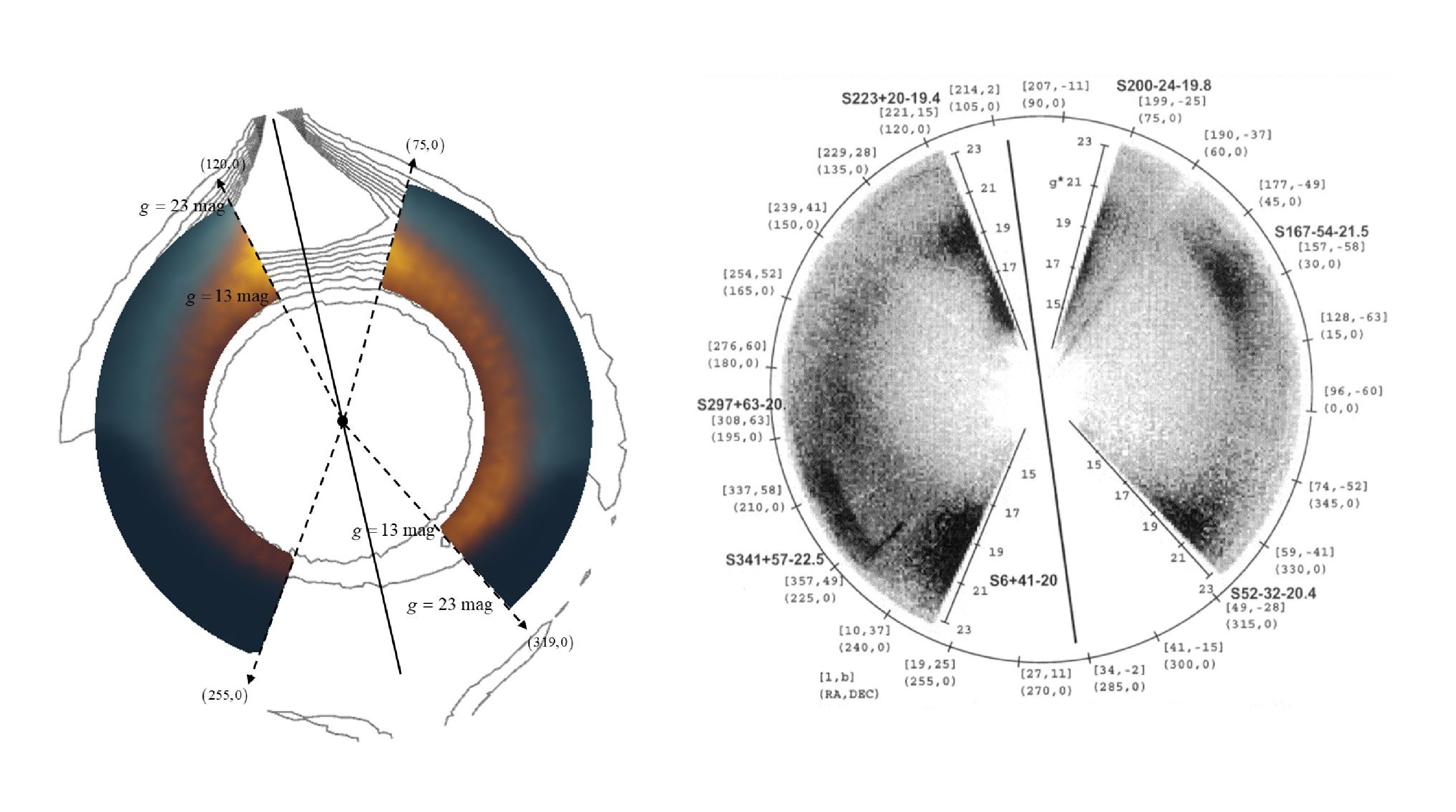}
	\caption{(left) GalMod realization of a density histogram in SDSS photometry of a FoV spanning the great circle along the celestial equator. Coordinates are added to the plot to show correspondence with Fig.1 of \citet[][right panel]{2002ApJ...569..245N}. The locations of the angular cuts (between   deg and   deg and between   deg and   deg) match the SDSS observations that avoid the Galaxy center (marked with a black line). The presence of non-axisymmetric features arising from the spiral arms and tilted bar in GalMod is indicated by the orange to yellow color scheme. The green color scheme is used for the other disk components. The magnitude (i.e. the radial coordinate) and angles are set to match the observed plot on the right approximately. To show the GalMod capabilities, we also under-plot the star-count contour of the other Galaxy components for zones not observed by the SDSS survey (bulge, halo, spiral arms and bar outside the observational SDSS limits). (right) The right-hand plot was reproduced with permission of ApJ. All the observed region are given in \citet[][]{2002ApJ...569..245N}.}
	\label{Greb01}
\end{figure*}

The realization of a FoV in GalMod is not limited in size, contrarily to the Besan\c{c}on model (current on-line ver. dated July 5, 2013, 9:46 CEST @ www.model.obs-Besan\c{c}on.fr) which constrains its use to 25 solid angles each of a sufficiently small size that the density gradients throughout the MW can be approximated as null, and the Trilegal approach which, in its on-line version (ver. 1.6 @ www.stev.oapd.inaf.it/cgi-bin/trilegal) provides a FoV of up to 10 deg$^2$.

\textit{In the era of surveys with extended sky coverage like the SDSS, 2MASS, Gaia, etc., there is a real necessity to have a model which can handle, with speed and precision, a FoV as wide as the entire sky, and representative of billions of stars. This is achieved by GalMod and Galaxia, which are capable of predicting the number of stars no matter the size of the FoV or the presence of density gradients within it}. We show these GalMod features through a qualitative comparison with a large-scale SDSS field. For example, concerning works of observational nature done with the SDSS survey, we can consider Fig.1 of \citet[][]{2002ApJ...569..245N} where a polar diagram $\left( \theta ,r \right)=\left( R.A.,g* \right)$ is plotted for stars down to magnitude $g*\cong 22$ covering stars out to a l.o.s. of 45 kpc. In Fig.\ref{Greb01} we considered a $\left( R.A.,g \right)$ projection in the standard SDSS photometric band $g$ (see paper by \citet[][]{2002ApJ...569..245N} for further details). We adopted similar cuts in $u-g>0.4\ \text{mag}$ and $g-r\in \left[ -1.0,2.5 \right]$ for a stripe centered at $\left( \alpha ,\delta  \right)=\left( 0,0 \right)$ with ${{\delta }_{\min /\max }}=\pm 1{}^\circ .26$ and spanning the whole plane in $\Delta \alpha$. The comparison is not meant to be quantitative; here we plotted just ${{10}^{6}}$ stars instead of $4\times {{10}^{6}}$ as in Fig.1 of \citet[][]{2002ApJ...569..245N} where the 7\% of stars added by the authors from SDSS stripes outside the equatorial plane are missing in our plot, which is limited only to in-plane directions.

Although the comparison is not straightforward, we can highlight the capabilities of GalMod in the context of large sky coverage surveys. Similar overdensities as seen in the SDSS data, arising from the MW's implemented components, are recovered with GalMod near the Galactic center directions (marked with the black line) as evidenced in the observational dataset. The overdensities due to remnants of external satellites evidenced by \citet{2002ApJ...569..245N} (e.g., the Sagittarius dwarf galaxy, other dwarfs and so forth) are not included in GalMod. Beside the central bar asymmetric overdensities we point out the bright overdensities at $g \cong 10$ mag that are due to (in order of relevance) the location of the Sun (here at the center of the plot), the asymmetric extinction model (due to spiral dust distribution adopted, see \citet{2016MNRAS.461.2383P}), and the asymmetric features implemented (i.e., a spiral arm marginally crossing the $\delta = 0$ deg plane). We implemented here also the error function assumed in \citet[][]{2002ApJ...569..245N} for the faintest magnitudes, which contributes to a major blurring effect on the most prominent signatures of these asymmetric features. While this model is not made to quantitatively measure any stripe of Table \ref{Table1} in \citet[][]{2002ApJ...569..245N}, and considering the on-line resolution limitation of the on-line version of GalMod with standard values of Table \ref{Table1} for the density profiles and extinction model, the ability of GalMod to approximate the observed Galactic features is impressive.

\subsection{Non-axisymmetric features: a 2MASS photometry based example of synthetic radial velocity generation}\label{2MASS}
\begin{figure}
	\includegraphics[width=9cm]{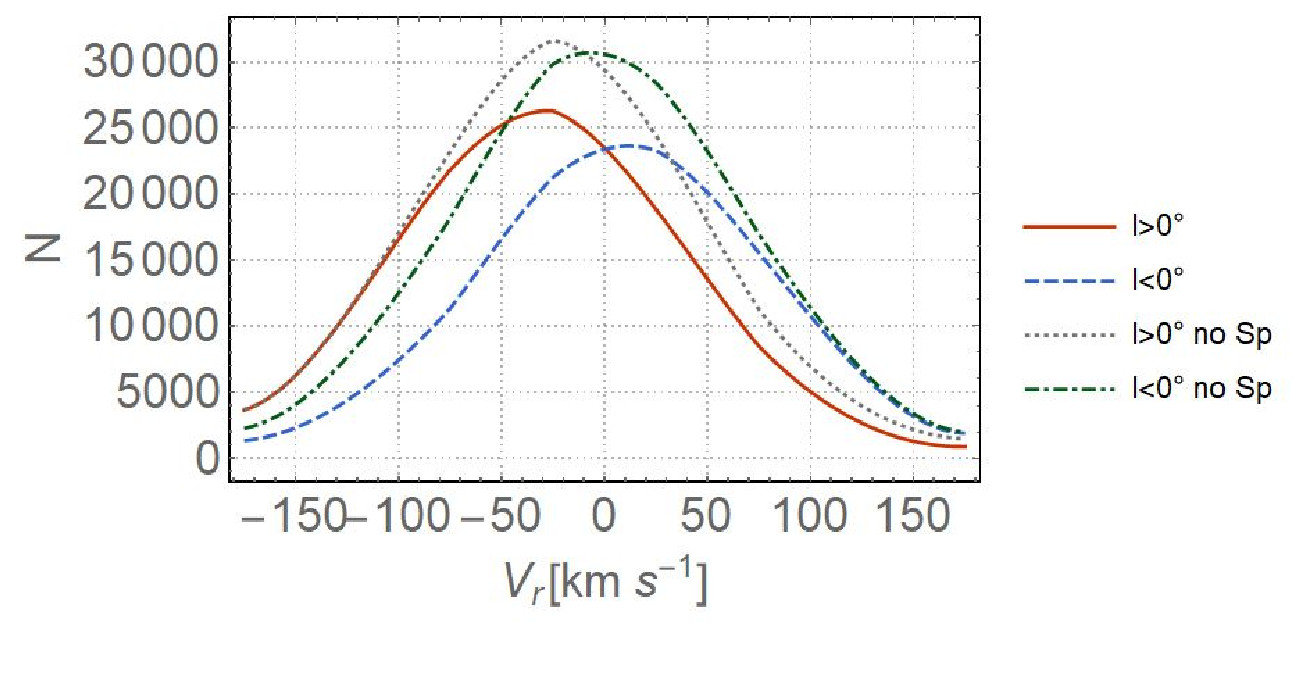}
	\caption{Radial velocity distribution for the field $l\in \left[ -10,10 \right]{}^\circ \times \left[ -4,4 \right]{}^\circ $ split in two samples for positive (red) and negative longitude (dashed blue). The same splitting is done for a mock catalog not including non-axisymmetric features (no bar/no spiral arms/no non-axisymmetric ISM) in gray (dotted) and green (dot-dashed) lines.}
	\label{VrBul}
\end{figure}
GalMod includes bulge, bar, and spiral arm structures, and a coherent description of the kinematics of these features is unique to GalMod. We selected in 2MASS photometry a field with coordinates $l\times b\in \left[ -10,10 \right]\times \left[ -4,4 \right]$ and
with $K<17\ \text{mag}$.
This represents a FoV of increasing interest  \citep[e.g.,][]{2016A&A...587L...6V} thanks to the VVV and the  GIRAFFE Inner Bulge Survey (GIBS) surveys.
 Adopting the parameters of  Table \ref{Table1}, we generate with GalMod the corresponding radial velocity distribution, as shown in Fig.\ref{VrBul}.
 We take this as an example to highlight the different radial velocity distributions of the stars that we can obtain by splitting the mock sample between positive and negative longitude. Note that we did not remove the spiral arms that the FoV is crossing along the l.o.s. toward the bulge. This exercise shows how \textit{two key GalMod ingredients, i.e., the photometric cut and the kinematics description, can be combined in order to obtain the feasibility of upcoming observations}. Once this simulation is convolved with an instrument response function, it can be used to predict the errors with which a plot as in Fig.\ref{VrBul} can be observationally obtained.
 
 For comparison, in the chart of Fig.\ref{VrBul} gray and green lines represent the same splitting realized with the symmetric mock catalog. Moreover, with GalMod the user can also artificially remove the modeling of the bar, the spiral arms, and assume a double exponential disk for stars and ISM (from which dust model and extinction is deduced). The result is again plotted for comparison in Fig.\ref{VrBul}. As is evident, the star count difference between positive and negative longitude is remarkable in the presence of the bar and spirals for the range of ${v_r} \in \left[ { - 150,150} \right]$ while the difference between red/blue and gray/green lines flattens at larger speeds. This example proves the potentiality  of GalMod in enhancing our comprehension of  Galactic observations.

\begin{figure*}
	\plotone{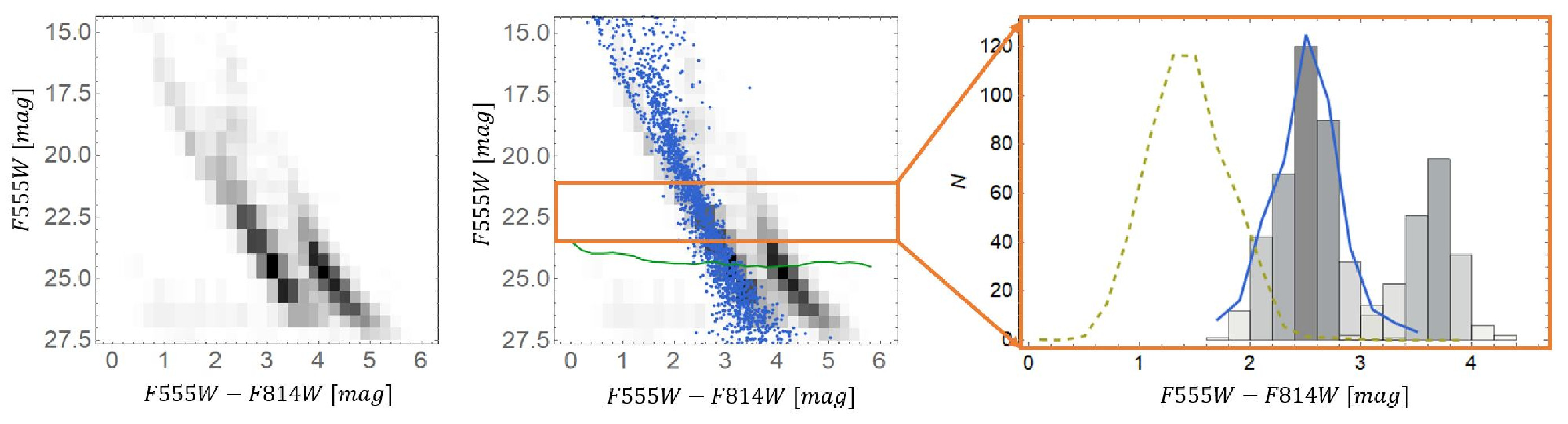}
	\caption{(Left) CMD of a field in the direction of the young Westerlund-2 star cluster. The Hess diagram (black)  shows the resulting stellar color-magnitude density distribution function HST imaging \citep[see ][]{2015AJ....150...78Z}. Two overdensities can be recognized, the redder one being the cluster populations, while the blues sequence stems from Galactic foreground stars. The central plot shows the same diagram compared with GalMod predictions (blue dots) for the Galactic field population, which nicely overlap with the observed foreground sequence. The green line shows the 50\% incompleteness limit. (Right) An example of a distribution in color of stars extracted in a two magnitudes wide luminosity bin in the orange range box from the central plot. The observational Hess diagram is represented as a histogram and the GalMod predictions as the blue line. The Green dashed line is the Trilegal model prediction.
	}
	\label{OpenCl}
\end{figure*}

\begin{figure*}
	\plotone{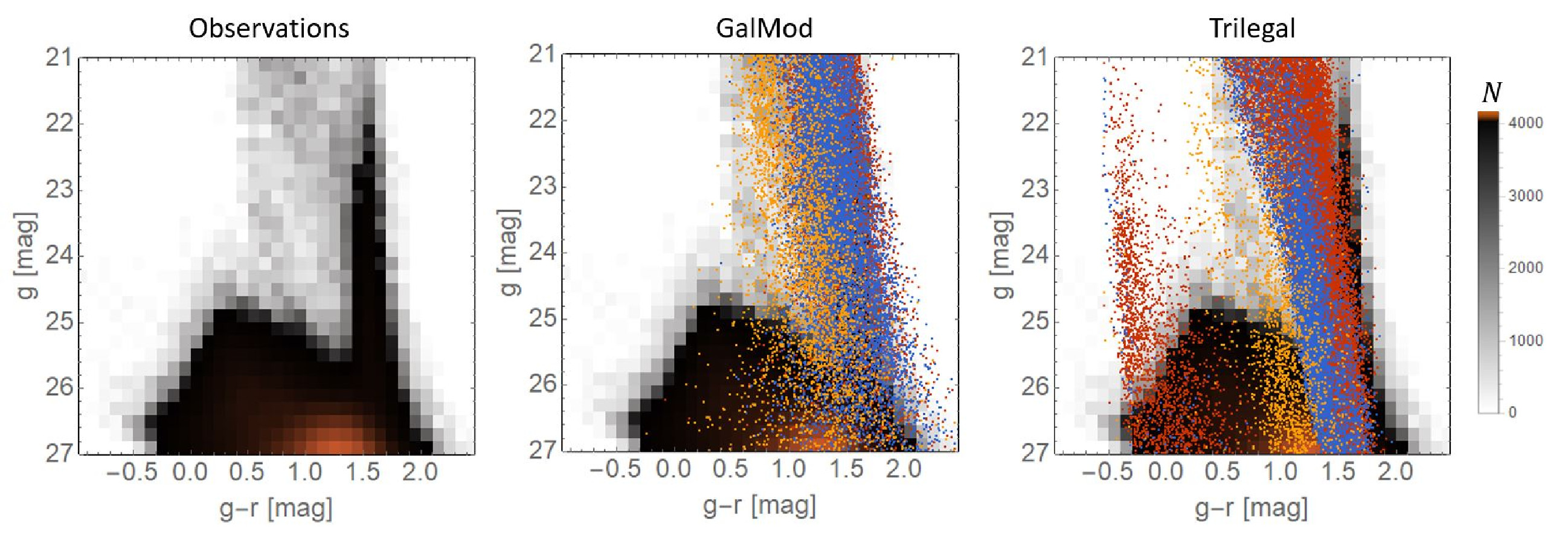}
	\caption{(Left panel) Hess diagram of the observations for a galaxy at 4 Mpc obtained within the PISCeS survey \citep[][]{2016ApJ...823...19C}. (Central panel) Same as the left panel with GalMod simulation of the Galactic foreground population overlaid. (Right panel) Same as the left panel, with the Trilegal model predictions. For all the three panels, the Hess diagram color scheme indicating the number of stars is presented by the right color-bar. Yellow dots represent the Galactic halo component, violet dots the thick disk, and red dots the thin disk component. GalMod overall provides the best match to the observed Galactic foreground sequences.}
	\label{CenA}
\end{figure*}

\subsection{Contamination FoV studies}\label{Contamination}
Beyond obvious GalMod case-studies such as the investigation of spiral galaxy models or the MW central areas, the matching of survey outputs, or the study of asymmetric modeling techniques, \textit{a goal of GalMod is to investigate the contamination due to MW stars in observations of external objects or towards, e.g., a Galactic star cluster}. A CSP of an object of interest, either belonging to the MW (e.g., a globular cluster, an open cluster, an association, a stream, etc.) or outside of the MW (an external galaxy) inevitably suffers from contamination by MW foreground populations. We present here two different examples of this kind.

\begin{itemize}
	\item \textbf{A young massive star cluster inside the MW plane.}
We show an example of contamination due to the MW in the field toward the Westerlund 2 cluster. The adopted dataset is the deep Hubble Space Telescope (HST) imaging, which is detailed in \citet[][]{2015AJ....150...78Z}. The cluster is located in the MW plane at 4.16 kpc from the Sun, and the l.o.s. crosses the Carina-Sagittarius spiral arm. The simulated FoV is centered in the proximity of the cluster at $\left( \alpha ,\delta  \right)=({{10}^{h}}{{23}^{m}},-57.5\deg )$ with an angular size of $12.5\times 12.5\ \text{arcmin}^2$ and extends up to ${{r}_{\text{hel}}}\le 5\ \text{kpc}$. Fig.\ref{OpenCl} shows the observed CMD and the CMD predicted by GalMod. A Hess diagram (in black) shows the observed data while the resulting DFs of the MW CSPs from GalMod are shown with blue dots.

The simulation and the observation agree even in this direction complicated by the presence of spiral arms and extinction. In the plot, the five thin-disk stellar populations of Table \ref{Table1} are grouped together and shown as blue dots. We omitted the thick disk component and the halo because of minor statistical importance. The agreement is not perfect because  the GalMod FoV is not finely tuned to the observed field: the mathematical representations we are using are just a rough approximation of Nature, and we do not expect to observe mathematically perfect exponential disks nor perfect logarithmic spiral arms. Additionally, the observed data are not corrected for incompleteness effects \citep[see green line Fig.\ref{OpenCl} and][]{2017AJ....153..122Z}.

A comparison with a model not equipped with spiral arms is in this example especially striking: on the right panel of Fig.\ref{OpenCl} the dashed line (realized by Trilegal) is compared with the corresponding blue line (realized by GalMod) to evidence the effect of the spiral arm stellar distribution and spiral arm extinction against a purely axisymmetric disk provided by Trilegal.  An even better result could be eventually achieved by searching for the best spiral arm pitch angle or scale length to match exactly the number of stars observed, a work that we consider to be beyond the goal of the present paper.

\item \textbf{Foreground Galactic sequences in extragalactic observations: the case of Cen A. }
To be able to estimate the stellar foreground contamination caused by our Galaxy is of paramount importance for the study of extragalactic objects resolved into stars, e.g., nearby galaxies within the Local Group or even the Local Volume. In such cases, the Galactic stellar populations along the adopted l.o.s. will have the role of foreground sequences contaminating the (more distant) target populations, for which for instance we want to estimate the properties from its CMD (e.g., total magnitude, distance, structural parameters, and so forth). As an example, we choose the recent wide-field Panoramic Imaging Survey of Centaurus and Sculptor (PISCeS), performed with the Magellan/Megacam imager (for more details, see \citealt[][]{2014ApJ...795L..35C}, \citealt[][]{2016ApJ...823...19C}, \citealt[][]{2014ApJ...793L...7S}, \citealt[][]{2016ApJ...830L..21T}). The survey targets two MW-mass like galaxies at 4 Mpc, i.e., the spiral galaxy Sculptor and the elliptical galaxy Centaurus A (Cen A). The final goal of PISCeS is to map the resolved stellar halo of these galaxies out to a galactocentric distance of about 150 kpc, to uncover substructures and faint satellites and compare them to predictions from cosmological simulations. As shown in the left panel of Fig.\ref{CenA}, for a galaxy at 4 Mpc only the brightest red giant branch (RGB) stars can be resolved with ground-based observations, which are found at $r>25\ \text{mag}$ and $g-r\cong 1.2\ \text{mag}$. The blue sequence at $g-r\cong 0.2\ \text{mag}$ is populated by unresolved background galaxies, while the sequences brighter than $g = 25$ mag (at $g-r\cong 0.5\ \text{mag}$ and $g-r\cong 1.5\ \text{mag}$) are Galactic. To correctly assess the presence and number of true Cen A RGB stars, we must statistically decontaminate this population from the foreground populations that have an overlapping color-magnitude distribution: therefore, an accurate Galactic field population model is crucial for such studies.

In Fig.\ref{CenA} we compare the predictions from GalMod and Trilegal.
In the GalMod synthetic realization shown in Fig.\ref{CenA} (central panel), the five thin disk populations are color-coded in red; violet is used for the thick disk and orange for the halo; the same color scheme is used for the Trilegal simulations in the right panel. Both sets of models have been convolved with photometric errors obtained from the observational dataset. The results from the two models are broadly comparable, except for the blue vertical sequence centered at $g-r\cong -0.3\ \text{mag}$ predicted by Trilegal, which is not seen in GalMod nor in the observed data. The GalMod populations   match the observed Galactic sequences significantly better, especially in color, while the Trilegal predictions have a systematically bluer color than the observed data. The realization is not fine-tuned to the FoV and could eventually be improved by searching for the best matching Galactic parameters from both models.

\end{itemize}
\subsection{GalMod as generator of N-body initial condition}\label{Nbody}
\begin{figure*}
	\plotone{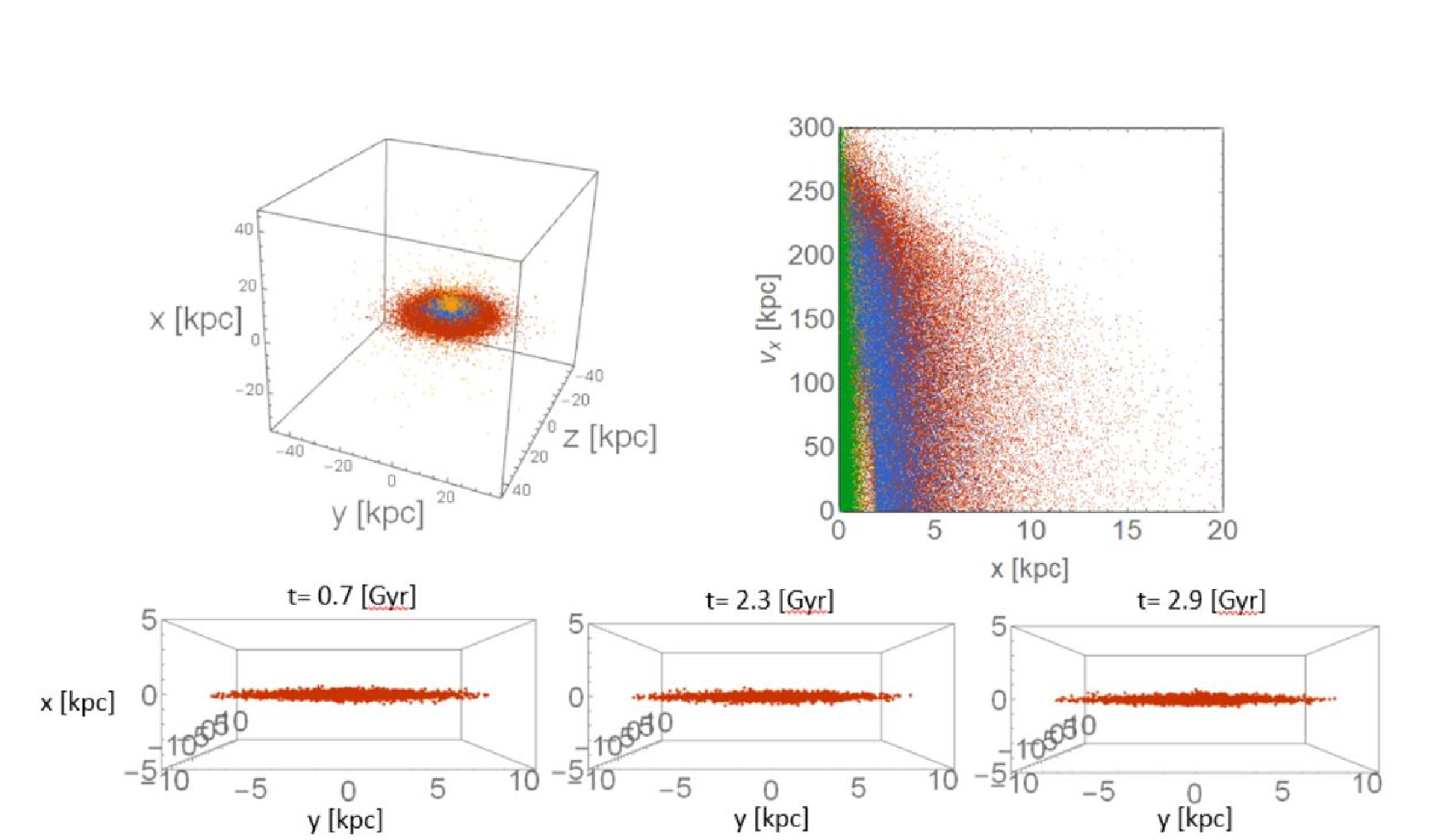}
	\caption{(Left panel) Configuration space of the spiral galaxy model in Table \ref{Table1} and (right panel) one Poincar\`{e} section. Red dots stand for thin disk stars, blue is used for thick disk stars, green for the bulge, and yellow for the halo. It is evident in the Poincar\`{e} section that there is a little gap (at $x = 1.6$ kpc, negative $x$ and $v_x$ are omitted) in the configuration space distribution for the thin and thick disk stars due to the bar/disk connection which is visible also in the contour plots of Fig.\ref{BarMW} as blue/red density contrasts. (Bottom panels) Three snapshots from the first 3 Gyr of a test evolution are shown for a galaxy disk phase-space alone where 100000 particles are embedded in a static analytical halo potential (see text for details) \citep[after ][]{2010A&A...514A..47P}}
	\label{NbodyIC}
\end{figure*}
One of the missing ingredients of the sophisticated modeling technique that we developed here is the time evolution. Currently, the only known techniques able to evolve the gas and stellar component in time in a MW-like galaxy simultaneously, are N-body integrators \citep[e.g.,][]{2017MNRAS.467.2430M,2003MNRAS.340..908K,2005MNRAS.364.1105S,2010A&A...513A..36M,1999A&A...348..371B}. Nevertheless, the value of these techniques is at the present stage purely theoretical in nature, because of the difficulty to tune them to match precisely an observational survey of the MW starting from high redshift. This limitation is in part due to the resolution problems that affect this N-body numerical integration and in part due to the missing phase-space map (location and velocity) of the MW gas distribution. Hence, because of the presence of a large gas fraction, spiral arms and giant molecular clouds that can scatter stars from their unperturbed orbit, techniques based on orbital integration are of limited practical use. GalMod, even though it includes spiral arms and bars, still misses the possibility to track in time the orbits of giant molecular clouds or the gas evolution in the configuration space.

Nevertheless, a small effort in the attempt of bridging this gap between N-body simulations and observations can be made.

GalMod comes with a complex Poisson-solver fine-tuned for the MW, able to generate, at least for the stellar components, fair i.c. for the MW phase-space or any other spiral galaxy. Hence it is natural to try to use GalMod to match perfectly a given survey of a portion of MW, and then to generate the phase-space for the whole Galaxy, with the same approach developed, e.g., by \citet[][]{2013MNRAS.430.1928H}, \citet[][]{Hunt2014}, \citet[][]{2013MNRAS.432.3062H}. The problem of accurate gas and dark matter maps will remain.

In GalMod kinematics is connected to the global galaxy potential only through the first order moments (see Appendix C) of a Boltzmann collisionless equation. Furthermore, the gas and dark matter distribution have to be added in agreement with the density profiles adopted in GalMod either through particle distribution, or through mass distribution on a mesh-grid, or through analytical external potential added to the N-body integrator. Keeping in mind these two requirements, GalMod can easily be used as a collisionless equilibrium structure generator.
This is in line with a quite long tradition of studies \citep[e.g.,][]{1993ApJS...86..389H} and these techniques are in constant development \citep[e.g.,][]{2014MNRAS.444...62Y,2006ARep...50..983R}. In Fig.\ref{NbodyIC} we present an example of i.c. of a galactic model tuned to match a disk galaxy. GalMod produces mass, metallicity, and phase-space for the input model as extensively explained in \citet[][]{2012A&A...545A..14P}. The stability of these equilibrium i.c. has been tested over a decade with different schemes for orbit integration: with MPI/parallel-Treecode based codes as \citet{2010A&A...513A..36M} in works such as \citet{1998MNRAS.297.1021C, 2000MNRAS.312..371B,2010A&A...514A..47P,2003A&A...405..931P}, with  GPU-integrator based codes as in \citet{1999A&A...348..371B} in works as those of \citet{2010A&A...514A..47P,2011A&A...525A..99P}, and with TreeSPH  based codes as in \citet{2003MNRAS.340..908K} in, e.g., \cite{2012A&A...545A..14P}. All these works have considered an i.c. generator for disk/dwarf galaxies in isolated/interacting systems and, independently from the ''engine'' (i.e., the numerical integrator) the i.c. used by GalMod always led to stable results(\footnote{The first work explicitly employing this type of i.c. condition generation probably dates back to \citet{1993ApJS...86..389H}.}).

The bottom panels in Fig.\ref{NbodyIC} show the evolution in the ${\bf{\gamma}}$ space (i.e., keeping the mass and metallicity, and the $M \times Z$ section of $\mathbb{E}$ artificially constant, and following only the dynamical evolution in $\bm{\gamma }=\bm{\gamma }\left( t \right)$) of the GalMod generated i.c.. Details on the library of i.c. generated are available in \citet{2010A&A...514A..47P}, with the only difference being that a fixed dark matter halo potential following Sec.4.1.3 of \citet[][]{2016MNRAS.461.2383P} or Appendix C is artificially implemented. Note that the allowed parameter space accessible through the GalMod web interface does not always lead to a dynamically stable structure. From the dynamical point of view bar/spiral arm instabilities are related to Safronov-Toomre criterion whose values are not provided by GalMod. From the numerical point of view, the stability is entirely dependent on the integrator scheme adopted. The bottom panels of Fig.\ref{NbodyIC} are just meant to show the correct treatment in a tree-code scheme \citep[e.g.,][]{2010A&A...513A..36M} of the numerical vertical heating that is avoided with i.c. equations implemented in GalMod \citep[the parameters of the simulation are exactly as in ][and reference therein]{2010A&A...514A..47P}. It is the responsibility of the user to compute the necessary indicators to realize a stable (or unstable) structure. Furthermore, once the gas treatment is accounted for, DM distribution and gas temperature dominate the evolution of the system entirely as seen, e.g., in \citet{2010A&A...514A..47P} where the energy feedback enhances the fluctuations of the DM gravitational potential and change the shape of $\rho_{\text{DM}}$ from cuspy to cored.

\subsection{M31 model}\label{M31}
\begin{figure*}
	\plotone{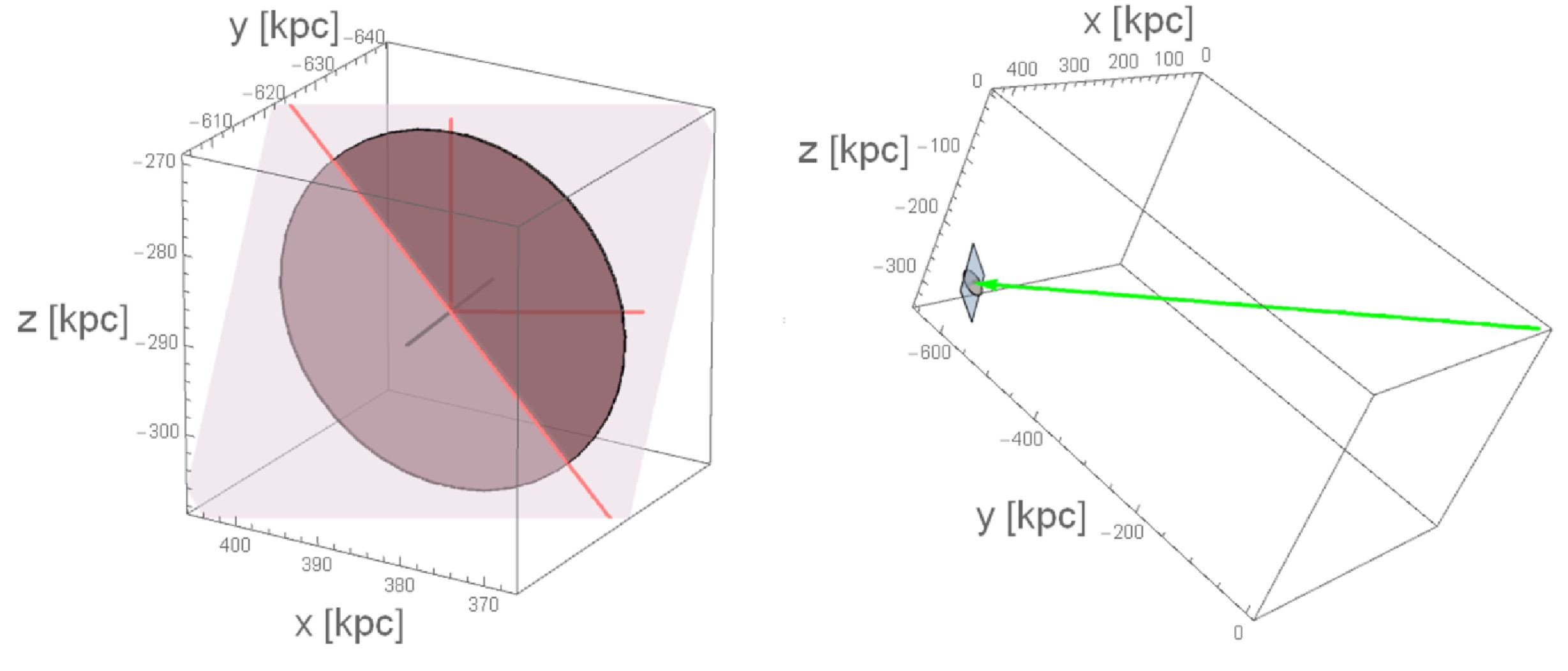}
	\caption{Geometrical model of M31. (Left) A solid disk has been plotted in place of the M31 disk density distribution, and a small plane represents a local approximation for the celestial sphere crossing the M31 barycentre at the M31 location. (Right) The l.o.s. is visible in green together with the MW (omitted) location at the origin of the system of reference. The solar location is at the basis of the green arrow.}
	\label{M31loc}
\end{figure*}
Because of the recent interest in the MW companion spiral M31, e.g., with the Pan-Andromeda Archaeological Survey (PAndAS, \citet[][]{2009Natur.461...66M}), we propose a more detailed model of the sky FoV in the direction of Andromeda by including Andromeda itself. The possibility comes naturally as a consequence of the wide parameter space allowed in GalMod (seen in Fig.\ref{ShapeFunk}, Fig.\ref{waveNum}, Fig.\ref{Amplit}) and of the possibility to arbitrarily move the observer position as shown in Eq.\eqref{(2.19)}.

Simply speaking, we need to shift and rotate the GalMod model to overlap the M31 position assuming the observer to be located at the site of the Sun. GalMod is equipped with a Poisson equation solver able to accommodate M31 scale parameters as large as in \citet{2002ApJ...573..597K} for the scale of the M31 disk and from \citet{2014ApJ...780..128I} for the halo.  The angle between the North Celestial Pole (NCP) and the projected major axis of M31 on the celestial sphere (CS) is $\theta \equiv \widehat{{{{\bm{\hat{e}}}}_{E}}{{{\bm{\hat{e}}}}_{N}}}$, counted from the northern direction ${{\bm{\hat{e}}}_{N}}$ positive toward the eastern direction ${{\bm{\hat{e}}}_{E}}$. We approximate the CS with a plane, $\Pi$, neglecting CS curvature at the M31 position. In this case both ${{\bm{\hat{e}}}_{E}}\in \Pi$ and ${{\bm{\hat{e}}}_{N}}\in \Pi$. The angle between the normal to the disk plane $\bm{\hat{n}}$ and the l.o.s.  is $i\equiv \widehat{l.o.s./\bm{\hat{n}}}$, i.e., the inclination. Finally, we need to account for the North-West (NW) edge of M31 being closest to us \citep[e.g.,][]{1977MNRAS.181..573N,1979A&A....75..311H}. We call the position of the Sun at ${{\left( R,\phi ,z \right)}_{\odot }}\equiv {{\bm{x}}_{\odot }}$ and the distance from the Sun to M31, ${{r}_{\text{hel,M31}}}=785\ \text{kpc}$ in the direction of $\left( l,b \right)_{\text{M31}}=\left( 121.6,-21.6 \right)\ \deg$ (in Galactic coordinates). In a right-handed system of reference, we consider a transformation $T$ given by
\begin{equation}\label{(4.4)}
	T:\left\{ \begin{aligned}
	& {{x}_{\text{M31}}}={{r}_{\text{hel,M31}}}\cos {{b}_{\text{M31}}}\cos {{l}_{\text{M31}}}+{{x}_{\odot }} \\
	& {{y}_{\text{M31}}}={{r}_{\text{hel,M31}}}\cos {{b}_{\text{M31}}}\sin {{l}_{\text{M31}}}+{{y}_{\odot }} \\
	& {{z}_{\text{M31}}}={{r}_{\text{hel,M31}}}\sin {{b}_{\text{M31}}}+{{z}_{\odot }}, \\
	\end{aligned} \right.
\end{equation}
to shift a galaxy model to the actual M31 position ${{\bm{x}}_{\text{M31}}}=\left\{ \text{386}\text{.5,-624}\text{.3, -288}\text{.9} \right\}\ \text{kpc}$. Hence, we define a vector that points from the location of M31 to the Sun as  ${{\bm{x}}_{\text{M31}\to \odot }}={{\bm{x}}_{\text{M31}}}-{{\bm{x}}_{\odot }}$.

We have then to consider that the inclination between the normal to the plane of M31, $\bm{\hat{n}}$, and the l.o.s. is $i=77.5{}^\circ$. With respect to the inclination of the l.o.s. to the plane of the disk (the plane $ Oxy $), there exists an angle equal to $b_{\text{M31}}$ that must be considered. Hence, we must tilt the disk by about $90{}^\circ -\left( i^\circ +b_{\text{M31}}  \right)$. This is performed with a rotation, say $R_1$, by this angle around the vector $\bm{k}$ given by the cross product of the vector ${{\bm{x}}_{\text{M31}\to \odot }}$ and the axis ${{\hat{e}}_{z}}$: $\bm{\hat{k}}\equiv \frac{{{\bm{x}}_{\text{M31}\to \odot }}\times {{{\hat{e}}}_{z}}}{\left\| {{\bm{x}}_{\text{M31}\to \odot }}\times {{{\hat{e}}}_{z}} \right\|}$  anchored at the fixed point ${{\bm{x}}_{\text{M31}}}$ (here $\left\| \bullet \right\|$ is the standard Euclidean norm). This inclination has a degree of freedom in its sign because the normal $\bm{\hat{n}}$ can perform an angle of $i$ in two directions but we choose an inclination of $90{}^\circ + b_{\text{M31}}$ because we want the edge of the M31 disk that is closest to us to point in the NW direction on the celestial sphere. The result is that the normal vector $\bm{\hat{n}}$ is tilted to point to the final position (we call it still $\bm{\hat{n}}$).

Now the tilt of the M31 major axis projected on the celestial sphere remains to be fixed. We need to find the intersection line between the disk plane of M31 and the plane of the celestial sphere (which so far is still coplanar with the plane $Oxy$). Using the Hessian equation for the planes, we want to solve the system
\begin{figure*}
	\plotone{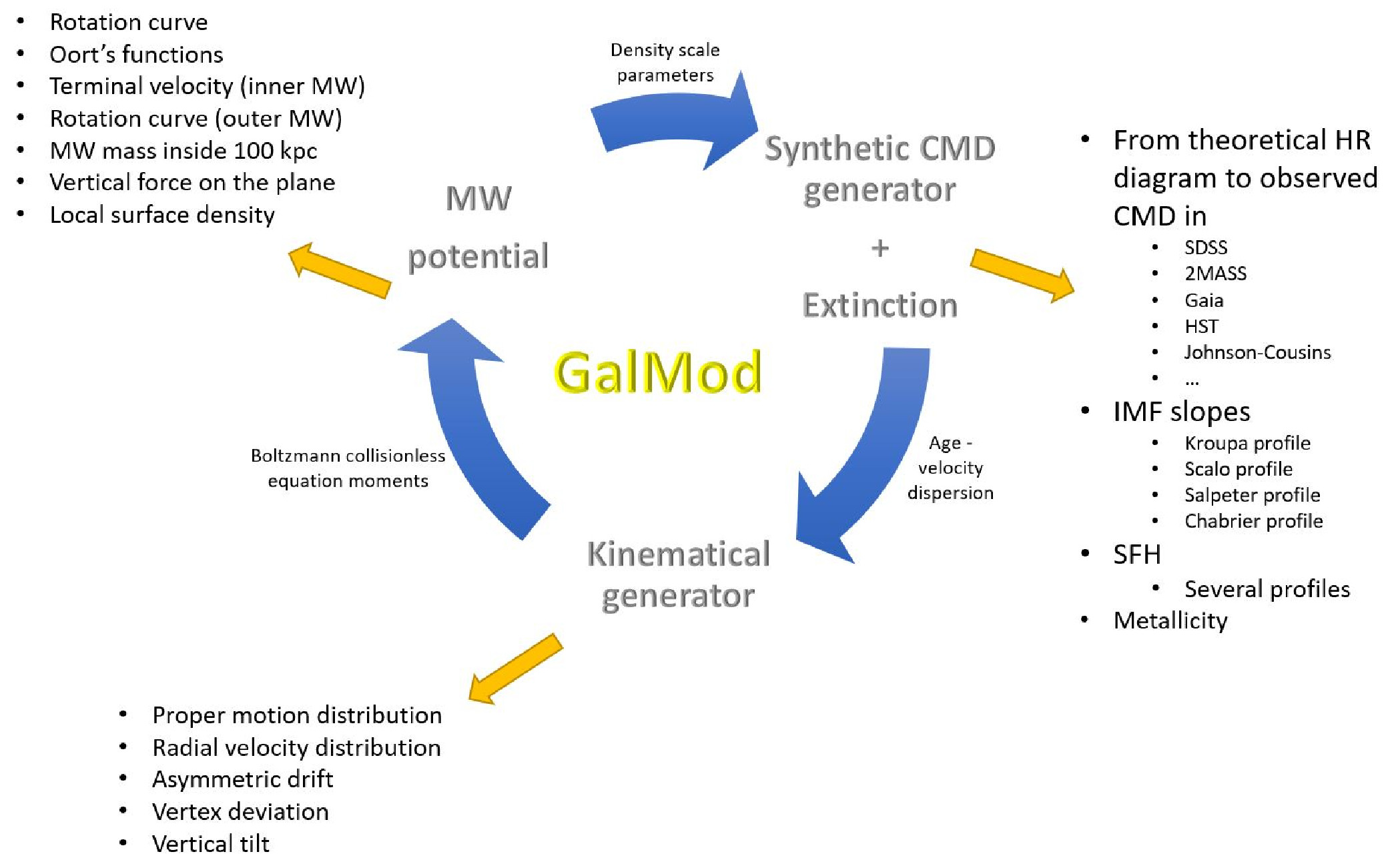}
	\caption{Consistency cycle of the GalMod model (\href{www.galmod.org}{www.GalMod.org}).}
	\label{Cycle}
\end{figure*}
\begin{equation}\label{(4.5)}
	\left\{ \begin{aligned}
	& \left\langle \bm{\hat{n}},\bm{x}-{{\bm{x}}_{\text{M31}}} \right\rangle =0 \\
	& \left\langle {{\bm{x}}_{\text{M31}\to \odot }},\bm{x}-{{\bm{x}}_{\text{M31}}} \right\rangle =0, \\
	\end{aligned} \right.
\end{equation}
where the second equation is the equation of the plane passing through the position of M31 with its normal pointing along the l.o.s., while the first is the equation of the plane of M31. The solution for the intersection line is found by numerical approximation ($l\in \mathbb{R}$ free parameter):
\begin{equation}\label{(4.6)}
	{{\alpha }_{1}}=\left\{ \begin{aligned}
	& x={{x}_{\text{M31}}}-n_x l \\
	& y={{y}_{\text{M31}}}+n_y l \\
	& z={{z}_{\text{M31}}+n_z l}. \\
	\end{aligned} \right.
\end{equation}
This equation gives the line ${{\alpha }_{1}}\in \Pi $ of the major axis of M31 in the celestial sphere not tilted, i.e. the projected major axis (pMA) direction ${{\bm{\hat{e}}}_{\text{pMA}}}$. Finally, we want to find the second line ${{\alpha }_{2}}\in \Pi $ representing the direction on the CS, $\Pi $, of the NCP. Recovering the orientation of this line means to solve the system:
\begin{equation}\label{(4.7)}
	\left\{ \begin{aligned}
	& \left\langle {{{\bm{\hat{e}}}}_{\text{pMA}}},\bm{\hat{n}} \right\rangle =\cos  \theta   \\
	& \left\langle {{{\bm{\hat{e}}}}_{\text{M31}\to \odot }},\bm{\hat{n}} \right\rangle =0, \\
	\end{aligned} \right.
\end{equation}
and of course, $\left\| {\bm{\hat{n}}} \right\|=1$, where the first equation of the system represents a cone having an angle of $\theta$ with the line through the projected major axis of M31 on the CS. The second equation of the system relates to the condition that the wanted normal has to belong to the CS in the direction of M31. We proceed numerically to obtain the line:
\begin{equation}\label{(4.8)}
	\left\{ \begin{aligned}
	& x={{x}_{\text{M31}}}+{{{\hat{n}}}_{x}}l \\
	& y={{y}_{\text{M31}}}+{{{\hat{n}}}_{y}}l \\
	& z={{z}_{\text{M31}}}+{{{\hat{n}}}_{z}}l, \\
	\end{aligned} \right.
\end{equation}
and we conclude.
To recap, we applied
\begin{itemize}
\item an initial translation to the M31 location of the MW-centered reference frame $T\left[ \bullet  \right]:T\left( {{\bm{x}}_{\odot \to M31}} \right)\left[ \bullet  \right]$;
\item a rotation on the plane orthogonal to the normal to the MW plane and the direction of the M31-sun anchored at M31 position as ${{R}_{1}}:{{R}_{1}}\left[ \bullet  \right]\left( i,{{\bm{x}}_{\odot \to \text{M31}}},{{\bm{x}}_{\text{M31}}} \right)\left[ \bullet  \right]$;
\item rotation of the position angle counted counterclockwise on the plane orthogonal to the l.o.s. anchored at the M31 position as ${{R}_{2}}:{{R}_{2}}\left( PA,{{\bm{x}}_{\odot \to \text{M31}}},{{\bm{x}}_{\text{M31}}} \right)\left[ \bullet  \right]$;
\end{itemize}
and the desired transformation matrix can be written in a compact form as:
\begin{equation}\label{(4.8)tris}
M\left( {{{\bm{x}}_ \odot },{{\bm{x}}_{{\text{M31}}}},i,PA} \right)\left[  \bullet  \right] = {R_2}\left( {{R_1}\left( {T\left( {{{\bm{x}}_ \odot },{{\bm{x}}_{{\text{M31}}}}} \right),i} \right),PA} \right)\left[  \bullet  \right]
\end{equation}
which takes any vector defined in the MW reference frame to a target galaxy reference frame (M31 in this case). The results of the translation and the rotations are represented in Fig.\ref{M31loc} (right and left panels respectively). All the values necessary for $M$  are available from public catalogs such as SIMBAD-astronomical database CDS(\footnote{http://simbad.u-strasbg.fr/simbad/}) or, e.g., from \citet{2006AJ....131.1163S}.
The plot of this transformation is given in Fig.\ref{M31loc}.

In addition to the configuration space for M31, we added the peculiar velocity vector ${{\bm{v}}_{\text{M31}}}$ for M31 obtained in \citet[][]{2007A&A...463..427P} as:
\begin{equation}\label{(4.9)}
\left( {{\mu }_{l}},{{\mu }_{b}} \right)_{\text{M31}}=-\left( 3.03,3.39 \right)\times {{10}^{-4}} \text{arcsec} \ \text{y}{{\text{r}}^{-1}}.
\end{equation}
These values are a consequence of the stationary point of an action, i.e., $\delta \int_{t}^{{{T}_{G}}}{L\left( \bm{x},\bm{\dot{x}};t \right)dt}=0$, suitably written for the evolution of the nearest group of galaxies, IC342, Maffei, Andromeda, M81, Cen A and Sculptor (see also Table 1 in \citet[][]{2009A&A...499..385P} for further details).

This phase-space transformation can be applied to every point of the phase-space and has been introduced in GalMod to obtain the FoV of M31(\footnote{It is worth to stress that the adaptation of the MW model to M31 (or any other spiral galaxy) is an oversimplification. We do not expect that M31 or even the MW are completely isolated systems, and it is well known that the interactions with their dwarf companions cause a morphological distortion of their spiral arms \citep{1998A&A...338L..33H,2006ApJ...638L..87G}. Interaction with external companions is indeed often advocated as a source of excitation for spiral density modes \citep{2014dyga.book.....B}.}).

In this example \textit{GalMod allows us to account for density gradients within the FoV of any chosen model of M31 without limits on the size and allows us to produce mock catalogs of the whole M31 in a single shot}.

\subsection{The extinction model}
In \citet[][]{2016MNRAS.461.2383P} we introduced an extinction model based on the one presented in the DART-ray radiation transfer code \citep{2017arXiv170903802N}. We assumed the dust model of \citet{2007ApJ...657..810D}, calibrated with the extinction curve, metal abundance depletion, and dust emission measurements in the local MW. From the extinction parameters and gas density, the optical depth crossed by the starlight is then numerically integrated, and the extinction derived. This procedure gives the GalMod user the possibility to directly tune the extinction both by adjusting the gas density and by modifying it through the spiral and bar density distribution profiles. We stress that no other codes allow a similar fine-tuning procedure through their web-page.

The methodology adopted by DART-ray allows one not only to compute the total flux of light from a star in a certain direction, but also the reflected light from the same direction due to the dust. This novel model, its underlying equations, and a comparison to different models of radiative transfer solutions are addressed in detail by \citet{2017arXiv170903802N}, and we refer the interested reader to that paper and the references quoted therein. In \citet{2016MNRAS.461.2383P} we limited ourselves to showing the impact that such an extinction model has on the final result of interest to GalMod users, the CMD and the ISM distribution. The fundamental dependence of the scattered light on the wavelength was already pointed out, e.g., in \citet{2004A&A...419..821T}, \citet{2004ApJ...617.1022P}, \citet{2001MNRAS.326..733B}. This point is further illustrated, e.g., by Fig.17 in \citet{2015MNRAS.449..243N}: the authors show that the fraction of scattered to total predicted stellar emission as a function
of wavelength can be as high as 25\%, depending on the galaxy inclination (referred to as $i$ in our previous Sec.\ref{M31}). For detailed discussion see, e.g., \citet{2017A&A...607A.125N, 2015MNRAS.449..243N,2014MNRAS.438.3137N} and references therein.  

As example of the importance of the spiral-geometry introduced in the extinction, in Fig.~8 of \citet{2016MNRAS.461.2383P} we have shown how the GalMod extinction model is entirely independent of geometry: no fixed geometry (i.e., a parametric function) or parametric cloud distribution is necessary. In Fig.10 of \citet{2016MNRAS.461.2383P}, we compared the GalMod extinction with a standard literature approach such as the double exponential ISM profile. Finally, the overall effect of the extinction model was compared to the Besan\c{c}on galaxy model in Fig.~9 of \citet{2016MNRAS.461.2383P}. These examples suffice to show the effect on the CMDs of the sophisticated extinction model we adopt in comparison with other literature standards, and we will not repeat them here. 

 GalMod aims to model not only the MW but also external galaxies(\footnote{Note, e.g., that Gadget has to be equipped with extra software to produce mock catalogs of an external galaxy in any photometric band, while Galaxia requires an external galaxy model, such as those from GalMod or any other N-body i.c. simulator, to produce mock catalogs in any available photometric band}). Hence, the potential of the GalMod extinction model should be evident after the considerations of the previous section. When GalMod is used to model M31, the extinction has to be computed not only in the foreground (i.e., in the MW) but also within M31 itself. This is because a star behind the bulge of M31 is less well visible with respect to a star at the closer edge of the M31 disk.  GalMod accounts for the 3D physical distribution of the external galaxy, e.g. M31, and it applies the extinction to the stars to automatically account for the magnitude and distance selection cuts of the user. This is done by accounting via numerical integration for the extinction parameters and the gas density of the external object, and for its optical depth crossed by the starlight from the target to the observer. The same can be done for any dwarf galaxy modeled with GalMod.

In the future we aim to introduce the possibility to model not only collisionless stellar systems but also CSPs of open/globular clusters where the GalMod extinction model will play a key role. 

Finally, we need to mention the following limit imposed on GalMod by the implemented extinction model. GalMod focuses on photometry, chemistry, and phase-space of a collisionless stellar system in the Local Group. A maximum value for the distance of the stars, ${{\bm{r}}_{\max }}$ , that we are allowed to model is imposed for Local Group objects: roughly for every arbitrary observer location, ${{\bm{r}}_{\odot }}$ , we imposed a limit of $\left\| {{\bm{r}}_{\odot }} \right\|<1$ Mpc and $\left\| {{\bm{r}}_{\odot }}-{{\bm{r}}_{\max }} \right\|<1$ Mpc. This is because higher l.o.s. column densities in the computation of the extinction can impact negatively on the GalMod performance even in empty intergalactic spaces. Furthermore, GalMod aims to model only the MW and Local Group galaxies, while more distant objects shall be modeled by accounting for their redshift in their CMDs, as well as for the Hubble expansion in their radial velocity. We reserve to develop the connection between resolved and integrated stellar populations and cosmological effects in future works.

\section{Conclusions}\label{Conclusions}
We have presented several features of GalMod, a versatile tool to model star-counts of stellar population surveys of the MW and other galaxies. Of these, the most important ones that we want to emphasize here are: GalMod
\begin{itemize}
	\item \textit{has no limits on the size of the field of view generated,}
	\item \textit{includes non-axisymmetric features such as spiral arms and bar,}
	\item \textit{offers a wide range of photometric systems,}
	\item \textit{includes an geometry-independent ray-tracing extinction model,}
	\item \textit{offers the possibility to simulate the M31 FoV,}
	\item \textit{offers the possibility to realize a collisionless semi-equilibrium model generator for N-body integrator i.c., }
	\item \textit{is freely accessible via a web interface at \href{www.galmod.org}{www.GalMod.org}.}
\end{itemize}

This work completes the description of GalMod started in \citet[][]{2016MNRAS.461.2383P} with the modeling of the thin and thick disk and ISM component and where we implemented spiral arm components including their photometry, chemical composition, and phase-space information.

We introduced in GalMod a sophisticated extinction model based on the DART-ray code \citep{2017arXiv170903802N}. We calibrated DART-ray for the dust to gas density, extinction curve, metal abundance depletion and dust emission measurements in the local MW following \citet{2007ApJ...657..810D}. From the extinction coefficients and gas density the optical depth passed by the star light is then numerically integrated and the extinction naturally derived. This procedure gives the GalMod user the possibility to directly control the extinction both by changing ruling the gas density and by modifying it through the spiral and bar density distribution profiles. No other codes allow through their web-page a similar fine-tuning procedure.

In this work, we completed the description of the features of GalMod by presenting a non-axisymmetric bulge component connected with spiral arm components and a second spherical component. We extended the number of photometric systems that are available to GalMod to include Gaia DR2, 2MASS, SDSS, HST, and many others thus giving the user a larger possibility to model data in the photometric bands of interest and avoiding the introduction of conversion formulae that risk introducing additional errors in the analysis.

The central part of the Galaxy as presented is the result of the superposition of a spherical exponential model and a bar model obtained directly from a fine-tuned bar instability model. The spherical model offers tunable parameters for the total mass and scale radius with a free spherical ellipsoid of velocities to control the kinematic temperature of the MW's central FoVs. The bar extends from the central region of the modeled galaxy naturally to the spiral arm structure and naturally it links to the pattern speed of the spiral arms with a solution of continuity: bar and spiral arms represent a structure connected by the global gravitational potential. 

The consistency cycle is represented graphically in Fig.\ref{Cycle}. GalMod is composed of four major blocks: a CMD generator, a Poisson solver, a kinematics generator, and a ray-tracing stellar extinction computer. Each block needs to satisfy independently theoretical and observational constraints and to connect with all the other boxes. All the blocks depend on the Poisson solver. This module produces the underlying information on which all the GalMod components rely. Once the density parameters are assigned, it computes the total axisymmetric potential and the leading derivatives from which constraints on the MW model can be easily obtained (e.g., rotation curve, Oort functions, terminal velocity, mass inside 100 kpc, vertical force on the plane, local surface density, relative density or mass ration in the solar neighborhood). The CMD module functions to realize a CMD from precomputed stellar models and is tightly connected to the galaxy Poisson solver because the density profiles are the major players in determining the number of stars per interval of color and magnitude in each FoV (together with the SFR, IMF, binaries, Z enrichment, and He enrichment). Hence, changing the density profile scale parameters will result not only in a different galaxy potential but also in a change in the CMDs. GalMod is offered with several SFR, IMF and Z profiles to cover extended parameter space possibilities beyond the canonical MW model. The kinematical module generator has the goal to provide the phase-space description for a mock survey with respect to proper motions or radial velocities. It includes treatment of non-axisymmetric features such as spiral arms and bar kinematics thus offering the vertex deviation in and out of the plane, the vertical tilt of the velocity ellipsoids, and the asymmetric drift. The kinematical module relates to the CMD generator using an age-velocity dispersion relation that ensures hotter kinematics for older stellar populations so that, e.g., an excess of old CSPs in the disk will result naturally in a hotter kinematical component. Finally, in the Jeans equations the total potential connects kinematics and MW potential and closes the circle (more implementation details are left for Appendix C).

Finally, it is worth to remind the reader of some of the limitations of GalMod, and of the directions of planned future improvements. The most interesting aspect to be further developed is the self-consistency in the treatment of the stellar evolution and dynamics. At the moment, the mock catalog provided by GalMod is the result of a parametric modulation of the existence space $\mathbb{E}$. In this respect:
\begin{itemize}
	\item stars do not form, evolve and die enriching the ISM with stellar winds or supernovae phase nor they move through phase-space in a self-consistent way. GalMod projects some parametric laws in the observable space. Therefore, a mock CMD is just the result of the adopted SFRs, IMF, chemical enrichment laws, and a set of stellar tracks that solution of the equation of stellar structure. GalMod includes neither the temporal evolution of the single stars, nor the enrichment of the ISM from stellar winds or from supernovae explosions, nor effect of binary stellar interaction, nor stellar rotation, nor magnetic fields.
	\item Another GalMod limitation in the phase-space is the decoupling of the vertical/radial kinematics: spiral arms vertical kinematics is treated as ordinary thin disk vertical-kinematics, but while radial and vertical kinematics find (at least historically) a justification in the epicycle approximation, for spiral arms we do not expect the treatment to be more than an oversimplification. Even in the case of non-spiral arm CSPs the radial/vertical kinematic treatment does not find a uniform consensus, see, e.g., Appendix C for GalMod implementation.
\end{itemize}
These are just a couple of very crude approximations that we adopt in GalMod and that are common to many other mock catalog generators. 

In this context, future efforts will aim at merging fully-hydrodynamical simulations and mock catalog generators. One of the most relevant features of GalMod is indeed the possibility to produce i.c. for the stellar component of a mock catalog directly tunable on real observations. In the future, we plan to push further the research with GalMod in this direction.

The model is available through the web interface at \href{www.galmod.org}{www.GalMod.org}, including the tutorial page that provides support to the user (the contact address is
galaxy dot model at yahoo dot com).

\acknowledgments
\textbf{Acknowledgments:} SP thanks J. Kollmeier for the fundamental support in the developing process of this paper. EKG gratefully acknowledges support from the Collaborative Research Center "The Milky Way System" (SFB 881) of the German Research Foundation (DFG), particularly via sub-projects A3 and A5.

\bibliographystyle{aasjournal}                    
\bibliography{Biblio}

\begin{appendix}
\section{Potential for the disk vertical  density  profiles}
A dependence of a disk's vertical density profile on height above or below the plane of the type ${{\rho }_{z}}\propto {{\operatorname{sech}}^{2}}z$ has its dynamical justification whenever we search for an equilibrium self-consistent solution of the 1D Poisson equation (in the vertical direction) of a CSP distribution with a thin axisymmetry density profile \citep[][]{1942ApJ....95..329S}. The density can be written as Eq.\eqref{(3.1)}. This is formally inconsistent with the implementation we adopted for the tilt of the vertical ellipsoid  \citep[][]{1991ApJ...368...79A,1991MNRAS.253..427C,2009A&A...500..781B} as this latter implies a coupling of radial and vertical direction that does not hold for  Eq.\eqref{(3.1)}, nevertheless, for the degree of precision of the observation so far, this formulation proved to be useful at past occasions (e.g., \citet[][]{2017arXiv170406274R}). If we adopt the formalism of Eq.\eqref{(3.1)} for our vertical profile, the numerical integration does not change with respect to what was implemented in \citet[][]{2016MNRAS.461.2383P} except for the vertical treatment of the derivative of the potential that we complete here. Starting from Eq.(A8) in \citet[][]{1989MNRAS.239..571K} we obtain:
\begin{equation}\label{A1}
\begin{aligned}
& {{\Phi }_{z}}\left( z \right)=\int_{-\infty }^{\infty }{d\zeta {{\rho }_{z}}\left( \zeta  \right){{e}^{-k\left| z-\zeta  \right|}}} \\
& =2{{e}^{-k\left| z \right|}}\int_{0}^{\left| z \right|}{d\zeta \cosh \left( k\zeta  \right){{\rho }_{z}}\left( \zeta  \right)}+2\cosh \left( kz \right)\int_{\left| z \right|}^{\infty }{d\zeta {{\rho }_{z}}\left( \zeta  \right){{e}^{-k\zeta }}} \\
& =-4\frac{{{e}^{-k\left| z \right|}}}{k}\int_{1}^{{{e}^{-k\left| z \right|}}}{dt\left( {{t}^{2}}+1 \right){{t}^{h_{z}^{-1}-2}}{{\left( {{t}^{h_{z}^{-1}}}+1 \right)}^{-2}}}-8\cosh \left( kz \right)\int_{{{e}^{-k\left| z \right|}}}^{0}{dtk{{\left( {{t}^{-\tfrac{1}{2}h_{z}^{-1}}}+{{t}^{\tfrac{1}{2}h_{z}^{-1}}} \right)}^{-2}}}.
\end{aligned}
\end{equation}
Here we consider the first integrand in the last row, and we write it as
\begin{equation}\label{A2}
\begin{aligned}
- 4\frac{{{e^{ - k\left| z \right|}}}}{k}\int_1^{{e^{ - k\left| z \right|}}} {dt\left( {{t^2} + 1} \right){t^{h_z^{ - 1} - 2}}{{\left( {{t^{h_z^{ - 1}}} + 1} \right)}^{ - 2}}}  &=  - 4\frac{{{e^{ - k\left| z \right|}}}}{k}\left( {\int_1^{{e^{ - k\left| z \right|}}} {dt\frac{{2{t^{h_z^{ - 1} - 2}}}}{k}{{\left( {{t^{h_z^{ - 1}}} + 1} \right)}^{ - 2}}} } \right. \\
&\left. { + \int_1^{{e^{ - k\left| z \right|}}} {dt\frac{{2{t^{h_z^{ - 1}}}}}{k}{{\left( {{t^{h_z^{ - 1}}} + 1} \right)}^{ - 2}}} } \right) \\
\end{aligned},
\end{equation}
where with $\psi \left( z \right)=\frac{{\gamma}'\left( z \right)}{\gamma\left( z \right)}$ is the polygamma function (the logarithmic derivative of the Euler Gamma function). In the same way,
\begin{equation}\label{A3}
	\int_{1}^{{{e}^{-k\left| z \right|}}}{dt\frac{2{{t}^{h_{z}^{-1}}}}{k}{{\left( {{t}^{h_{z}^{-1}}}+1 \right)}^{-2}}}=\frac{{{h}_{z}}}{k}\left( -2{{e}^{k\left| z \right|}}{{}_{\left( 1-{{h}_{z}} \right)}}{{{\hat{F}}}_{\left( 1-{{h}_{z}} \right)}}+\frac{2{{e}^{h_{z}^{-1}\left( {{h}_{z}}+1 \right)k\left| z \right|}}}{{{e}^{\frac{k\left| z \right|}{hz}}}+1}+{{h}_{z}}\left( \psi \left( -\frac{{{h}_{z}}}{2} \right)-\psi \left( \frac{1-{{h}_{z}}}{2} \right) \right)-1 \right).
\end{equation}
Finally, by considering the second integral in Eq.(A1) we obtain
\begin{equation}\label{A4}
	\int_{{{e}^{-k\left| z \right|}}}^{0}{dtk{{\left( {{t}^{-\tfrac{1}{2}h_{z}^{-1}}}+{{t}^{\tfrac{1}{2}h_{z}^{-1}}} \right)}^{-2}}}=\frac{8{{h}_{z}}\cosh (kz){{e}^{-\frac{\left( h_{z}^{-1}+1 \right)kz}{sgn (z)}}}}{k}\left( 1-{{h}_{z}}\gamma({{h}_{z}}+1){{}_{\left( 1,{{h}_{z}}+1 \right)}}{{{\tilde{\hat{F}}}}_{\left( {{h}_{z}}+2 \right)}}-\frac{1}{{{e}^{\frac{kz}{{{h}_{z}}sgn (z)}}}+1} \right).
\end{equation}
If we collect all the previous terms, and consider that
\begin{equation}\label{A4bis}
- \psi \left( {\frac{{{h_z}}}{2}} \right) + \psi \left( {\frac{{{h_z} + 1}}{2}} \right) + \psi \left( { - \frac{{{h_z}}}{2}} \right) - \psi \left( {\frac{1}{2} - \frac{{{h_z}}}{2}} \right) = 2\left( {\frac{1}{{{h_z}}} + \pi \csc (\pi {h_z})} \right),
\end{equation}
and consider all the multiplicative factors, we finally get:
\begin{equation}\label{A5}
	\begin{aligned}
& {{\Phi }_{z}}\left( z \right)=\frac{4{{h}_{z}}{{e}^{\left( -h_{z}^{-1}-3 \right)k\left| z \right|}}}{\left( {{h}_{z}}+1 \right)k\left( {{e}^{k\left| z \right|h_{z}^{-1}}}+1 \right)}\left( \left( {{h}_{z}}+1 \right){{e}^{\left( h_{z}^{-1}+1 \right)k\left| z \right|}}\times  \right. \\
& -\left( {{e}^{\frac{kz}{{{h}_{z}}sgn (z)}}}+1 \right)\left( _{\left( 1,-{{h}_{z}} \right)}{{{\hat{F}}}_{\left( 1-{{h}_{z}} \right)}}{{e}^{\frac{2kz}{sgn (z)}}}+{{\ }_{\left( 1,{{h}_{z}} \right)}}{{{\hat{F}}}_{\left( {{h}_{z}}+1 \right)}} \right) \\
& \left. +\left( {{e}^{2k\left| z \right|}}+1 \right){{e}^{k\left| z \right|h_{z}^{-1}}}+{{e}^{k\left| z \right|}}\left( \pi {{h}_{z}}\csc (\pi {{h}_{z}})\left( {{e}^{k\left| z \right|h_{z}^{-1}}}+1 \right)+2\cosh (kz) \right) \right) \\
& \left. -2{{h}_{z}}{{e}^{2k\left| z \right|}}\cosh (kz)\left( {{e}^{k\left| z \right|h_{z}^{-1}}}+1 \right){{\ }_{\left( 1,{{h}_{z}}+1 \right)}}{{{\hat{F}}}_{\left( {{h}_{z}}+2 \right)}} \right)
\end{aligned}
\end{equation}
which is regular everywhere. The integration of Eq.(A6) does not present any difficulty and proceeds exactly as in \citet[][]{2016MNRAS.461.2383P}. The computing of the necessary derivatives is trivial. As shown in \citet[][]{2016MNRAS.461.2383P} the use of a hypergeometric functions integrator is convenient in terms of speed and precision of the integral because the integral is reduced to one dimension and because a significant amount of literature is available to help with the implementation (Galmod simply implements LAPack, the publically available Linear Algebra PACKage, see GitHub repository).

\section{Spherical bulge model}
As mentioned in Sec.\ref{Theoretical}, there can be situations in which a spherical model needs to be investigated, and we want to equip GalMod with more flexible instruments to investigate the widest range of topics. We added a spherical bulge component formulated as a couple of density-potential families given by solving the Poisson  equation for the density in parametric form:
\begin{equation}\label{B1}
	\rho ={{\rho }_{\text{0,blg}}}{{e}^{-\frac{r}{{{h}_{\text{blg}}}}}}.
\end{equation}
The literature is rich in more sophisticated solutions whose applicability in the context of GalMod is deferred to future studies. For the Poisson equation in spherical coordinates we write:
\begin{equation}\label{B2}
	\begin{aligned}
	& \frac{1}{r}\frac{\partial }{\partial r}\left( r\frac{\partial \Phi \left( r \right)}{\partial r} \right)=4\pi G{{\rho }_{\text{0,blg}}}{{e}^{-\frac{r}{{{h}_{\text{blg}}}}}}\Leftrightarrow  \\
	& {{\Phi }_{\text{blg}}}\left( r \right)={{c}_{1}}\lg r+{{c}_{2}}-4\pi Gh_{\text{blg}}^{2}{{\rho }_{\text{0,blg}}}\text{Ei}\left( -\frac{r}{{{h}_{\text{blg}}}} \right)+4\pi Gh_{\text{blg}}^{2}{{\rho }_{\text{0,blg}}}{{e}^{-\frac{r}{hr}}}  \\
	\end{aligned}
\end{equation}where we made use of the exponential integral function $\text{Ei}\left( z \right)\equiv -\int_{-z}^{\infty }{dt\frac{{{e}^{-t}}}{t}}$ and, requiring the potential to be null for $r\to \infty $ with ${{c}_{1}}={{c}_{2}}=0$ we obtain:
\begin{equation}\label{B3}
{{\Phi }_{\text{blg}}}\left( r \right)=4\pi Gh_{\text{blg}}^{2}{{\rho }_{\text{0,blg}}}\left( {{e}^{-\frac{r}{{{h}_{\text{blg}}}}}}-\text{Ei}\left( -\frac{r}{{{h}_{\text{blg}}}} \right) \right).
\end{equation}
The use of the exponential integral is convenient in this context because  it is formally an easily tabulated 1D integral, but also because we can prove that it easily cancels out in the computing of the constraints on the MW potential that we are going to evaluate. In the computation of the constraints implemented in the MW Poisson solver, ${{\Phi }_{\text{blg}}}$ is entering only through its derivatives. We report here the equations that are going to substitute  for the contribution to the circular speed in Sec.4.1.4 of \citet[][]{2016MNRAS.461.2383P}:
\begin{equation}\label{B4}
	v_{c,\text{blg}}^{2}=4\pi G{{\rho }_{\text{0,blg}}}{{\text{e}}^{-\frac{R}{{{h}_{\text{blg}}}}}}{{h}_{\text{blg}}}\left( {{h}_{\text{blg}}}+R \right),
\end{equation}
for the vertical force on the plane:
\begin{equation}\label{B5}
	{{F}_{z,\text{blg}}}\left( R,z \right)=\frac{4\pi G{{\rho }_{\text{0,blg}}}{{\text{e}}^{-\frac{\sqrt{{{R}^{2}}+{{z}^{2}}}}{{{h}_{\text{blg}}}}}}{{h}_{\text{blg}}}z\left( {{R}^{2}}+{{z}^{2}}+{{h}_{\text{blg}}}\sqrt{{{R}^{2}}+{{z}^{2}}} \right)}{{{\left( {{R}^{2}}+{{z}^{2}} \right)}^{3/2}}},
\end{equation}moreover, for the total mass up to a maximum radius ${{r}_{\max }}$:
\begin{equation}\label{B6}
	{{M}_{\text{blg}}}\left( {{r}_{\max }} \right)=4{{\rho }_{\text{0,blg}}}\pi {{h}_{\text{blg}}}\left( 2h_{\text{blg}}^{2}-{{\text{e}}^{-\frac{{{r}_{\max }}}{{{h}_{\text{blg}}}}}}\left( 2h_{\text{blg}}^{2}+2{{h}_{\text{blg}}}{{r}_{\max }}+r_{\max }^{2} \right) \right).
\end{equation}
Finally, the radial velocity dispersion reads:
\begin{equation}\label{B7}
	\begin{aligned}
	\sigma _{rr}^2\left( R \right) &= \pi {\rho _{0,{\text{blg}}}}G\frac{{\beta \left( r \right)h_{{\text{blg}}}^2}}{{{r^2}}}{{\text{e}}^{ - \frac{r}{{{h_{{\text{blg}}}}}}}}\left( {h_{{\text{blg}}}^2\left( {8{{\text{e}}^{\frac{r}{{{h_{{\text{blg}}}}}}}} - 7} \right) + 8r_a^2{{\text{e}}^{\frac{{2r}}{{{h_{{\text{blg}}}}}}}}\left( {\gamma\left( { - 1,\frac{r}{{{h_{{\text{blg}}}}}}} \right) - 2\gamma\left( { - 1,\frac{{2r}}{{{h_{{\text{blg}}}}}}} \right) - \gamma\left( {0,\frac{{2r}}{{{h_{{\text{blg}}}}}}} \right)} \right)} \right.\\
	&\left. { - 6{h_{{\text{blg}}}}r - 2\left( {{r^2} + r_a^2} \right)} \right),
	\end{aligned}
\end{equation}
with $\beta \left( r \right) = \frac{{{r^2}}}{{r_a^2 + {r^2}}}$ being the anisotropy parameter and $r_a$ the anisotropy radius.

This completes the presentation of the Galaxy model potential.

\section{Implementation notes}\label{ImplNotes}
In this section, we sketch a brief technical resume of the GalMod platform contents (ver. 16.2). Nevertheless, we warn the reader that the platform "GalMod" is continuously updated and the web page (the tutorial page: \href{www.galmod.org/gal/tutorial.html}{www.galmod.org/gal/tutorial}) is the most up-to-date place where to search for the last refinements, bug corrections, and tested implementations that might differ from what is presented in this section. A cycle of 12 upgrades per year is planned.

We organized GalMod in modules, i.e., independent sub-units of the source code that interact through interfaces (e.g., the Poisson solver modules, the CMD generator modules, the extinction integrator modules, and so forth). This approach facilitates the use of multiple language codes/libraries with different types of licenses.

\begin{figure*}
\centering
\includegraphics[width=12cm]{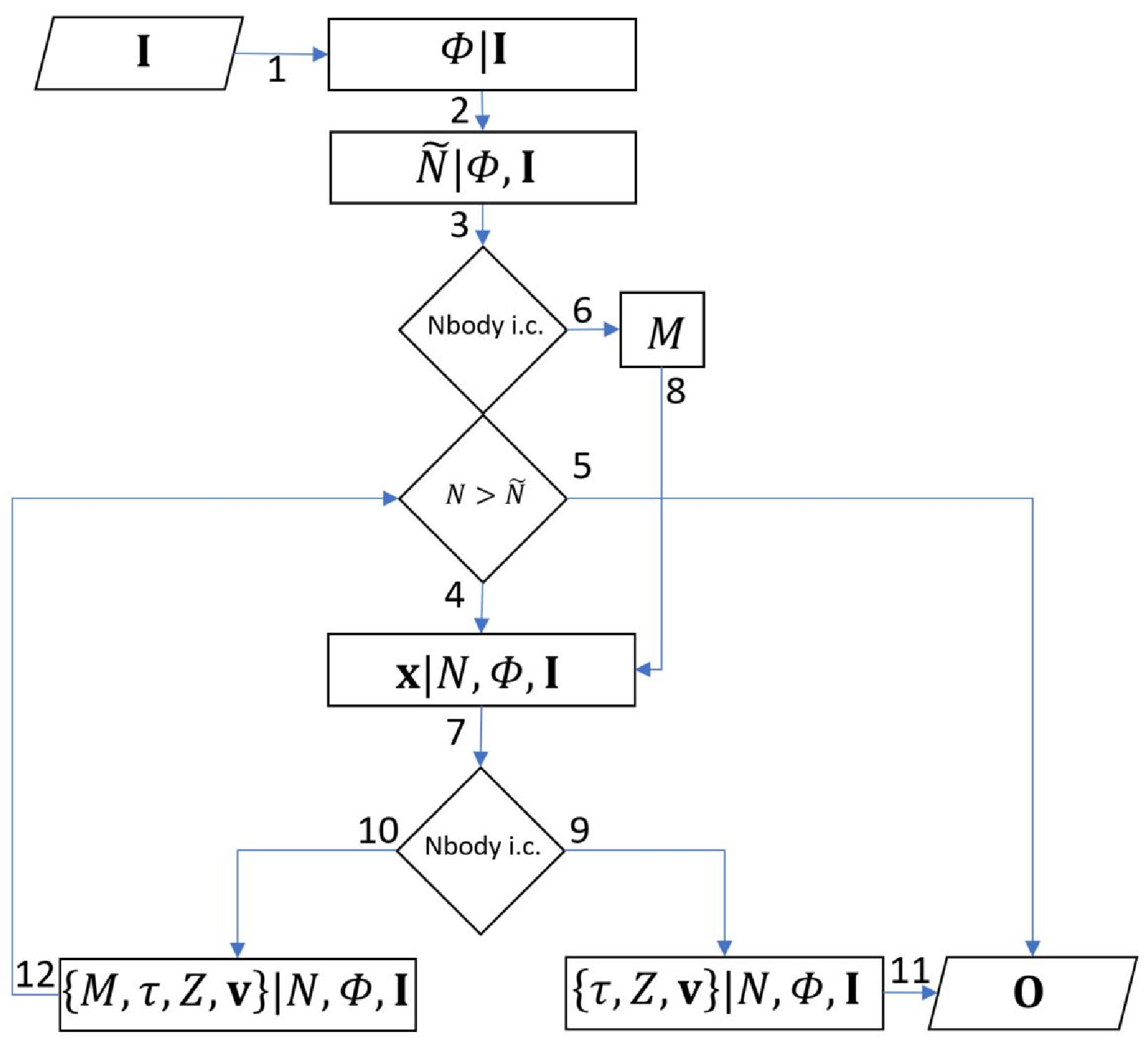}
\caption{The superstructure of GalMod. Each square box represents a series of modules of GalMod where some computation is performed (either in serial or parallel mode). Numbered blue arrows represent the ideal flux from the input parameters $\bm{I}$, to the delivery of the mock catalog (or the output) $\bm{O}$. The modules $\Phi \left( \rho  \right)|\bm{I}$ are the Poisson-solver modules that, once the density/potential parameters have been adopted, provide $\Phi$, its gradients and Laplacian, and a series of kinematics indicators (rotation curve, total mass, etc.) relative to $\bm{I}$. $\tilde{N}|\Phi ,\bm{I}$ are the modules computing the maximum number of stars for the galaxy considering $\Phi \left( \rho  \right)$ and the input parameters $\bm{I}$. $\tilde{N}$ can be simply an input parameter depending on the type of simulation. $M$ are the modules assigning the star-like particle mass in the case GalMod is used to generate N-body i.c.. $N$ is the number of stars generated so far. $\bm{x}|N,\Phi ,\bm{I}$ are the modules distributing the stars according to the density profiles, $\Phi \left( \rho  \right)$, their number $N$, and the input condition $\bm{I}$. The age/metallicity/velocity dispersion modules $\left\{ \tau ,Z,\bm{v} \right\}|N,\Phi ,\bm{I}$ compute the distribution of the stars or star-like particles in the age $\tau$, metallicity $Z$,  and velocity $\bm{v}$ space. They include the mass generation of the star and the extinction modules in the case of mock stellar catalogs only, while a separate module provides the same information with a preassigned stellar particle mass in the case of i.c. generation.}
\label{FlowC}
\end{figure*}

A rough scheme of the GalMod infrastructure is in Fig.\ref{FlowC}. We presented each of these modules in dedicated papers. The technique to generate photometry and chemistry of the stars is introduced in \citet[][]{2012A&A...545A..14P} and references therein, the kinematics and potential solver in \citet[][]{2016MNRAS.461.2383P} and references therein, and the extinction based on DART-ray in \citet[][]{2017arXiv170903802N}. The web-interface system and cloud computing service is realized and maintained by Clover-lab$^{\text{TM}}$. These references collect the results of several years of work done by several researchers and programmers. They are not meant to be an exhaustive list.

GalMod produces a stochastic realization (i.e., stellar parameters) sampled from a given (multi-dimensional) distribution function. While the technique to realize this in practice is very common and straightforward for DFs in the phase-space $\bm{\gamma}=\mathbb{E}\cap \left( M\times Z \right)$ at any given age $t$ of a CSP \citep[see, e.g., Sec 7.3.6 in][]{2002nrca.book.....P}, the same treatment in the sub-manifold $M\times Z=\mathbb{E}\cap \bm{\gamma}$ is not. Hence, in what follows we will review how GalMod implements the DFs for the $M\times Z$ plane following \citet[][]{2012A&A...545A..14P}   while limiting ourselves to review the equations currently implemented to sample the DFs for the $\bm{\gamma}$ space  \citep[see][and reference therein, for a complete presentation of the equations]{2016MNRAS.461.2383P}.

\begin{figure}
\centering
\includegraphics[width=12cm]{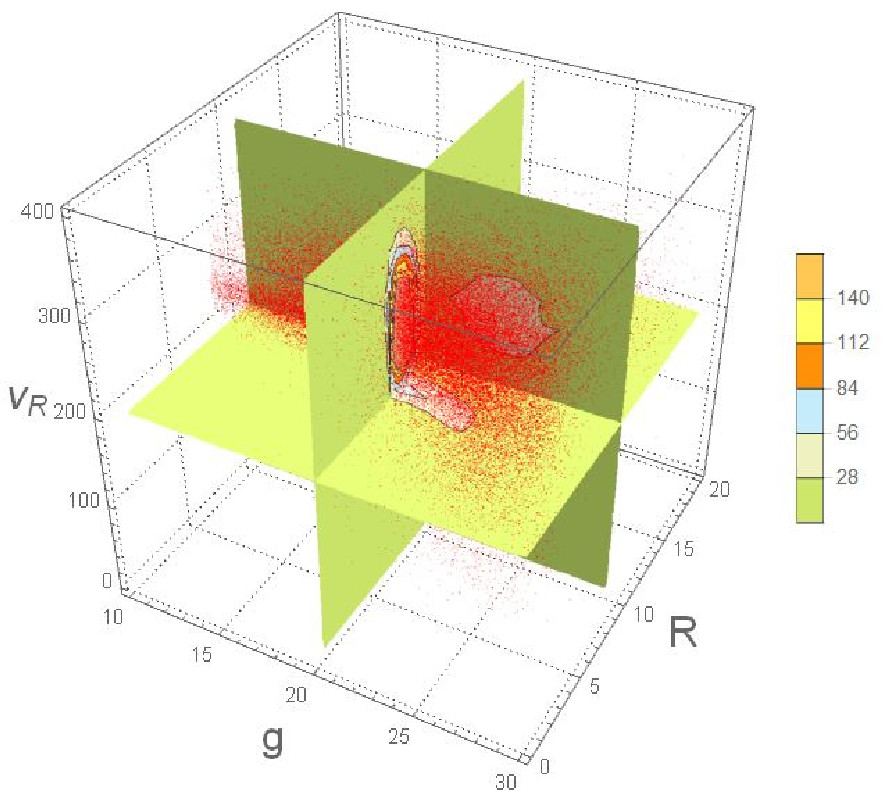}
\caption{An example of a section of a CSP DF before the selection cuts are applied. The parameters are from Table 1, the number of stars is in units of $10^{10}$. Red dots represent the sampled point at the moment of the snapshot during the process of sampling. In Fig.\ref{CSPt01} the DF of the SSP is taken for the $\bm{DB}$ of the SSP with an IMF from Eq.\eqref{(2.17)} and the SDSS $g$ band. $R$ is the radial distance from the galactic center and $v_R$ the velocity along the $R$ direction.} 
\label{CSPt01}
\end{figure}

The $M\times Z$ plane at a given age, $t$, is usually represented as a 2D scatter plot with Hertzsprung-Russell coordinates, i.e., effective temperature ${{T}_{\text{eff}}}$, luminosity $L$, and origin in $O$ at the instant $t$, say $\left( O,{{T}_{\text{eff}}},L;t \right)$, or its observational counterpart the $\left( O,c,m;t \right)$ plane, i.e., the CMD with  color $c$, and magnitude $m$. Because any CSP is a superposition of several SSPs (see Sec.2 and Fig.\ref{EvolTemp}), we generate first a database ($\bm{DB}$) of SSPs to cover all the parameter space of interest, i.e., spanning all metallicities, IMFs, bolometric corrections of interest and for all the ages of interest. The $\bm{DB}$ is generated to cover $Z\in \left[ 0.0001,0.004 \right]\text{dex}$, $M\in \left[ 0.2,20.0 \right]{{M}_{\odot }}$ for an age range spanning the interval $\tau \in \left[ {{10}^{6}},1.3\times {{10}^{9}} \right]\text{yr}$ and covering a range of effective temperature ${T_{{\text{eff}}}} \in \left[ {3.5,50} \right] \times {10^3}$K and gravity ${\log _{10}} \in \left[ { - 2.5,0.5} \right]$ with stellar models from \citet[][]{2014MNRAS.445.4287T}, \citet[][]{2008A&A...484..815B} or \citet[][]{2009A&A...508..355B} (any other set of stellar models can eventually be considered upon request if the user is interested in the effects of the stellar rotation, He enrichment, binary fraction, etc.).  No preset interval bins are assumed for the $\bm{DB}$ (every time a new set of input parameters, say $I$, is required and computed, it is added to the $\bm{DB}$) but they are for stellar models: stellar tracks or isochrones are precomputed for the metallicity values $Z = \left\{ {0.0001,{\text{ }}0.0004,{\text{ }}0.001,{\text{ }}0.002,{\text{ }}0.004,{\text{ }}0.008,{\text{ }}0.017,{\text{ }}0.040,{\text{ }}0.070} \right\}$. They are then interpolated with a (linear) interpolating scheme as in \citet{2008A&A...484..815B}, e.g., their  Fig.7. Any other scheme can be considered equally valid. As shown in \citet[][]{2012A&A...545A..14P} and seen here in Fig.\ref{CSPt01}, it is convenient to include in the $\bm{DB}$ both the photometric systems and the IMF considered to speed up the computation.

Despite the rapidity in sampling the $\bm{\gamma}$ section, where the DFs are entirely analytic, we also included in the $\bm{DB}$ a $\bm{\gamma}$-section of the SSPs. This approach is adopted to avoid the computation of the gravitational potential ${{\Phi }_{\text{tot}}}$ each time a user opts for an already pre-evaluated set of parameters (e.g., the values of Table \ref{Table1}).
Once the DFs are available, the sampling can proceed as in Fig.\ref{FlowC}(\footnote{We stress that this flow-chart is just an example, GalMod is in continuous development to grant a better response to the user, and any other scheme is equally valid.}).

Through the web interface, GalMod obtains the basic set of parameters $\bm{I}$ that define the total potential ${{\Phi }_{\text{tot}}}\left( \rho \left( \bm{I} \right) \right)$. If the parameters have never been previously computed, $\bm{I}\notin \bm{DB}$, GalMod computes the resulting potential ${{\Phi }_{\text{tot}}}$ and adds it to the $\bm{DB}$ (arrow 1, in Fig.\ref{FlowC}). The potential is obtained by numerical integration of Eq.(15) (and all the necessary derivatives) in \citet[][]{2016MNRAS.461.2383P} by means of collocation techniques \citep[e.g.,][]{LEVIN199695}:
\begin{equation}\label{I1}
{{\Phi }_{D}}\left( R,\phi ,z \right)=-4\pi G{{\rho }_{0}}h_{R}^{-1}\int_{0}^{\infty }{{{J}_{0}}\left( kR \right){{\left( h_{R}^{-2}+{{k}^{2}} \right)}^{-3/2}}}\frac{h_{z}^{-1}{{e}^{-k\left| z \right|}}-k{{e}^{-h_{z}^{-1}\left| z \right|}}}{h_{z}^{-2}-{{k}^{2}}}dk,
\end{equation}
with ${{\Phi }_{D}}$ being the potential of a disk CSP, ${{\rho }_{0}}$ the density of each disk CSP, ${{h}_{R}}$ the scale length, ${{h}_{z}}$ the scale high, and ${{J}_{0}}$ the Bessel function; together with Eq. (20) and (21) in \citet[][]{2016MNRAS.461.2383P}:
\begin{equation}\label{I2}
	{{\Phi }_{H*}}\left( r \right)=\left\{ \begin{aligned}
  & 4\pi G{{\rho }_{0H*}}\frac{r_{\odot }^{-\alpha }}{r}\frac{\left( \alpha +2 \right)h_{rH*}^{\alpha +3}+{{r}^{\alpha +3}}}{\left( \alpha +2 \right)\left( \alpha +3 \right)}\wedge r>{{h}_{rH*}} \\
 & -2\pi G{{\rho }_{0H*}}\frac{3h_{rH*}^{2}-{{r}^{2}}}{3}{{\left( \frac{{{h}_{rH*}}}{{{r}_{\odot }}} \right)}^{\alpha }}\wedge r{{h}_{rH*}}, \\
\end{aligned} \right.
\end{equation}
\begin{equation}\label{I3}
	{{\Phi }_{\text{DM}}}\left( R,\phi ,z \right)\equiv \frac{v_{0\text{DM}}^{2}}{2}\log \left( h_{R\ \text{DM}}^{2}+{{R}^{2}}+{{q}^{-2}}{{z}^{2}} \right),
\end{equation}
or our Eq.\eqref{(3.3)} where Eq.\eqref{A5} is considered, with ${{\Phi }_{H*}}$ being the stellar halo potential, ${{\rho }_{0H*}}$ the density of the stellar halo CSP, ${{r}_{\odot }}$ the observer location in spherical coordinates, $\alpha$ the slope of the density profile, ${{\Phi }_{\text{DM}}}$ the dark matter halo potential, and ${{v}_{0,\text{DM}}}$ the asymptotic circular velocity of the dark matter profile with flattening factor $q$ and scale length ${{h}_{R\text{DM}}}$. A hot coronal gas model with the density profile
\begin{equation}\label{I4}
{{\rho }_{\text{HCG}}}=\frac{{{\sigma }^{2}}}{2\pi G{{r}^{2}}\left( 1+\eta \frac{r}{{{r}_{v}}} \right)},
\end{equation}
for $r<{{r}_{v}}\frac{{{e}^{\eta }}-1}{\eta }$ (and null otherwise) is added following \citet{2012A&A...542A..17P} for the sake of modeling external galaxies around the MW (only if their barycenter is within 200 kpc from the MW center), where $\sigma ={{2}^{-1/2}}{{v}_{c}}$, ${{r}_{v}}={{r}_{200}}$ (i.e., 200 times the Einstein-de Sitter mean mass density) and $\eta =1$. We evaluate the potential derivative either analytically \citep[see Pasetto's Ph.D. thesis in ][]{2006A&A...451..125V,2016MNRAS.461.2383P} or numerically from the ${{\Phi }_{\text{tot}}}$ \citep[standard finite-difference derivatives are used with back/forward schemes of different orders where needed, e.g.,][]{2002nrca.book.....P}. Here we assume that the total potential ${{\Phi }_{\text{tot}}}$ is the linear superposition of the potential of the CSPs considered (i.e., no modified-Newtonian dynamics is considered). From ${{\Phi }_{\text{tot}}}$, the dynamical constraints discussed in Sec.4.2 of \citet[][]{2016MNRAS.461.2383P} are computed. Their values, e.g., the resulting rotation curve, Oort constants, etc., see also Fig.\ref{Cycle},  are obtained from the set of parameters $\bm{I}$, and provided to the users together with the mock catalogs. More details about the output format and column headers are in a dedicated tutorial page.

GalMod proceeds then (arrow 2, in Fig.\ref{FlowC}) to compute the number of stars $\tilde{N}=\sum\limits_{c}^{{}}{{{N}_{c}}}$ that the FoV contains for each CSP in accordance with Eq.\eqref{(2.3)} to Eq.\eqref{(2.5)} and with Eq.(11) of \citet[][]{2016MNRAS.461.2383P}, which we report here:
\begin{equation}\label{I5}
{{N}_{c}}\text{ }=\int_{{{\mathbb{R}}^{2}}}^{{}}{d\overset{\scriptscriptstyle\frown}{\Omega }}\int_{\mathbb{R}}^{{}}{d{{r}_{\text{hel}}}\mathfrak{J}{{\rho }_{c}}\left( \bm{x};t \right),}
\end{equation}
where Eq.(12) in \citet[][]{2016MNRAS.461.2383P} has been now updated with:
\begin{equation}\label{(2.19)}
		T:\left\{ \begin{aligned}
	& x={{R}_{\odot }}\cos {{\phi }_{\odot }}-{{r}_{\text{hel}}}\cos b\cos \left( l+{{\phi }_{\odot }} \right) \\
	& y={{R}_{\odot }}\sin {{\phi }_{\odot }}-{{r}_{\text{hel}}}\cos b\sin \left( l+{{\phi }_{\odot }} \right) \\
	& z={{z}_{\odot }}+{{r}_{\text{hel}}}\sin b. \\
	\end{aligned} \right.
\end{equation}
$\overset{\scriptscriptstyle\frown}{\Omega }$ is the solid angle obtained from the input form $\bm{I}$, and $\mathfrak{J}$ the Jacobian of the transformation in Eq.\eqref{(2.19)}. This transformation now accounts for the possibility to move the Solar location (i.e., the observer) inside (or outside) the modeled galaxy, e.g., the MW. In this transformation, we assumed that the reference systems in Galactic coordinates (i.e., $\left( \odot ,{{r}_{\text{hel}}},l,b \right)$) keep the Galactic plane (the plane for $b=0$) always parallel to the MW Galactic plane in Galactocentric coordinates (i.e., the plane $z=0$ in $\left( O,\bm{x} \right)$) with the origin of longitudinal coordinates toward the Galaxy center. Hence, in any new solar/observer location chosen by the GalMod user, the Galactic center will retain ${{\left( \alpha ,\delta  \right)}_{\text{GC}}}=\left( {{17}^{h}}{{42}^{m}},-28{}^\circ 5{5}' \right)$ as well as the new North Galactic Pole coordinates  ${{\left( \alpha ,\delta  \right)}_{\text{NGP}}}=\left( {{12}^{h}}{{49}^{m}},27{}^\circ 2{4}' \right)$.
The computation of the integral in Eq.\eqref{I5} is performed with standard recursive multidimensional Monte Carlo integration \citep[e.g.,][]{1978JCoPh..27..192L}. Alternatively, $\tilde{N}$ can be fixed by the user in need to generate mock catalogs with a specific number of stars or i.c. for N-body simulations with a precise number of star-particles. 

GalMod proceeds then on different paths (arrow 3 in Fig.\ref{FlowC}) depending on $\bm{I}$ required to generate N-body equilibrium models or stellar mock catalogs. In the case of i.c. generation (arrow 6, Fig.\ref{FlowC}), the mass of the star-like particles is immediately obtained by splitting the total mass (set by the different density profiles using Eq.(43), (45) in \citet[][]{2016MNRAS.461.2383P} and our \eqref{B6}) between the $\tilde{N}$ stellar-particles:
\begin{equation}\label{I6}
\begin{aligned}
  & {{M}_{D}}=4\pi \sum\limits_{D}^{{}}{\frac{{{\rho }_{0,D}}{{e}^{-{{R}_{\max }}h_{R,D}^{-1}}}}{h_{R,D}^{-2}h_{z,D}^{-1}}\left( {{e}^{{{R}_{\max }}h_{R,D}^{-1}}}-{{R}_{\max }}h_{R,D}^{-1}-1 \right)}, \\
 & {{M}_{H*}}=\frac{4\pi {{r}_{\odot }}^{-\alpha }}{3\left( \alpha +3 \right)}\sum\limits_{{{H}^{*}}}^{{}}{\frac{{{\rho }_{0,{{H}^{*}}}}}{{{d}_{0,{{H}^{*}}}}}\left( 3{{r}_{{{\max }^{\alpha +3}}}}+\alpha h_{r,{{H}^{*}}}^{\alpha +3} \right)}, \\
\end{aligned}
\end{equation}
with an index $D=1,...,{{N}_{D}}$ ranging from the number of disk SSPs, ${{H}^{*}}=1,...,{{N}_{H*}}$ to the number of stellar halo CSPs, and where we limited the maximum extent of all the CSPs to ${{r}_{\max }}=\left\| {{r}_{O}}-{{r}_{\max ,\text{gal}}} \right\|<50$ kpc (i.e., ${{r}_{\max }}$ is the maximum distance of all the stars of a galaxy from the center of the galaxy).
GalMod proceeds then to distribute the stars in agreement with the density profiles of Eq.(14), (19), (22), (56) of \citet{2016MNRAS.461.2383P} (arrow 8, in Fig.\ref{FlowC}):
\begin{equation}\label{I7}
\begin{aligned}
  & {{\rho }_{D}}\left( R,\phi ,z \right)={{\rho }_{\odot }}{{e}^{-\frac{R-{{R}_{\odot }}}{{{h}_{R}}}-\frac{z-{{z}_{\odot }}}{{{h}_{z}}}}}, \\
 & {{\rho }_{H*}}\left( r \right)=\frac{{{\rho }_{0H*}}}{{{r}_{\odot }}}\left\{ \begin{matrix}
   h_{r{{H}^{*}}}^{\alpha } & r{{h}_{rH*}}  \\
   {{r}^{\alpha }} & r>{{h}_{rH*}},  \\
\end{matrix} \right. \\
\end{aligned}
\end{equation}
together with Eq. \eqref{(3.1)}, \eqref{(3.8)} and \eqref{B1} for the bar, spirals and the bulge respectively (the meaning of the symbols is as in Sec.\ref{BarStruct} and see Appendix A of \citet{2016MNRAS.461.2383P} for the hypergeometric formulation of the reduction factor). Resonance locations are also provided  to the user for spiral arms with $m=2$ and $m=4$. 

If $\bm{I}$ requires the generation of i.c. (arrow 7, in Fig.\ref{FlowC}) we are left to sample the age/velocity-dispersion/metallicity relation (arrow 9, in Fig.\ref{FlowC}). The age/velocity-dispersion/metallicity relation is implemented as in \citet[][]{2016MNRAS.461.2383P}: because of the size in mass of the stellar-particles, to each particle a SSP is assigned (in agreement with IMF and SFH read from $\bm{I}$) of a given metallicity Z, and the velocity space is initialized with moments of order one from Eq.(16), (64), (65) of \citet[][]{2016MNRAS.461.2383P} that we again report here for completeness:
\begin{equation}\label{I8}
{{\mathbf{\bar v}}_\phi } = \left[ {\begin{array}{*{20}{c}}
	{\left| {{{\bar v}_\phi }} \right|\frac{{{r_{{\text{hel}}}}\cos b\cos l}}{R}} \\ 
	{\left| {{{\bar v}_\phi }} \right|\frac{{{R_ \odot } - {r_{{\text{hel}}}}\cos b\cos l}}{R}} \\ 
	0 
	\end{array}} \right]
\end{equation}
where the mean stream velocity relates to the Jeans equations, $\frac{\partial \mathbf{\bar{v}}}{\partial t}+\mathbf{\bar{v}}\cdot \frac{\partial }{\partial \mathbf{r}}\mathbf{\bar{v}}+\frac{\partial {{\Phi }_{tot}}}{\partial \mathbf{r}}=-\left( \frac{\partial \ln N}{\partial \mathbf{r}}+\frac{\partial }{\partial \mathbf{r}} \right)\cdot {{\mathbf{\sigma }}^{\otimes 2}}$, that in the adopted approximation read:
\begin{equation}\label{I9}
\left\{ \begin{aligned}
& {{{\bar{v}}}_{R}}{{\partial }_{R}}{{{\bar{v}}}_{R}}+\frac{{{{\bar{v}}}_{\phi }}}{R}{{\partial }_{\phi }}{{{\bar{v}}}_{R}}-\frac{\bar{v}_{\phi }^{2}}{R}+{{\partial }_{R}}{{\Phi }_{\text{tot}}}+\sigma _{RR}^{2}{{\partial }_{R}}\ln \rho +\frac{1}{R}\sigma _{R\phi }^{2}{{\partial }_{\phi }}\ln \rho +\sigma _{Rz}^{2}{{\partial }_{z}}\ln \rho +{{\partial }_{R}}\sigma _{RR}^{2}+\frac{{{\partial }_{\phi }}{{\sigma }_{R\phi }}+{{\sigma }_{RR}}-{{\sigma }_{\phi \phi }}}{R} \\ 
&+{{\partial }_{z}}{{\sigma }_{Rz}}=0 \\
& {{{\bar{v}}}_{R}}{{\partial }_{R}}{{{\bar{v}}}_{\phi }}+\frac{{{{\bar{v}}}_{\phi }}}{R}{{\partial }_{\phi }}{{{\bar{v}}}_{\phi }}+\frac{{{{\bar{v}}}_{\phi }}{{{\bar{v}}}_{R}}}{R}+\frac{1}{R}{{\partial }_{\phi }}{{\Phi }_{\text{tot}}}+\sigma _{R\phi }^{2}{{\partial }_{R}}\ln \rho +\frac{1}{R}\sigma _{\phi \phi }^{2}{{\partial }_{\phi }}\ln \rho +{{\partial }_{R}}{{\sigma }_{R\theta }}+\frac{{{\partial }_{\phi }}{{\sigma }_{\phi \phi }}+2{{\sigma }_{R\phi }}}{R}=0 \\ 
& {{\partial }_{z}}{{\Phi }_{\text{tot}}}+\sigma _{Rz}^{2}{{\partial }_{R}}\ln \rho +\sigma _{zz}^{2}{{\partial }_{z}}\ln \rho +{{\partial }_{R}}{{\sigma }_{Rz}}+\frac{{{\partial }_{\phi }}{{\sigma }_{\phi z}}+{{\sigma }_{Rz}}}{R}+{{\partial }_{z}}{{\sigma }_{zz}}=0 \\ 
\end{aligned} \right.
\end{equation}
This is a complex partial differential system of equations in which all the equations retain a dependence on  $\left( R,\phi ,z \right)$. In these equations enters the total potential ${{\Phi }_{\text{tot}}}={{\Phi }_{\text{tot}}}\left( R,\phi ,z \right)$, hence all the terms maintain their azimuthal dependence. This translates in a practical impossibility to easily approach this system numerically. We approach the GalMod kinematics in a simplified way leaving a rigorous treatment to future developments of GalMod. For the case of the dynamical equilibrium (${{\partial }_{t}}\bullet =0$) axisymmetric CSPs (${{\partial }_{\phi }}\bullet =0$) of the disks, we assume ${{\bar{v}}_{R}}={{\bar{v}}_{z}}=0$ and we obtain from the previous 
\begin{equation}
\left\{ \begin{aligned}
& v_{c}^{2}+\frac{R}{\rho }{{\partial }_{R}}\left( \rho \sigma _{RR}^{2} \right)+\frac{R}{\rho }{{\partial }_{z}}\left( \rho \sigma _{Rz}^{2} \right)+{{\sigma }_{RR}}-{{\sigma }_{\phi \phi }}=\bar{v}_{\phi }^{2} \\ 
& \frac{R}{\rho }{{\partial }_{R}}\left( \rho \sigma _{R\phi }^{2} \right)+\frac{R}{\rho }{{\partial }_{z}}\left( \rho \sigma _{\phi z}^{2} \right)+2{{\sigma }_{R\phi }}=0 \\ 
& -R{{\partial }_{z}}{{\Phi }_{\text{tot}}}=\frac{1}{\rho }{{\partial }_{R}}\left( R\rho \sigma _{Rz}^{2} \right)+\frac{R}{\rho }{{\partial }_{z}}\left( \rho \sigma _{zz}^{2} \right), \\ 
\end{aligned} \right.
\end{equation}
and from the first, we extract the mean stream velocity for the thin disk CSPs as:
\begin{equation}
{{\bar{v}}_{\phi }}\left( R,z \right)=\sqrt{v_{c}^{2}+\frac{R}{\rho }{{\partial }_{R}}\left( \rho \sigma _{RR}^{2} \right)+\frac{R}{\rho }{{\partial }_{z}}\left( \rho \sigma _{Rz}^{2} \right)+{{\sigma }_{RR}}-{{\sigma }_{\phi \phi }}},
\end{equation}
and for the thick disk CSP as
\begin{equation}
{{\bar{v}}_{\phi }}\left( R,z \right)=\sqrt{v_{c}^{2}+\frac{R}{\rho }{{\partial }_{R}}\left( \rho \sigma _{RR}^{2} \right)+{{\sigma }_{RR}}-{{\sigma }_{\phi \phi }}},
\end{equation}
while the halo CSP is not rotating. In these equations ${{v}_{c}}={{v}_{c}}\left( {{\Phi }_{\text{tot}}} \right)$ is the circular velocity, and ${{\sigma }_{ij}}$ are the elements of the second order moment of the distribution function, ${{\rho }_{c}}$ the density of the CSP considered, and $\left\{ {{r}_{\text{hel}}},l,b \right\}$ the observed distance, latitude, and longitude of the stars distributed accordingly with the ${{\rho }_{c}}$. In the case of the spiral arm CSPs alone, GalMod considers the following mean streams \citep[][]{2016MNRAS.461.2383P}:
\begin{equation}\label{I9}
\begin{aligned}
  & {{{\bar{v}}}_{R}}=\frac{{{\Sigma }_{0}}}{\Sigma }\frac{{{\Phi }_{1}}}{{{\sigma }_{RR}}}\frac{X}{2}\text{sin}{{\text{c}}^{\text{-1}}}\left( \nu \pi  \right)\left( _{\left( \tfrac{1}{2},1 \right)}{{{\hat{F}}}_{\left( -\nu ,2+\nu  \right)}}{{-}_{\left( \tfrac{1}{2},1 \right)}}{{{\hat{F}}}_{\left( \nu ,2-\nu  \right)}} \right), \\
 & {{{\bar{v}}}_{\phi }}=\frac{{{\Sigma }_0}}{\Sigma }{{v}_{c}}-\frac{{{\Sigma }_0}}{\Sigma }{{\Phi }_{\text{sp}}}\frac{{{v}_{c}}}{\sigma _{RR}^{2}}\left( 1-\text{sin}{{\text{c}}^{\text{-1}}}\left( \nu \pi  \right){{\ }_{\left( \tfrac{1}{2},1 \right)}}{{{\hat{F}}}_{\left( 1-\nu ,1+\nu  \right)}} \right). \\
\end{aligned}
\end{equation}
The only non-null moments considered are the non-null terms of the following matrix:
\begin{equation}\label{I10}
{{\bm{\sigma }}_{c}}\equiv {{\left( \bm{v}-\bm{\bar{v}} \right)}^{\otimes 2}}=\left[ \begin{matrix}
   {{\sigma }_{RR}^{2}} & {{\sigma }_{Rz}^{2}} & {{\sigma }_{R\phi }^{2}}  \\
   {{\sigma }_{Rz}^{2}}^2 & {{\sigma }_{\phi \phi }^{2}} & 0  \\
   {{\sigma }_{R\phi }^{2}} & 0 & {{\sigma }_{zz}^{2}}  \\
\end{matrix} \right],
\end{equation}
where ${{\sigma }_{R\phi }}$ is usually referred as to the “vertex deviation” and ${{\sigma }_{Rz}}$ as the “vertical tilt” of the velocity ellipsoid of each CSP \citep[e.g.,][]{2012A&A...547A..71P,2012A&A...547A..70P}. The velocity distribution resulting from the superposition of several CSPs of this type is highly complex: an example of the degree of complexity can be captured by the vertex deviation map resulting from the ${{\sigma }_{R\phi }}$ term of Fig. 13 in \citet[][]{2016MNRAS.461.2383P}. Observational evidence for the trend of $\sigma_{Rz}$  outside the solar neighborhood can be seen in \citet{2012A&A...547A..71P}. If the trends of $\bm{\sigma }=\bm{\sigma }\left( R,\phi ,z \right)$ is one of the challenges of the modern stellar dynamics, the slopes of these elements in the configuration space,  ${{\nabla }_{\bm{x}}}\bm{\sigma }$, are even more difficult to determine from observations to date. In GalMod (ver 16.2) we implemented the following radial-azimuthal dependence for spiral arms CSPs \citep[see also Eqs.(17), (18) and (70) in][]{2016MNRAS.461.2383P}:
\begin{equation}\label{I11}
\begin{aligned}
  & \sigma _{RR}^{2}=\overline{v_{R}^{2}}-\bar{v}_{R}^{2}, \\
 & \sigma _{\phi \phi }^{2}=\overline{v_{\phi }^{2}}-\bar{v}_{\phi }^{2}, \\
 & \sigma _{R\phi }^{2}=\overline{{{v}_{R}}{{v}_{\phi }}}-{{{\bar{v}}}_{\phi }}{{{\bar{v}}}_{R}}, \\
 \end{aligned}
\end{equation}
with
\begin{equation}\label{I12}
\begin{aligned}
  & \overline{v_{R}^{2}}=\frac{{{\Sigma }_{0}}}{\Sigma }\sigma _{RR}^{2}-\frac{{{\Sigma }_{0}}}{\Sigma }{{\Phi }_{1}}\left( 1-\text{sin}{{\text{c}}^{-1}}\left( \nu \pi  \right)\left( _{\left( \tfrac{1}{2},1 \right)}{{{\hat{F}}}_{\left( 1-\nu ,1+\nu  \right)}}-2{{X}^{2}}\left( _{\left( \tfrac{3}{2},2 \right)}{{{\hat{F}}}_{\left( 2-\nu ,2+\nu  \right)}}-{{3}_{\left( \tfrac{5}{2},3 \right)}}{{F}_{\left( 3-\nu ,3+\nu  \right)}} \right) \right) \right), \\
 & \overline{v_{\phi }^{2}}=\frac{{{\Sigma }_{0}}}{\Sigma }v_{c}^{2}\left( 1-\frac{{{\Phi }_{1}}}{\sigma _{RR}^{2}}\left( 1-\text{sin}{{\text{c}}^{-1}}\left( \nu \pi  \right){{\ }_{\left( \tfrac{1}{2},1 \right)}}{{{\hat{F}}}_{\left( 1-\nu ,1+\nu  \right)}} \right) \right), \\
 & \overline{{{v}_{R}}{{v}_{\phi }}}=\left. \frac{{{\Phi }_{\text{sp}}}}{{{\sigma }_{RR}}}\frac{{{\Sigma }_{0}}}{\Sigma }\frac{\text{sin}{{\text{c}}^{\text{-1}}}\left( \nu \pi  \right)}{2}\frac{X}{\gamma }\left( \gamma {{v}_{c}}\left( _{\left( \tfrac{1}{2},1 \right)}{{{\hat{F}}}_{\left( -\nu ,\nu +2 \right)}}-{{\ }_{\left( \tfrac{1}{2},1 \right)}}{{{\hat{F}}}_{\left( \nu ,2-\nu  \right)}} \right)+ \right.\iota X{{\sigma }_{RR}}\left( _{\left( \tfrac{3}{2},2 \right)}{{{\hat{F}}}_{\left( 1-\nu ,\nu +3 \right)}}{{-}_{\left( \tfrac{3}{2},2 \right)}}{{{\hat{F}}}_{\left( \nu +1,3-\nu  \right)}} \right) \right), \\
\end{aligned}
\end{equation}
where the notation is as in the text and $\partial_{\bullet}$ is a compact notation for the partial derivative, $\iota$ is the complex units, and real part of the equations is taken when necessary \citep[see, e.g.,][for further details]{2016MNRAS.459.3182P}. The in/out of plane dependence of Eq.\eqref{I10} for spiral CSPs (and the in/out dependence for symmetric CSPs) is carried by the following functional forms:
\begin{equation}\label{I13}
\begin{gathered}
{\sigma _{ii}^2}\left( {R,z} \right) = \left\{ {\begin{array}{*{20}{c}}
	{{\sigma _{zz}^2}\left( {R,0} \right) + {{\left. {\frac{{\lambda \left( R \right)\left( {\sigma _{RR}^2 - \sigma _{zz}^2} \right)}}{{2{h_R}R}}} \right|}_{\left\{ {R,0} \right\}}}{z^2}}&{\frac{{{{\left\| {\mathbf{\sigma }} \right\|^2}}}}{{\bar v_\phi ^2}} \ll 1} \\ 
	{{\sigma _{ii}^2}\left( {R,0} \right)\left( {1 + \sqrt {\frac{{\left| z \right|}}{{2{h_z}}}} } \right)^2}&{i = R,\phi } 
	\end{array}} \right. \hfill \\
\sigma _{RR}^2\left( {R,0} \right) = \sigma _{RR,0}^2 \times \left\{ {\begin{array}{*{20}{c}}
  {{e^{ - \frac{R}{{{h_R}}}}}}&{} \\ 
  {{R^2}{e^{ - \frac{{2R}}{{{h_R}}}}}}&{{\text{i}}{\text{.c}}{\text{.}}} 
\end{array}} \right. \hfill \\ 
\sigma _{\phi \phi }^2\left( {R,0} \right) = \left( {1 + \frac{{\partial \ln {v_c}}}{{\partial \ln R}}} \right)\frac{{\sigma _{RR}^2\left( {R,0} \right)}}{2} \hfill \\ 
\sigma _{zz}^2\left( {R,0} \right) = \sigma _{zz,0 }^2{e^{ - \frac{R}{{{h_R}}}}} \hfill \\
\end{gathered}
\end{equation}
and
\begin{equation}\label{I131}
\begin{gathered}
  \sigma _{Rz}^2 = {\left. {{\partial _z}\sigma _{Rz}^2} \right|_{z = 0}}z \hfill \\
  \left\{ \begin{gathered}
  {\partial _z}\sigma _{Rz}^2 = \lambda \left( R \right){\left. {\frac{{\sigma _{RR}^2 - \sigma _{zz}^2}}{R}} \right|_{z = 0}} \hfill \\
  \lambda \left( R \right) = {\left. {\frac{{{R^2}{\partial _{R,z,z}}\Phi }}{{3{\partial _R}\Phi  + R{\partial _{R,R}}\Phi  - 4R{\partial _{z,z}}\Phi }}} \right|_{z = 0}}. \hfill \\ 
\end{gathered}  \right. \hfill \\ 
\end{gathered} 
\end{equation}
The second equation of \eqref{I13}(a) plays a role every time any of the dispersion tensor elements at the observer position $\sigma$ pass the 50\% of the value of the rotation curve. The first equation of \eqref{I13}(b) together with \eqref{I13}(d) offer a constant anisotropy radial profile adopted by GalMod in all the situations apart for the N-body i.c. generation mode (see Fig.\ref{FlowC}), where a constant-$Q$ model is assumed, with $Q$ being the Safronov-Toomre criterion \citep[][]{1960AnAp...23..979S,1964ApJ...139.1217T}. 
This approach is meant to be a "work in progress" formulation and information is automatically provided to the user on the formulation adopted.

The mock catalog is finally written to a file with standard MPI-IO techniques. A mail server contacts the user to provide a password and temporary access coordinates to download the mock catalog.

The GalMod computing flow proceeds in an almost identical manner in the case a mock catalog of stars (instead of star-like particles) is requested by the user through $\bm{I}$. If $\tilde{N}>0$, i.e., the selection cuts are not inconsistent (e.g., as consequence of the modeled galaxy being located outside the FoV, or because the magnitude/radial velocity cuts are incompatible with the distance), GalMod proceeds to distribute the stars randomly in accordance with the underlying DFs, i.e., extracting a random number from the DF representative of the density profiles as mentioned above (arrow 9, in Fig.\ref{FlowC}). In this case, the placeholders for $\tilde{N}>0$ point mass stars are sampled without any preselected cuts in mass, age-metallicity or velocity space. This number of stars can exceed the user request $\bm{I}$, $N$ (or be lower than it) because $N$ depends on the selection cuts. GalMod proceeds then (arrow 7, in Fig.\ref{FlowC}) to "inflate" these placeholders with true stars by assigning them the mass, the age, and the metallicity in agreement with the profiles of Eqs. \eqref{(2.6)}, \eqref{(2.8)}, \eqref{(2.10)}, \eqref{(2.12)}, \eqref{(2.14)}, \eqref{(2.16)}, \eqref{(2.18)} that were coherently used at the point of arrow 2, in Fig.\ref{FlowC}, and in agreement with the request $\bm{I}$. Velocity dispersions are then initialized with the same equations as in the i.c. generation case (arrow 10, in Fig.\ref{FlowC}).

Finally, post-processing cuts are applied to the mock catalog in agreement with the request $\bm{I}$. If the number of stars $\tilde{N}>0$ is not reached the procedure iterates until the star generated $N$ exceeds in number the maximum galactic number of stars expected $\tilde{N}$ for the FoV, i.e., $N \ge \tilde{N}>0$ or the number required by the user is reached.

We remark that $\bm{O}$ does not contain only a mock catalog, but it provides diagnostics of the underlying gravitational potential ${{\Phi }_{\text{tot}}}\left( \bm{I} \right)$ resulting from $\bm{I}$ (as the rotation curve profile, the Oort constants, the total mass, resonance location in the case spiral arms are required, etc.). These parameters are unique to GalMod (in comparison with Besan\c{c}on, and other models) and represent valuable information on the correctness of the resulting underlying potential. Furthermore, GalMod provides the user with the number $\tilde{N}$ that would be necessary to complete the DF sampling for each CSP, thus allowing the user to recover the expected number of stars for each arbitrary bin of interest, e.g., in color $\Delta c={{c}_{1}}\left( \bm{I} \right)-{{c}_{2}}\left( \bm{I} \right)$, radial velocity $\Delta {{v}_{r}}$, stellar gravity $\Delta {{\log }_{10}}g$, and so forth. The reason for this procedure is twofold. On the one hand, it avoids forcing GalMod to generate an enormous number of dots in the scatter plot diagrams as a representation of the stars for any interval of $c$ and $m$ when dealing with a very large FoV. The user is left free to generate even unrealistic surveys such as, e.g., 2MASS full-sky surveys down to $K=35$ mag. Such an unrealistic mock catalog would occupy several Petabyte just by filling any bin of, say, color and proper motion  $\left\{ \Delta c,\Delta {{\mu }_{b}} \right\}$ with “dots over dots.” This approach is useless and impossible to handle by the MySQL database of GalMod. Vice-versa, by providing the DF and the numbers $\tilde{N}$ for each CSP we allow the user to know the number of stars expected in each $\left\{ \Delta c,\Delta {{\mu }_{b}} \right\}$ without  generating them (and then counting them) or without any pre-imposed bin interval in color $\Delta c$, magnitude $\Delta m$, radial velocity $\Delta {{v}_{r}}$ or any other parameter of interest. On the other hand, any machine-learning technique nowadays requires computing CMDs (representative of billions of stars) billions of times to span huge parameter spaces: to generate scatter plots with millions of dots - billions of times - would then result in an impossible practicality \cite[see also Sec. 5 of ][]{2016MNRAS.461.2383P} (\footnote{The technique that allows us to generate CMDs synthetically for billions of stars is detailed extensively in a dedicated paper \citet[][]{2012A&A...545A..14P}, where a few examples of SSPs distribution function and synthetic CMDs representative of billions of stars are shown in their Figs.~1, 2, and 5. The code is freely available upon request to the authors.}).

Finally, when the mock catalog is complete, the IO procedures and delivery continue as in the i.c. generation case.

\end{appendix}

\end{document}